\documentclass{article}
\usepackage{jheppub} 
\usepackage{graphicx}
\usepackage{float}
\usepackage{caption}
\usepackage{amsmath}
\usepackage{amssymb}
\usepackage{amsthm}
\usepackage{mathabx}
 \usepackage{url}
 \usepackage{subcaption}
 \usepackage{cancel}
 \usepackage{appendix}
 \usepackage{tikz} 
 \usepackage{xparse}
 \usetikzlibrary{fit}
 \usetikzlibrary{fit,shapes.geometric}
 \usetikzlibrary{arrows.meta}
\usetikzlibrary{decorations.markings}
\usetikzlibrary{calc}
\usetikzlibrary{positioning}

\def\beq{\begin{equation}}
\def\eeq{\end{equation}}
\def\beal{\begin{align}}
\def\eal{\end{align}}
\usepackage{xcolor}

\def\Eq#1{Eq.~(\ref{#1})}

\tikzset{
  double arrow/.style args={#1}{
    postaction={
      decorate,
      decoration={
        markings,
        mark=at position 0.84 with {
          \arrow[scale=#1]{Latex}
        },
        mark=at position 0.2 with {
          \node[scale=#1,transform shape,rotate=180] {
            \pgfarrowdraw{Latex}
          };
        }
      }
    }
  },
  double arrow-/.default=1.5
}

\tikzset{
  double arrow2/.style args={#1}{
    postaction={
      decorate,
      decoration={
        markings,
        mark=at position 0.9 with {
          \arrow[scale=0.7]{Latex}
        },
        mark=at position 0.15 with {
          \node[scale=0.7,transform shape,rotate=180] {
            \pgfarrowdraw{Latex}
          };
        }
      }
    }
  },
  double arrow-/.default=1.5
}

\tikzset{
  double arrow-/.style args={#1}{
    postaction={
      decorate,
      decoration={
        markings,
        mark=at position 0.57 with {
          \arrow[scale=#1]{Latex}
        },
        mark=at position 0.43 with {
          \node[scale=#1,transform shape,rotate=180] {
            \pgfarrowdraw{Latex}
          };
        }
      }
    }
  },
  double arrow-/.default=1.5
}

\tikzset{
  ->-/.style args={#1}{
    postaction={
      decorate,
      decoration={
        markings,
        mark=at position 0.65 with {\arrow[scale=#1]{Latex}}
      }
    }
  },
  ->-/.default=1
}
\tikzset{
  ->--/.style args={#1}{
    postaction={
      decorate,
      decoration={
        markings,
        mark=at position 0.52 with {\arrow[scale=#1]{Latex}}
      }
    }
  },
  ->-/.default=1
}

\tikzset{
  -->-/.style args={#1}{
    postaction={
      decorate,
      decoration={
        markings,
        mark=at position 0.76 with {\arrow[scale=#1]{Latex}}
      }
    }
  },
  ->-/.default=1
}

\newcommand{\Tube}[2][]{%
\node[
    draw,
    thick,
    rounded corners=6pt,
    inner sep=6pt,
    fit=#2,
    #1
] {};
}

\NewDocumentCommand{\TriGraph}{O{} m m m}{%
\begin{scope}[shift={(#2)},scale=#3]

\coordinate (B) at (0,0.35);
\coordinate (A) at (-0.3,-0.17);
\coordinate (C) at (0.3,-0.17);

\fill (A) circle (0.7pt);
\fill (B) circle (0.7pt);
\fill (C) circle (0.7pt);

#3 
#4 

#1 

\end{scope}
}

\NewDocumentCommand{\BoxGraph}{ O{black} O{0.7} O{0,0} m m }{%
\begin{scope}[shift={(#3)},scale=#4]

\def\boxcolor{#1}
\def\circlesize{#2}

\coordinate (A) at (-0.3,0);
\coordinate (B) at (0,0.3);
\coordinate (C) at (0.3,0);
\coordinate (D) at (0,-0.3);

\fill[\boxcolor] (A) circle[radius=\circlesize pt];
\fill[\boxcolor] (B) circle[radius=\circlesize pt];
\fill[\boxcolor] (C) circle[radius=\circlesize pt];
\fill[\boxcolor] (D) circle[radius=\circlesize pt];

#5
\end{scope}
}

\newcommand{\ThreeTree}[4]{%
\begin{scope}[shift={(#1)},scale=#2]
\coordinate (A) at (-0.5,0);
\coordinate (B) at (0,0);
\coordinate (C) at (0.5,0);
\fill (A) circle (1pt);
\fill (B) circle (1pt);
\fill (C) circle (1pt);
\draw [thick] (A)--(B);
\draw [thick] (B)--(C);
#3 
#4 
\end{scope}
}

\tikzset{
  double arrow/.default=1,
  double arrow2/.default=1,
  double arrow-/.default=1,
  ->-/.default=1,
  ->--/.default=1,
  -->-/.default=1
}

\theoremstyle{definition}

\theoremstyle{remark}

\newcommand{\valencia}{Instituto de F\'{\i}sica Corpuscular, Universitat de Val\`{e}ncia -- Consejo Superior de Investigaciones Cient\'{\i}ficas, Parc Cient\'{\i}fic, E-46980 Paterna, Valencia, Spain.}

\title{The geometry of multiloop Feynman Integrals from the Scattering Facet}
\author[a]{Jos\'e R\'{\i}os-S\'anchez}
\author[a]{and Germ\'an Rodrigo}
\affiliation[a]{\valencia}

\emailAdd{jose.rios@ific.uv.es}
\emailAdd{german.rodrigo@csic.es}

\abstract{The Scattering Facet (SF) of the Cosmological Polytope is a positive geometry defined from a Feynman diagram and encoding its combinatorics. Its canonical differential form reproduces the result of the integration over the energy components of loop momenta in scalar Feynman integrals, and is therefore closely related to the Loop-Tree Duality (LTD). In this article, we present a novel description of the SF as a base-fiber geometry. The fibers are a family of polytopes called deformed graphical zonotopes, which have a natural interpretation in terms of the causal evolution of the underlying Feynman diagram, and capture interesting information on scattering amplitudes. This decomposition of the SF induces a description of the canonical form leading to the so-called causal representation. Moreover, it provides new mathematical tools that allow us to characterize all its regular triangulations. Canonical forms obtained from these triangulations correspond to spanning tree representations, matching the original LTD amplitudes expressed as a sum over spanning trees. This geometrical viewpoint provides us with new methods to obtain the causal and spanning tree representations, without the need of performing any integration.
}

\date{\today}
\preprint{\today}

\begin{document}

\maketitle
\section{Introduction}

Computing higher order perturbative contributions to scattering amplitudes in Quantum Field Theory has become essential in modern High-Energy Physics (HEP) to test the Standard Model (SM) of particle physics and unveil potential signals of new physics. Motivated by the excellent performance of the CERN's Large Hadron Collider (LHC), this has led to a deep theoretical study of Feynman integrals, as well as the development of alternative mathematical frameworks, in the quest to overcome current bottlenecks. The recent program of positive geometries has opened the door for a completely new way to calculate physical observables based purely on geometry.

The standard method with Feynman diagrams involves introducing virtual off-shell degrees of freedom, which are not detectable in the final state and need to be integrated out over all possible configurations. This leads to complicated intermediate expressions which one could avoid by working with purely on-shell states~\cite{Bern:1994zx,Britto:2005fq}. The search for formulations based purely on on-shell states has a long history, and positive geometries~\cite{Arkani-Hamed:2017tmz,Arkani-Hamed:2013jha,Arkani-Hamed:2017mur} provide a geometric realization of this idea. These formulations often bypass intermediate virtual and gauge-redundant configurations and admit descriptions that do not rely explicitly on spacetime locality.

The connection between positive geometries and scattering amplitudes is provided by top-dimensional differential forms in the space of external kinematics variables, called canonical forms~\cite{Arkani-Hamed:2017tmz}. They are in one-to-one correspondence with a given positive geometry. The Amplituhedron was first introduced in Ref.~\cite{Arkani-Hamed:2013jha} to describe scattering amplitudes in $\mathcal{N}=4$ SYM, and the canonical form of the ABHY Associahedron~\cite{Arkani-Hamed:2017mur} encodes the tree-level biadjoint scalar amplitude as a sum over cubic graphs. 

Recently, a geometry called the Cosmological Polytope (CP) was introduced in Ref.~\cite{Arkani-Hamed:2017fdk}. A CP is related to a Feynman diagram $\Gamma$ representing the cosmological correlations. It is of dimension $E+V-1$, where $E$ and $V$ are the edges and vertices of $\Gamma$, respectively. Its canonical form gives the so called wavefunction of the universe in a flat space-time, which encodes correlations between quantum states in cosmology. The coordinates of the ambient space $Z$ are given by the external energies $x_i$ together with variables $y_i$ associated with the on-shell energies of internal propagators. In a flat space-time, the wavefunction of the universe has denominators that are
linear functions of these variables. This linearity geometrically realizes the wavefunction of the universe as the canonical form of a polytope. See Refs.~\cite{Benincasa:2022gtd,VicenteVazao:2025cwb} for an introduction to the CP and the relation to the wavefunction of the universe.

The canonical form of the SF does not directly give the fully integrated Feynman amplitude, but is rather the result of integrating only the energy components of loop four-momenta in a Feynman integral. We call these integrals Feynman energy integrals for short. The result therefore depends on the on-shell energies of internal particles $y_i$, which are functions of the loop three-momenta and their mass. This expression represents a first step for the evaluation and computation of scattering amplitudes, which then consists on integrating the remaining spatial three-momenta. In fact, it exhibits certain numerical advantages over the traditional Feynman integration. This has been intensively studied in the Loop-Tree Duality (LTD) literature. 

LTD, first introduced in Ref.~\cite{Catani:2008xa,Bierenbaum:2010cy,Bierenbaum:2012th}, is based on integrating in the complex plane one component of each loop four-momentum, yielding thus as a result the Feynman energy integral if the energy components are integrated, i.e. the same expression obtained from the canonical form of the SF. The different ways of taking iterated residues to perform the complex integration correspond to the different choices of setting on shell or cutting one propagator in each loop of the Feynman diagram, each choice corresponding to a different spanning tree of the graph. Then, the result is written as a sum over spanning trees. The expression for scalar LTD has a nice interpretation in terms of triangulations of the SF that we will study in detail in this article. Another way of getting this same amplitude is old-fashioned perturbation theory~(OFPT), based on the Lippmann-Schwinger formalism~\cite{Weinberg:1995mt}. 

The main inspiration for the results obtained in this article come from the causal properties of LTD~\cite{Buchta:2014dfa,
Tomboulis_2017,Runkel:2019yrs,aguileraverdugo2019causalityunitaritythresholdsanomalous,Aguilera-Verdugo:2020set,snowmass2020, Aguilera_Verdugo_2021,Ram_rez_Uribe_2021,capatti2020manifestlycausallooptreeduality,JesusAguilera-Verdugo:2020fsn,TorresBobadilla:2021ivx,Sborlini_2021,Kromin:2022txz,Ramirez-Uribe:2024rjg,LTD:2024yrb,Imaz:2025buf}. The causal representation in LTD is obtained by explicitly summing analytically over all nested residues~\cite{Aguilera-Verdugo:2020set}, which makes their singular structure manifestly causal and, in particular, suitable for numerical integration as it has no spurious poles of noncausal origin. This causal LTD expression can further be simplified by identifying those internal configurations that are causal compatible. In physical terms, this corresponds to acyclic settings~\cite{Ramirez-Uribe:2021ubp,Clemente:2022nll,Ramirez-Uribe:2024wua,Ochoa-Oregon:2025opz}, in which particles in the graph propagate along specific orientations, and the momentum flow does not decribe closed cycles. The final causal LTD expression is preceded by a simple prefactor, which is the measure that guarantees Lorentz-invariance of the amplitude. 

This decomposition of Feynman diagrams into a prefactor, which happens to be the canonical form of a simplex, and the rest of the causal LTD amplitude, corresponding to the canonical form of a family of polytopes called deformed graphical zonotopes, reveals a novel description of the SF as having a base-fiber structure. This has a mathematical explanation in terms of Cayley embeddings and Cayley polytopes~\cite{ubt_eref1116} that we will explain in detail, and gives useful tools for the study of the SF.

The outline of the article is the following. In Section~\ref{sec:PolytopesIntegrals}, we explain the relation between the SF and Feynman energy integrals, involving the study of their singular structure. Section~\ref{sec:GraphicalZonotope} presents the graphical zonotope and its deformations, which is closely related to the SF, and whose combinatorics and face structure encode interesting physical properties of scattering amplitudes. Then, we explain and prove in Section~\ref{sec:CayleyPolytope} the base-fiber structure of the SF by treating it as a Cayley polytope. This geometric structure translates into a decomposition of the canonical form that leads to the causal representation, studied in detail in Section~\ref{sec:CausalRepresentation}. We give a geometrical interpretation of this representation in terms of subdivisions of dual polytopes and explain how to obtain it without the need of the complex integration and partial decomposition machinery appearing in causal LTD. We also discuss a way to express the causal representation free of any spurious pole so that it becomes optimal for numerical integration. Finally, Section~\ref{sec:SpanningTree} shows how the representation of the canonical form as a sum of spanning trees corresponding to the original LTD expression is simply a sum over simplices in a triangulation of the SF. We also obtain this spanning tree representation with two different methods, one using graphs, and the other involving linear algebra and cones, which can be reduced to operations with matrices.

Much of what we will treat in this article lies at the intersection of the CP literature and the LTD literature, which use different notation. In what follows, 
\begin{itemize}
    \item We note $\ell_i$ the loop four-momenta in the Feynman integral and $\ell_{i,0}$ their energy components, which are the integration variables of the Feynman energy integral. 
    \item We note $x_i$ the total energy of the external particles entering vertex $i$, where the label $i$ is often given to the vertex of the corresponding graph. In LTD, the notation used is $p_{i,0}$.
    \item We note $y_e$ the on-shell energy of the internal propagator $e$, i.e. $y_e = \sqrt{{\bf q}_e^2+m_e^2}$, where the three-momentum ${\bf q}_e$ is a linear combinations of the loops' and external particles' three-momenta. The mass of the corresponding internal propagator is $m_e$. In the integral representation of the SF a complex prescription  $\imath 0$ is added to the denominators to make the integral well-defined. We incorporate this prescription inside the on-shell energy by defining a variable $q_{e,0}^{(+)} = \sqrt{{\bf q}_e^2+m_e^2-\imath 0}$, appearing in the LTD literature. Then, $y_e$ is the real part of $q_{e,0}^{(+)}$.
    \item We note $\lambda_{\dots}$ the denominators appearing in both the canonical form of the SF and in the Feynman energy integrals. They depend on the on-shell energies, which in the integral representation are $q_{e,0}^{(+)}$, as they carry the complex prescription, while in the canonical form they are the real variables $y_e$. We use $\lambda_{\dots}$ to denote both, since it is clear from the context whether or not it carries the complex prescription. The notation $\lambda_{\dots}$ is standard in LTD literature and is sometimes denoted by $E_{\dots}$ in the CP literature.
    \item We use the words loop and cycle interchangeably in this article to refer to a cycle in a graph, since we dig in the mathematical literature of graph combinatorics at some points. The term "loop" should therefore not be confused with its mathematical meaning. 
\end{itemize}

\section{Polytopes and contour integrals \label{sec:PolytopesIntegrals}}

The variables parametrizing Feynman energy integrals are the external energies at vertices $x_i$, and the on-shell energies of virtual particles $y_e$. Then, the SF lives in the space $Z$ of external parameters of dimension $d$, having $x_i$ and $y_e$ as coordinates.

The SF is defined from a graph $\Gamma$, and is denoted $\mathcal{S}_\Gamma$. We will show that the corresponding Feynman energy integral is, up to a global factor, the \textit{canonical function} of $\mathcal{S}_\Gamma$. Ref.~\cite{Arkani-Hamed:2017tmz} gives a rigorous mathematical description of canonical forms of positive geometries. The canonical form is a top-dimensional differential form. We have
\begin{equation}
    \Omega(\mathcal{S}_\Gamma)=\tilde{\Omega}(\mathcal{S}_\Gamma)\langle Z d^d Z\rangle 2^E/d!,
    \label{eq:CanonicalFunction}
\end{equation}
where $\langle Z d^d Z\rangle \equiv \mathrm{det}(ZdZ\dots dZ)$ is the standard projective top-form measure, and $E$ is the number of edges of $\Gamma$. The function $\tilde{\Omega}(\mathcal{S}_\Gamma)$ is called the \textit{canonical function}. We did not include the $2^E/d!$ factor in the canonical function, as opposed to other works on positive geometries, in order not to carry it in the rest of this article. The canonical function is closely related to the Feynman energy integral as it will be defined in Eq. (\ref{eq:FeynmanIntegral}), i.e.
\begin{equation}
    \tilde{\Omega}(\mathcal{S}_\Gamma)=(-1)^E\mathcal{I}_\Gamma.
    \label{eq:IntegralCanonicalForm}
\end{equation}

For the canonical form to be projectively invariant, $\tilde{\Omega}(\mathcal{S}_\Gamma)$ must be a homogeneous function of degree $-(d+1)$. In the following, we will usually employ the canonical function instead of the canonical form in order to drop the differential measure. We now briefly explain canonical forms and their main properties.

\subsection{Canonical forms of polytopes}
\label{sec:canonicalforms}

We are interested in the geometrical properties of \textit{convex polytopes} (that we will often simply call polytopes, $P$). Convex polytopes are formed from the convex hull of $n$ points in $d$ dimensions, with $n>d$. We focus on positive geometries, which are equipped with a unique top-degree differential form, the canonical form $\Omega(P)$, such that~\cite{Arkani-Hamed:2017tmz} it is meromorphic, it has simple poles on the boundaries of $P$, the residue of $\Omega$ on any boundary is the canonical form of that boundary and there are no poles elsewhere.

In many cases, as for the SF, it is more natural to work projectively. In fact, the facets are described by homogeneous quantities. Because of their homogeneous nature, they define a convex cone $\mathcal{C} \subset \mathbb{R}^{d+1}$. Since points related by positive rescaling are identified, the natural space is a projective space $\mathbb{P}^d$.

\paragraph{Projective polytopes} A \textit{projective polytope} $P$ is obtained by choosing an affine chart (or "gauge"), given by a vector $a$. It defines a hyperplane which intersects the cone $\mathcal{C}$ transversely giving a convex polytope. The intersection
\[
P^{(a)} := \mathcal{C} \cap \{X \mid a\cdot X =1\}
\]
is an ordinary $d$-dimensional polytope. Different charts correspond to different affine representatives of the same projective geometry. The canonical form is naturally projective and transforms covariantly under changes of chart, so the underlying projective geometry is chart-independent.
In this article, we choose the chart defined by the vector $a=(1,1,\dots,1)$.

Let $v_i \in \mathbb{R}^{d+1}$, $i=\{1,\dots,n\}$, be the generators of extremal rays of the cone $\mathcal{C}$, normalized so that $\sum_{k=1}^{d+1}(v_i)_k=1$. Any point $p$ in $\mathcal{C}$ can be written as $p=\sum_{i=1}^n C_i v_i \in \mathbb{R}^{d+1}$ for some $C_i \geq 0$, $i=\{1,\dots,n\}$. The choice of $C_i$ is not unique if the cone is not simplicial. The choice of chart given by the vector $a=(1,1,\dots,1)$ imposes $\sum_{i=1}^{n}C_i=1$ if we want to have $p\in P^{(a)}$. In other words, any $p\in P^{(a)}$ can be written as $p=\sum_{i=1}^{n} C_i v_i$, since it belongs to the desired chart given by $a$, i.e. $\sum_{i=1}^{d+1}(p)_i=1$. One can think of these $C_i$ coefficients as barycentric coordinates.

We define the $(d+1) \times n$ matrix $V$ with columns $v_i$ and the corresponding linear map $V:\mathbb{R}^n\rightarrow \mathbb{R}^{d+1}$. Because $p_i=\sum_{j=1}^n V_{ij} C_j$ belongs to the hyperplane $\sum_{i=1}^{d+1}(p)_i=1$, we have $P^{(a)}=V(\Delta ^{n-1})$, where $\Delta ^{n-1}$ is the $(n-1)$-dimensional simplex $\Delta^{n-1}=\left\{C_i \geq 0, \sum_i C_i=1\right\}$.

\paragraph{Triangulations} A practical way to compute canonical forms is by triangulation. Suppose the polytope $P$ is decomposed into simplices:
\[
P=\bigcup_{\alpha} \Delta_\alpha,
\]
with disjoint interiors. Then the canonical form is additive:
\[
\Omega(P)=\sum_\alpha \Omega(\Delta_\alpha).
\]
This is particularly useful because the canonical form of a simplex is known explicitly and straightforward to write. Spurious poles introduced by internal triangulation boundaries cancel in the sum, leaving only poles on the true boundaries of $P$.

\paragraph{Dual polytopes} The canonical form of a polytope $P$ evaluated at a point $z \in \mathrm{Int}(P)$ has a natural interpretation in terms of the volume of a related polytope, called \textit{dual polytope}, which depends on the evaluation point $z$, denoted by $P_z^\ast$. It is defined by~\cite{Mazzucchelli:2025gyg}
\begin{equation}
P_z^\ast=(P-z)^\circ=\left\{y \in \mathbb{R}^d:\left\langle x-z, y\right\rangle \geqslant-1, \forall x \in P\right\},
\end{equation}
Each facet of $P$ corresponds to a vertex of $P_z^\ast$, and each vertex of $P$ corresponds to a facet of $P_z^\ast$. Inclusion relations of their face lattices are reversed.

A remarkable fact is that the canonical function associated with $P$ can be interpreted as the normalized volume of this dual polytope with respect to a point $z$. Then,
\begin{equation}
    \tilde{\Omega}(P)(z) \propto \mathrm{Vol}(P_z^\ast).
\end{equation}
The proportionality symbol appears because $\tilde{\Omega}(P)$ can be defined up to some factor for convenience. This is the case in Eq. (\ref{eq:CanonicalFunction}), where the canonical function of the SF absorbs a factor $d!/2^E$.

Because volumes are additive too, one can also get the desired canonical function by triangulating the dual polytope, which does not always correspond to a triangulation of the original polytope. This is the way one gets the causal representation, as it will be explained in Section~\ref{sec:CausalRepresentation}. In fact, all of this is closely related to scattering amplitudes: in the case of the SF, boundaries encode the physical singularity structure and factorization channels, logarithmic singularities reflect locality and unitarity, and triangulations/subdivisions of the polytope correspond to different ways in which one can write the scattering amplitude.

The relation between Feynman energy integrals and the SF given by Eq.~(\ref{eq:IntegralCanonicalForm}) is not obvious a priori. We will see how one can get this polytope from the singularities of the integral. Conversely, Feynman energy integrals are simply contour integral representations of the SF. 

\subsection{The scattering facet's boundary as a Landau variety \label{sec:Landau}}

Consider a connected graph $\Gamma$ with $N$ vertices, $E$ edges, so that $L=E-N+1$ is the number of loops. We denote by $E(\Gamma)$ and $V(\Gamma)$ its edge set and vertex set, respectively. Let us call $\ell_{i}$, with $i=\{1,\dots,L\}$, the loop four-momenta, $\ell_{i,0}$ their energy components and $p_i$ for $i=\{1,\dots,N\}$ the total external four-momentum entering vertex $i$, with energy components $x_i \equiv p_{i,0}$. This corresponds to a momentum assignment with respect to a reference graph orientation $\vec{\Gamma}$. The mass of propagator $e$ is $m_e$, and its four-momentum is given by the linear combination
\beq
    q_{e}\equiv \sum_{j=1}^L \beta_{j,e} \ell_{j}+\sum_{k=1}^N \gamma_{e,k} p_k
    \label{eq:MomentaAssig}
\eeq
for $\beta_e\in \{-1,0,1\}^L$, $\gamma_e \in \{-1,0,1\}^N$. Let $q_{e,0}$ be their energy components, and $y_e\equiv \sqrt{\vec{q}_{e}^2+m_{e}^2}$ their on-shell energies. The imaginary prescription is added for the integration contour to avoid the poles of the integrand. The Feynman propagator associated to the propagator with momentum $q_e$ can be written
\begin{equation}
    G_{\rm F}(q_e)=\frac{1}{q_e^2-m_e^2 + \imath 0}=\frac{1}{\left(q_{e,0}-q_{e,0}^{(+)}\right)\left(q_{e,0}+q_{e,0}^{(+)}\right)} \,,
    \label{eq:DenomGf}
\end{equation}
where $q_{e,0}^{(+)} = \sqrt{{\bf q}_e^2+m_e^2-\imath 0}$ is the positive energy pole, obtained by including the Feynman complex prescription inside the square root of $y_e$. We define
\begin{equation}
     G_{\rm F}^{(\pm)}(q_{e})\equiv \frac{1}{q_{e,0}\mp q_{e,0}^{(+)}} . 
    \label{eq:DenomGf+-}
\end{equation}
Then, the scalar Feynman energy integral is
\begin{equation}
    \mathcal{I}_\Gamma = \int_{\ell_{1,0},\dots,\ell_{L,0}} \prod_{e\in E(\Gamma)} G_{\rm F}^{(+)}(q_{e})G_{\rm F}^{(-)}(q_{e})~, \quad \int_{\ell_{i,0}} = \frac{1}{2\pi \imath} \int d\ell_{i,0}.
    \label{eq:FeynmanIntegral}
\end{equation}

The Landau equations~\cite{Landau:1959fi} determine the singularities of the integral. There can be values of the external parameters for which the integrand is singular, but the integration contour can be deformed away from the real axis, leading to a finite result. Only values for which there is a \textit{pinch} of the integration contour that makes its deformation impossible yield a singularity of the integral. 

The geometric picture drawn by the singularity structure of the integral is better understood with the mathematical framework explained in Ref.~\cite{Pham2011Singularities}. The space $Z\subset\mathbb{R}^{E+N-1}$ of external parameters, with coordinates $x_i\in \mathbb{R}$, $i\in\{1,\dots,N-1\}$ (because of energy conservation), $y_e\in \mathbb{R}_{\geq 0}$, $e\in\{1,\dots,E\}$, constitutes the base space. The total space $M$ includes additionally the space $\mathbb{R}^L$ of integration variables $\ell_{i,0}$. The function $f(M)\equiv \prod_{e\in E(\Gamma)} G_{\rm F}^{(+)}(q_{e})G_{\rm F}^{(-)}(q_{e})$, the integrand of $\mathcal{I}_\Gamma(Z)$, is defined in it. Then there is a natural projection $\pi:M\rightarrow Z$. The fiber over each point of $Z$ is parametrized by the integration variables $\ell_{i,0}$.

Therefore, integrating $f(M)$ over the fiber $\pi^{-1}((\mathbf{x},\mathbf{y}))$ over a certain point $(\mathbf{x},\mathbf{y})\in \mathbb{R}^{E+N-1}$ of $Z$ yields the result of the integral $\mathcal{I}_\Gamma(\mathbf{x},\mathbf{y})$. Projecting to $Z$ the space of solutions to the Landau equations in $M$ yields the so-called \textit{Landau variety}, which gives the set of singularities of the integral. This just means that if a point in $Z$ is a pole of the integral then there is at least one set of values for the integration variables $\ell_{i,0}$ which satisfies the Landau equations for the value of the external parameters given by $Z$\footnote{For the scalar Feynman energy integrals, the facet hyperplanes of the SF coincide with the real codimension-one loci of the Landau variety. In the case of a generic integral, being a solution to the Landau equations is only a necessary condition for a point to be a singularity of the integral.}. 

The case of the Feynman energy integral is particularly simple. It is called a \textit{linear pinching}, i.e. a pinching that happens at an intersection of $n+1$ hypersurfaces in $n$ integration variables. The resulting singularities are then simple poles~\cite{Pham2011Singularities}, giving a rational function, and so the integral has no multi-valuedness. This matches the logarithmic behaviour of canonical forms. Remarkably, the real part of the Landau variety given by the integral coincides with the boundary arrangement of a convex projective polytope: the SF.

Now, let us study the loci of Landau singularities. This will give us hints
of the underlying polytope. It is
convenient to introduce the signed linear denominator functions
\begin{equation}
D_e^{(+)}\equiv y_e-q_{e,0},
\qquad
D_e^{(-)}\equiv y_e+q_{e,0},
\end{equation}
so that, up to the irrelevant overall sign of the first factor,
\begin{equation}
-G_{\rm F}^{(+)}(q_e)\sim\frac{1}{D_e^{(+)}-\imath0},
\qquad
G_{\rm F}^{(-)}(q_e)\sim\frac{1}{D_e^{(-)}-\imath0}.
\end{equation}

We associate the real parameters $\alpha_e^{(+)} \geq 0$ and $\alpha_e^{(-)} \geq 0$ to these denominators. All $\alpha$ cannot be zero simultaneously. The Landau equations applied to $\mathcal{I}_\Gamma$ are
\begin{align}
\alpha_e^{(\kappa)} \big(q_{e,0} - \kappa y_e \big) &= 0, \quad e=1,\dots,E, \label{eq:LandauOS}\\[6pt]
\sum_{e\in E(\Gamma)} (-\alpha_e^{(+)}+\alpha_e^{(-)}) \beta_{j,e} &= 0, \quad j=1,\dots,L,
\label{eq:LandauDeriv}
\end{align}
with $\kappa=\{+,-\}$. Pinches are related to solutions of these equations for non-zero values of some of the $\alpha_e^{(\kappa)}$ parameters. 
There are two families of minimal solutions, i.e. inclusion-minimal subsets of non-zero $\alpha_e$ solving these equations: vanishing on-shell energies and vanishing causal thresholds

The vanishing of on-shell energies is a family of obvious minimal solutions to the Landau equations : setting $\alpha_e^{(+)}=\alpha_e^{(-)}>0$ for any non-bridge edge $e$ (for a bridge this solution would not be minimal). We call bridge an edge such that deleting it disconnects the graph. Then Eqs.~(\ref{eq:LandauOS}) force $y_e=0$ and require $q_{e,0}=0$, which can be satisfied for any edge participating in a loop. Therefore, the integral has singularities at $y_e=0$ for any edge $e$ that is not a bridge. 

For the other family of solutions, set $\alpha_e=-\alpha_e^{(+)}+\alpha_e^{(-)}$. Then, nonzero solutions of $\sum_{e\in E(\Gamma)} \alpha_e \beta_e=0$, $j=1,\dots,L$, where $\beta_e$ is treated as a vector in $\mathbb{R}^L$, are solutions to (\ref{eq:LandauDeriv}). This set of equations imposing linear dependences where we keep track of the sign of the coefficients (to know which of the $\alpha_e^{(+)}$ or $\alpha_e^{(-)}$ needs or not to be strictly positive) is controlled by an oriented matroid (see Appendix \ref{app:OrientedMatroids} for a brief introduction to oriented matroids and the graphic and cographic oriented matroids). Note that the matrix $\beta$, defined as the $L \times E$ matrix with components $\beta_{j,e}$, is the \textit{cycle matrix} of the directed graph $\vec{\Gamma}$. The rows of $\beta$ form a basis of $\mathrm{ker} B$, the kernel of the incidence matrix of $\vec{\Gamma}$. Therefore, the underlying oriented matroid is the cographic oriented matroid.

We now consider solutions for which we do not have, for any $e \in E(\Gamma)$, $\alpha_e^{(+)}>0$ and $\alpha_e^{(-)}>0$ simultaneously. Then, for every edge, the solution is associated to either the sign "+" if $\alpha_e^{(+)}>0$, "-" if $\alpha_e^{(-)}>0$ and "0" if $\alpha_e^{(+)}=\alpha_e^{(-)}=0$. Therefore, the set of nonzero $\alpha$ such that solutions to the Landau equations exist can be written as signed sets on the edge set of the graph : $\{+,-,0 \}^{E(\Gamma)}$. It can itself be represented by a partial orientation of the graph, where "0" means an unoriented edge and "+" and "-", respectively, tell us whether to keep or reverse a reference edge orientation of $\Gamma$.

Minimal solutions to Eq.~(\ref{eq:LandauDeriv}) are signed circuits of the cographic oriented matroid. This just means that we look at the inclusion-minimal subsets of propagators whose $\beta_e$ vectors are linearly dependent, and we keep track of the signs. The \textit{oriented bonds}, or signed circuits of the cographic oriented matroid correspond to oriented minimal cuts of the graph. These are proper oriented binary partitions of the vertex set of the graph where both resulting partition subsets induce connected subgraphs. In fact, if it was not the case, it could be decomposed into the union of at least two different proper partitions satisfying this, and would not be minimal. Non-minimal solutions to these stationary Landau equations involve unions of oriented bonds\footnote{By union of oriented bonds we mean a solution to the Landau equations for which the subset of non-vanishing $\alpha_e^{(\kappa)}$ parameters is the union of the subsets of non-vanishing $\alpha_e^{(\kappa)}$ parameters of these minimal solutions.}. Graphically, they correspond to partitions of the vertex set which need not be binary and whose parts need not induce connected subgraphs. More precisely, their sign vectors are covectors of the graphic oriented matroid.

The set of Eqs. (\ref{eq:LandauDeriv}) determines which $\alpha_e^{(\kappa)}$ are non-zero. Call $\mathcal{D}$ the set of denominators with non-zero $\alpha$, and $\chi(e)$ the sign associated to the edge $e$. The denominators in $\mathcal{D}$ correspond to the oriented edges that are cut by the oriented minimal cut and their orientation follows that of the cut. Eq.~(\ref{eq:LandauOS}) requires these denominators to vanish, thereby setting the corresponding propagators on shell, while keeping track of the sign, and solve the resulting set of equations. We get
\begin{equation}
    q_{e,0}=\chi(e) y_e, \quad \forall e \in \mathcal{D}.
    \label{eq:landau2}
\end{equation}

Because a minimal cut involving $k$ edges fixes $k-1$ independent linear combinations of loop momenta (this is at the origin of a linear pinching), this system of equations fixes every values of the loops' energies involved in the cut, and there is an additional equation imposing relations between the external parameters $x_i$ and $y_e$. Define $S$ to be the subset of vertices that characterizes the minimal cut (the partition subset that is first in an ordered partition, and $\delta(S)$ the non-empty subset of edges cut. Then Eq. (\ref{eq:landau2}) imposes 
\begin{equation}
    \lambda_S=\sum_{e\in \delta(S)} y_e + \sum_{i \in S} x_i  = 0.
    \label{eq:causalthreshold}
\end{equation}

These $\lambda_S$ are often called \textit{energy denominators}, or \textit{causal thresholds} in the LTD literature, as their vanishing is related to threshold singularities. In practice, we will often note them $\lambda_{(i,j,\dots)}$, where $i,j,\dots$ are the vertices belonging to the subset $S$. From now on, we also define $L_{S}$ to be the facet of the SF corresponding to the oriented bond labeled by $S$:
\begin{equation}
    L_S= \{\lambda_{S}=0\} \cap \mathcal{S}_\Gamma.
    \label{eq:DefLS}
\end{equation}
Consequently, we expect the lowest order singularities of the integral $\mathcal{I}_\Gamma$ to appear when $y_e=0$, for any edge $e$ that is not a bridge, and when $\lambda_S=0$, for any vertex subset $S$ where the subgraph induced by it and its complement $V \backslash S$ are connected. These equations define the facets of the SF.

Looking at non-minimal solutions to the Landau equations introduces some difficulties, and complicated combinatorics. It may happen that the union of $k$ minimal solutions to the Landau equations imposes $k+1$ or more independent relations between external parameters. Geometrically, the singular locus is therefore not the intersection of the $k$ hyperplanes defined by the minimal solutions, it is of lower dimension. This is related to linear dependencies between the normals of the facets intersecting at a vertex of the polytope. Consequently,  a collection of candidate facet equations can have an intersection that is empty in the positive geometry or can meet only in unexpectedly high codimension. We will see later how this is captured by the graphical zonotope. It is quite remarkable that this Landau variety in the space $Z$ (real $x_i$ and $y_e$ positive) happens to be the boundary of a convex projective polytope, the SF.

The study of the Landau equations has revealed the (co)graphic oriented matroid controlling it. It is a useful tool to help us understand the singular structure of Feynman energy integrals even beyond minimal solutions, which can be encoded by signed sets on the edges of the graph. Understanding this would give us the SF face structure. We will discuss it in Sec.~\ref{sec:FaceStructure}.

\subsection{Contour integral representation of the scattering facet \label{sec:ContourIntegralRep}}

In Sec.~\ref{sec:canonicalforms}, we defined the map $V$ which sends coordinates $C_i$ associated to each vertex to points of the projective polytope $P$, in the chart given by the vector $a=(1,1,\dots,1)\in \mathbb{R}^{d+1}$. Applying it to the SF $\mathcal{S}_\Gamma$, which has $2E$ vertices~\cite{Benincasa:2021qcb} (self-loops do not count for this counting), gives a surjective linear map $V: \Delta^{2E-1} \rightarrow \mathcal{S}_\Gamma^{(a)}$, where $\mathcal{S}_\Gamma^{(a)}$ is the realization of $\mathcal{S}_\Gamma$ in the chart given by $a$ (from now on, we simply denote it by $\mathcal{S}_\Gamma$). This realization lives in the hyperplane
\beq
    H=\left\{ \sum_{i=1}^E y_i=1 \right\},
\eeq
so, it is of dimension $E+N-2$. Here, we used energy conservation $\sum_{i=1}^{N} x_i=0$. Working in the section $\sum_e y_e=1$ does not impose a physical restriction, since $\mathcal{S}_\Gamma$ is projective: a generic point can be brought to this section by an overall rescaling, and the result for arbitrary $x_i$ and $y_e$ is recovered from the homogeneity of the canonical function.

Now, let us characterize the matrix $V$. Let $\mathbf{x_i}$, $i=1,\dots,N$ and $\mathbf{y_e}$, $e=1,\dots,E$ be the standard basis vectors in $\mathbb{R}^{N+E}$. Then the Cosmological Polytope $\mathcal{P}_\Gamma$ of the graph $\Gamma$ can be defined as~\cite{Arkani-Hamed:2017fdk}
\beq
    \mathcal{P}_\Gamma=\operatorname{conv}\left(\bigcup_{e=\{i, j\} \in E}\left\{-\mathbf{y_e}+\mathbf{x_i}+\mathbf{x_j},\mathbf{y_e}-\mathbf{x_i}+\mathbf{x_j}, \mathbf{y_e}+\mathbf{x_i}-\mathbf{x_j} \right\} \right) .
    \label{eq:VerticesCP}
\eeq
We easily see that only the last two of the set of three vertices defined by each edge of $\Gamma$ lie in $\mathcal{S}_\Gamma$, since they live in the hyperplane satisfying energy conservation. The vertex definition of the SF is then~\footnote{This satisfies $\sum_{k=1}^{N+E}(v_i)_k=1$ for all vertices $v_i$, and so lives in the desired affine chart.}
\beq
    \mathcal{S}_\Gamma=\operatorname{conv}\left(\bigcup_{e=\{i, j\} \in E} \left\{\mathbf{y_e}+\mathbf{x_j}-\mathbf{x_i},\mathbf{y_e}+\mathbf{x_i}-\mathbf{x_j} \right\} \right) . 
    \label{eq:VerticesSF}
\eeq

We get $V$ by grouping in columns the vectors of each vertex $\mathbf{v_i}$ of $\mathcal{S}_\Gamma$, with coordinates $(x_1, \dots, x_{N},$ $ y_1, \dots, y_E)$. For the ordering, we choose a reference orientation $\vec{\Gamma}$ of $\Gamma$. If the edge $e$ (where $e$ should be seen as a number indexing an edge of the graph, after ordering of edges) is directed from vertex $i$ to vertex $j$ in $\vec{\Gamma}$, then the vector $\mathbf{y_e}+\mathbf{x_j}-\mathbf{x_i}$ goes in the $e$-th column, and the vector $\mathbf{y_e}+\mathbf{x_i}-\mathbf{x_j}$ goes in the $(E+e)$-th column. One can then see that 
\beq
    V=
    \begin{pmatrix}
        \hat{B} & -\hat{B} \\
        Id_{E} & Id_{E}
    \end{pmatrix},
    \label{eq:VMatrix}
\eeq
where $\hat{B}$ is a $(N-1)\times E$ reduced incidence matrix. It is obtained by deleting a row of the incidence matrix $B$ of the directed graph $\vec{\Gamma}$. This is due to momentum conservation, which makes $\rm{rank}(B)=N-1$ for a connected graph. $V$ is a $(E+N-1)\times 2E$ matrix of rank $E+N-1$.

Ref.~\cite{Arkani-Hamed:2017tmz} showed that canonical forms of convex projective polytopes have a representation as contour integrals over the coefficients $C_i \in \mathbb{R}_{\geq 0}$. We define $C_i^+=C_i$ and $C_i^-=C_{i+E}$, so $C=(C_1^+,\dots,C_E^+,C_1^-,\dots,C_E^-)$. From Ref.~\cite{Arkani-Hamed:2018bjr}, the contour integral representation of the SF gives the corresponding Feynman energy integral. Let us show this explicitly.

To evaluate the canonical form of the SF at a point $p$, we need to eliminate (indicated by a hat symbol) one coordinate $x_i$ because of momentum conservation. We define $\hat{p}=(x_1,\dots,\hat{x}_i,\dots,x_{N},y_1,\dots,$ $y_E)\in \mathcal{S}_\Gamma$, for $i \in \{1,\dots,N\}$. Then,
\begin{equation}
    \tilde{\Omega}(\mathcal{S}_\Gamma)(\hat{p}) =\frac{1}{2^E(2 \pi i)^{L}} \int_{\mathbb{R}^{2E}} \frac{d^{2E} C}{\prod_{j=1}^{2E} (C_j-\imath\epsilon_j/2)} \delta^{E+N-1}\left(\hat{p}-VC\right),
    \label{eq:ContourIntegral}
\end{equation}
with $\epsilon_j>0$ small constants. We choose $\epsilon_e=\epsilon_{e+E}>0$ for all $e=1,\dots,E$. This complex prescription in every denominator factor defines the contour of integration, which is moved away from the singularities of the integrand with the imaginary prescription in order to have a well-defined integral. Also, following \cite{Arkani-Hamed:2017tmz}, at finite $\epsilon_j$ the contour deformation $C_j \rightarrow C_j- \imath \epsilon_j/2$ is accompanied by an infinitesimal displacement of the evaluation point, $\bar{p}\equiv p+\frac{\imath}{2}V \epsilon $, with $\bar{p}\rightarrow p$ as $\epsilon_j \rightarrow 0^+$. In the following we suppress the bar, since $\bar{p} \rightarrow$ p when $\epsilon_j \rightarrow 0^+$.

Let us first look at what happens when $\mathcal{S}_\Gamma$ is a simplex. This means $2E=E+N-1$ which implies $E=N-1$, so the underlying graph is a tree (we assume connectedness of $\Gamma$) that we call $\mathcal{T}$. $\mathcal{S}_\mathcal{T}$ is of dimension $E+N-2$ and the map $V: \Delta^{2E-1} \rightarrow \mathcal{S}_\mathcal{T}$ is bijective. To simplify Eq.~(\ref{eq:ContourIntegral}), we want to find the solution to the system of equations $\hat{p}=VC$. This gives 
\begin{align}
    x_i&=\sum_{e \text{ incoming to }  i} \tilde{C_e}-\sum_{e' \text{ outgoing from }  i} \tilde{C_{e'}}, \\
    y_e&=C_e^++C_{e}^-,
\end{align}
where we defined $\tilde{C_e} \equiv C_e^+-C_{e}^-$, and $e=1,\dots,E$.

To solve for $\tilde{C_e}$, one defines the subset of vertices $S_e$ of the graph such that only the edge $e$ is cut, outgoing from $S_e$ and incoming to $V\backslash S_e$. Then one easily sees that $\sum_{i \in S_e} x_i=-\tilde{C_e}$. Therefore,
\begin{align}
    C_e^+&=\frac{1}{2}\left( y_e-\sum_{i \in S_e} x_i \right)=\frac{1}{2} \lambda_{V\backslash S_e} \\
    C_{e}^-&=\frac{1}{2}\left( y_e+\sum_{i \in S_e} x_i \right)=\frac{1}{2} \lambda_{S_e},
\end{align}
where the causal thresholds $\lambda$ were defined in \Eq{eq:causalthreshold}. In fact, the barycentric coordinate $C_e^\kappa$ of $\mathcal{S}_\mathcal{T}$ controls how far the point $p$ is from the facet opposed to its associated vertex in the simplex. Then, $C_e^\kappa=0$ means $p$ lies in a facet, which translates in a pole of the canonical form. Integrating over the delta functions in \Eq{eq:ContourIntegral}, and recalling the relation to the canonical function in \Eq{eq:CanonicalFunction}, we get
\beq
    \tilde{\Omega}(\mathcal{S}_\mathcal{T})(p) =\frac{1}{\prod_{e\in E(\Gamma)}\lambda_{S_e} \lambda_{V\backslash S_e}},
\eeq
where no imaginary prescription was needed because there is no integration contour. $\tilde{\Omega}(\mathcal{S}_\mathcal{T})$ is the canonical function of the tree graph $\mathcal{T}$.

Now let us study the integral representation for a general graph $\Gamma$, for which, in general, the map $V$ is not injective anymore. Using the result on tree graphs, we can deduce a particular solution $\hat{p}=VC^{(p)}$. Choose any spanning tree $\mathcal{T}$ of $\Gamma$. Then we can set $C^{+(p)}_{e}=C^{-(p)}_{e}=y_e/2$ for $e \notin E(\mathcal{T})$ and the map $V$ becomes injective when we fix $C^{+}_e-C^{-}_e=0$ for $e \notin  E(\mathcal{T})$. We define the cut $S_e^{(\mathcal{T})}$ to be the minimal cut of $\Gamma$ for which the only cut edge of $\mathcal{T}$ is $e$. Note that after the choice of a spanning tree, this subset is also defined in a loop graph. The particular solution is then 
\begin{align}
    C^{+(p)}_e&=\frac{1}{2}\left( y_e-\sum_{i \in S_e^{(\mathcal{T})}} x_i \right), \quad \text{ if } e \in E(\mathcal{T}), \\
    C^{-(p)}_{e}&=\frac{1}{2}\left( y_e+\sum_{i \in S_e^{(\mathcal{T})}} x_i \right), \quad \text{ if } e \in E(\mathcal{T}),\\
    C^{+(p)}_{e}&=C^{-(p)}_{e}=y_e/2, \quad \text{ if } e \notin E(\mathcal{T}).
\end{align}

A general solution to the system of equations $VC=0$ is given by the kernel of $V$:
\beq
    \rm ker(V)=\left\{\binom{\mathbf{-\beta}}{\mathbf{\beta}}: \mathbf{\beta} \in \rm ker (B)\right\}.
\eeq
The kernel of the incidence matrix of $\vec{\Gamma}$, of dimension $L$, encodes the cycle space of the graph. One can choose a basis for $\rm ker (B)$ such that each basis element $\mathbf{\beta}_i$ corresponds to a choice of cycle $\mathcal{C}_i$ in $\Gamma$, and a cyclic orientation $\vec{\mathcal{C}_i}$ of it. Then, 
\begin{equation}
    \mathbf{\beta}_{i,e}\equiv(\mathbf{\beta}_i)_e=\left\{\begin{aligned}
    1 & \text{ if } e \in \mathcal{C}_i \text{ and } e \text{ has a common orientation in } \vec{\mathcal{C}_i} \text{ and } \vec{\Gamma}, \\
    -1 & \text{ if } e \in \mathcal{C}_i \text{ and } e \text{ has a different orientation in } \vec{\mathcal{C}_i} \text{ and } \vec{\Gamma}, \\
    0 & \text{ if }e \notin \mathcal{C}_i.
    \end{aligned}\right.
\end{equation}

A basis of $\rm  ker (B)$ then corresponds to the set of vectors $\{\beta_1,\dots, \beta_L\}$ such that $\{\mathcal{C}_1,\dots, \mathcal{C}_L\}$ is a basis of cycles in $\Gamma$. Note that the choice of a spanning tree gives a cycle basis of the graph, so we use the same spanning tree $\mathcal{T}$ used to obtain the particular solution $C^{(p)}$. The assignment of signs to each edge of the graph, given by the map 
\begin{align}
    E(\Gamma) &\rightarrow \{1,-1,0\}^L\\
    e &\mapsto (\mathbf{\beta}_{1,e},\dots,\mathbf{\beta}_{L,e})
\end{align}
depends on the choice of basis of cycles, their orientations and the reference orientation. It corresponds physically to an assignment of loop momenta to the graph. A general solution to $VC^{(h)}=0$ is then 
\begin{align}
    C^{+(h)}_e&=-\frac{1}{2}\sum_{i=1}^L \ell_{i,0} \mathbf{\beta}_{i,e},\\
    C^{-(h)}_{e}&=\frac{1}{2}\sum_{i=1}^L \ell_{i,0} \mathbf{\beta}_{i,e}, \quad e=1,\dots,E
\end{align}
for some real coefficients $\ell_{i,0}$ for $i=1,\dots, L$.

The sum of the particular and homogeneous solution gives a general solution to $\hat{p}=VC$:
\begin{align}
    C_e^+&=\frac{1}{2}\left(  y_e-q_{e,0} \right),\\
    C_{e}^-&=\frac{1}{2}\left(  y_e+q_{e,0} \right), \quad e=1,\dots,E,
\end{align}
where the $q_{e,0}$ are the energy components of the internal propagators' momenta defined in Eq. (\ref{eq:MomentaAssig}). The expression for the $q_{e,0}$ depends on the choice of the spanning tree $\mathcal{T}$. In fact, there exists a choice of loop momenta assignment to the Feynman diagram such that $\gamma_{e,i}= 1$ if $i \in S_e^{(\mathcal{T})}$ and $e \in \mathcal{T}$, and $\gamma_{e,i}= 0$ otherwise, matching the particular solution we found.

Finally, note that the Jacobian of the transformation from the $(\mathbf{x},\mathbf{y})$ variables to the C variables gives a global $1/2^E$ factor upon integration over the delta functions in Eq. (\ref{eq:ContourIntegral}). Then, we write the canonical function as
  \begin{equation}
     \tilde{\Omega}(\mathcal{S}_\Gamma)(p) =  \int_{\ell_{1,0}, \cdots, \ell_{L,0}} \prod_{e \in E(\Gamma)}\frac{1}{ (q_{e,0}^{(+)}-q_{e,0})(q_{e,0}^{(+)}+q_{e,0})}=(-1)^E \mathcal{I}_\Gamma(p),
\label{eq:ContourFeynmanIntegral}
 \end{equation}
where $q_{e,0}^{(+)}$, defined after \Eq{eq:DenomGf}, was obtained by adding the complex prescription inside the $y_e$ variables.

 The complex prescription of the integral representation of the canonical form exactly matches the one used in LTD~\cite{Catani:2008xa}.

The final result for the canonical form of the SF matches the scalar Feynman energy integral (see \Eq{eq:FeynmanIntegral}) up to a sign. Note that the initial integration form (\ref{eq:ContourIntegral}) was dependent on the incidence matrix $B$ of a directed graph. Its columns define a vector configuration $\mathcal{V}$ which encodes the graphic oriented matroid. The linear dependencies between these vectors correspond to oriented cycles in the graph, i.e. signed circuits of the matroid. Upon integration of the delta functions, we work with the dual configuration of vectors $\mathcal{V}^\perp$: the $E$ vectors in $\mathbb{R}^L$ such that the rows of the corresponding $L\times E$ matrix form a basis of $\rm ker (B)$. This vector configuration is called its \textit{Gale transform}~\cite{MR85552}. Gale transforms encode the dual matroid. In our case, the linear dependencies of $\mathcal{V}^\perp$ relate to oriented cuts of the graph (we made use of it in Sec.~\ref{sec:Landau}), and so correspond to the cographic oriented matroid. The Gale transform, appearing as an integration of delta functions in the integral representation, explains why the graphic and cographic oriented matroid are both present in the study of the SF.

\section{The graphical zonotope \label{sec:GraphicalZonotope}}

We will show in this Section that the SF is deeply related to a polytope obtained from its defining Feynman diagram called the \textit{graphical zonotope}. The canonical form of the SF is obtained in a simple way from the canonical forms of deformations of the graphical zonotope. This is motivated by the causal LTD representation, which decomposes into a prefactor and the so-called reduced LTD expression, which is simply the canonical form of a deformed graphical zonotope. We define graphical zonotopes and their deformations and present some interesting properties.

The graphical zonotope has appeared very recently in the physics literature in relation to the CP. In Refs.~\cite{Baumann:2025qjx,Glew:2025ypb} the graphical zonotope encodes patterns appearing in the differential equations that give the wavefunction coefficients of conformally coupled scalars in power-law cosmologies. In Ref.~\cite{Glew:2026von}, they appear in the geometrical structure of a more complicated polytope that describes cosmological correlators in flat spacetime.

\subsection{Definitions and combinatorics \label{sec:DefCombinatoricsGZ}}

A \textit{zonotope} is a convex polytope built as a Minkowski sum of line segments. We can write 
\begin{equation}
    \mathcal{Z}=w_1 + \dots + w_n ,
\end{equation}
where $w_i\subset \mathbb{R}^d$ are line segments, called generators of the $d$-dimensional zonotope. Here, the plus signs denote the Minkowski sum of polytopes.

The \textit{graphical zonotope} is a zonotope defined from the edges of a graph $\Gamma=(V,E)$ with $N$ vertices and $E$ edges. The realization of the graphical zonotope we will use can be written as 
\begin{equation}
    \mathcal{Z}_\Gamma=\sum_{\{i j\} \in E}\left[\mathbf{\mathbf{x_j}}-\mathbf{\mathbf{x_i}}, \mathbf{\mathbf{x_i}}-\mathbf{\mathbf{x_j}} \right],
\end{equation}
where the $\mathbf{x_i}$ are the standard basis vectors in $\mathbb{R}^N$, with coordinates $x_i$. There is one independent coordinate $x_i$ for each vertex of the graph. Nonetheless, the graphical zonotope of a connected graph is always ($N-1$)- dimensional. From the form of the generators $\mathbf{x_j}-\mathbf{x_i}$ it can be seen that it lives in the hyperplane $\sum_{i=1}^{N} x_i =0$. Physically, $x_i$ represents the sum of energies of the external particles leaving vertex $i$ and this hyperplane sets energy conservation of the scattering process. Note that the definition of graphical zonotopes also applies to non-simple graphs\footnote{Simple graphs are graphs with at most one edge between any pair of vertices and no self-loops.}: multiple edges correspond to repeated generators and therefore affect the zonotope, while self-loops contribute the zero vector and can be discarded.

For a reference orientation $\vec{\Gamma}$ of a graph $\Gamma$, there is a vector $\mathbf{x_j}-\mathbf{x_i}$ associated to every edge of the graph, where the edge is oriented from vertex $i$ to vertex $j$. This defines a vector configuration, that can be encoded in a $\mathbb{R}^{N\times E}$ matrix, the incidence matrix $B$ of $\vec{\Gamma}$, as explained in Sec.~\ref{sec:ContourIntegralRep}. The linear dependencies between these vectors are encoded by the graphic (oriented) matroid (see Appendix~\ref{app:OrientedMatroids}) that plays an important role in the combinatorics of $\mathcal{Z}_\Gamma$. More generally, every zonotope arises from a vector configuration and therefore determines an oriented matroid. As any other polytope, $\mathcal{Z}_\Gamma$ can be defined as the convex hull of a set of vertices, and also by the equations of the hyperplanes corresponding to its facets.

\paragraph{Vertices} The vertices of $\mathcal{Z}_\Gamma$ correspond to being simultaneously at one endpoint of all of its defining generators $\left[\mathbf{x_j}-\mathbf{x_i}, \mathbf{x_i}-\mathbf{x_j} \right]$. Each endpoint of one of these segments corresponds to a choice of an orientation for that edge of $\Gamma$. But cycles in an oriented graph give linearly dependent generators. The sum of the vectors corresponding to the oriented edges in a directed cycle vanish:
\begin{equation}
    \sum_{\{i \rightarrow j\} \in C} \left(\mathbf{x_j}-\mathbf{x_i}\right)=0 \quad \Longleftrightarrow \quad C \text { is a directed cycle } .
\end{equation}
Therefore, vertices are in bijection with acyclic orientations of the graph. In the language of oriented matroids, the choice of an orientation for each edge of $\Gamma$ corresponds to a maximal signed set, and orientations without directed cycles translate into signed covectors. Then, the vertices of $\mathcal{Z}_\Gamma$ are the topes of the graphic oriented matroid.

The $i^{th}$ component of the vertex $v^{\vec{\Gamma}}$ corresponding to the acyclic orientation $\vec{\Gamma}$ of $\Gamma$ is 
\begin{equation}
    v_i^{\vec{\Gamma}}=\operatorname{indeg_{\vec{\Gamma}}(i)}-\operatorname{outdeg_{\vec{\Gamma}}(i)},
\end{equation}
where indeg and outdeg are, respectively, the indegree and outdegree of a vertex with respect to the orientation $\vec{\Gamma}$.

\paragraph{Faces} We define an oriented cut of order $k$ to be a partition of the vertex set of a graph
\begin{equation}
    V(\Gamma)=S_1 \sqcup \dots \sqcup S_{k+1}
\end{equation}
into $k+1$ non-empty subsets inducing connected subgraphs, together with an acyclic orientation of the quotient graph obtained by contracting each $S_a$ to a vertex. This is equivalent to orienting only the edges cut by the partition, in a way that for every pair of adjacent subsets $S_a$ and $S_b$, all edges connecting them are oriented in the same direction, and in an acyclic way. Then, $k$-codimensional faces of the graphical zonotope are in bijection with oriented cuts of order $k$~\cite{stanley2007introduction}.

To show that, we consider a point in $\mathcal{Z}_\Gamma$ obtained by weighting each vector $\mathbf{x_i}-\mathbf{x_j}$ by a factor $\rho_{ij} \in [-1,1]$. The faces of a convex polytope are given by the subset of points that maximize a linear functional $\phi$. One has by linearity
\beq
    \phi\left( \sum_{\{ij\}\in E(\Gamma)} \rho_{ij} (\mathbf{x_i}-\mathbf{x_j} )\right)=\sum_{\{ij\}\in E(\Gamma)} \rho_{ij} (\phi(\mathbf{x_i})-\phi(\mathbf{x_j}) ).
\eeq
Therefore, this linear function can be seen as a potential on the vertices of the graph, labelled by the variables $\mathbf{x_i}$. To maximize the potential, we choose $\rho_{ij}=1$ (resp. $\rho_{ij}=-1$) when $\phi(\mathbf{x_i})>\phi(\mathbf{x_j})$ (resp. $\phi(\mathbf{x_i})<\phi(\mathbf{x_j})$). These edges are assigned to a "$+$" sign (resp. "$-$" sign), while a "$0$" is assigned when $\phi(\mathbf{x_i})=\phi(\mathbf{x_j})$. Note that the resulting signed set depends on a reference orientation, coming from the choice $\mathbf{x_i}-\mathbf{x_j}$ instead of $\mathbf{x_j}-\mathbf{x_i}$ as generator of the zonotope. This maximization of the potential induces a partial orientation of the graph, where an edge $\{ij\}$ is oriented from $i$ to $j$ if $\phi(\mathbf{x_j})>\phi(\mathbf{x_i})$, and there is no orientation for all edges such that $\phi(\mathbf{x_i})=\phi(\mathbf{x_j})$. Since it comes from a potential, this partial orientation must be acyclic. It corresponds to a face of the graphical zonotope. The dimension of the face is  the dimension of $S(\phi)$, where 
\beq
S(\phi)=\rm span\{ \mathbf{x_i}-\mathbf{x_j} : \phi(\mathbf{x_i})=\phi(\mathbf{x_j}), \{ij\} \in E(\Gamma) \}.
\eeq

All of this has a nice interpretation in terms of oriented matroids~\cite{z-lop-95,Gruji2016CountingFO}: faces of zonotopes are in bijection with signed covectors of $\mathcal{M}_\Gamma$. In fact the signs "+", "-" and "0" defined above are a signed set of the oriented matroid, and the fact that they come from a potential implies that these signed sets describing faces cannot contain any directed cycle: they are signed covectors. A face $F$ encoded by a signed covector $X_F$ is the zonotope
\begin{equation}
    \mathcal{Z}_{F}=\sum_{\{ij\}\in X_F^0} \left[\mathbf{\mathbf{x_j}}-\mathbf{\mathbf{x_i}}, \mathbf{\mathbf{x_i}}-\mathbf{\mathbf{x_j}} \right] + \sum_{\{ij\}\in X_F^+} \mathbf{\mathbf{x_i}}-\mathbf{\mathbf{x_j}}+ \sum_{\{ij\}\in X_F^-} \mathbf{\mathbf{x_j}}-\mathbf{\mathbf{x_i}},
     \label{eq:FaceGZ}
\end{equation}
where $X_F^\kappa$, $\kappa\in\{+,-,0\}$, is the subset of edges with sign $\kappa$ in the covector $X_F$. Then, the geometry of $F$ corresponds to the graphical zonotope of the graph having $X_F^0$ as edge set. 

One can easily see that the minimal non-trivial case is for the linear functional $\phi$ to take only two distinct values, giving an oriented cut of order $1$. Let $A$ and $B$ be two complementary vertex subsets in $V$, both inducing connected subgraphs in $\Gamma$. Let $\phi(a_1)=...=\phi(a_{|A|})>\phi(b_1)=...=\phi(b_{|B|})$ with $a_i \in A$ and $b_j \in B$. This corresponds to orienting from $A$ to $B$ only the edges with exactly one endpoint in $A$ and one endpoint in $B$. It is a cocircuit of the graphic oriented matroid. It describes facets of $\mathcal{Z}_\Gamma$ since $\rm dim(S(\phi))=dim(\mathcal{Z}_\Gamma)-1=N-2$. We will usually represent facets by a tube, which induces a partition into two connected subgraphs, and encircles the first of them in the order given by the ordered partition defining the facet.

For vertices, i.e. $d$-codimensional faces with $d=N-1$, every partition subset of the cut of order $d$ is a singleton, so the quotient graph is just $\Gamma$, and the orientation assigned to it has to be acyclic by definition. Therefore, vertices of the graphical zonotope are in bijection with acyclic orientations of the graph. In fact, they are encoded by maximal signed covectors (topes), i.e. acyclic orientations.

\paragraph{Hypercube projection} There is an alternative definition of zonotopes as projections of hypercubes~\cite{z-lop-95}. Consider the $p$-hypercube
\beq
    C_p=\{  \textbf{y} \in \mathbb{R}^p : -1 \leq y_i \leq 1 \text{ for all } i \},
\eeq
and any set of vectors $\mathbf{b_i} \in \mathbb{R}^d$, $i \in \{1,\dots,p\}$ such that the matrix of this vector configuration, $\hat{B}=(\mathbf{b_1},\dots,\mathbf{b_p}) \in \mathbb{R}^{d\times p}$, defines a surjective projection. Any linear projection of the hypercube into $d$ dimensions gives a zonotope $\mathcal{Z}(\hat{B})$ :
\beq
    \hat{B} . C_p = \{ \hat{B} \mathbf{y} : \mathbf{y}\in C_p  \}.
\eeq
Conversely, every zonotope is, up to translation, the image of a hypercube $C_p$ under a linear projection given by the vector configuration $\hat{B}$. 

Every hypercube is written as a Minkowski sum of line segments $[-\mathbf{u_i},\mathbf{u_i}]$, for all $i\in \{1,\dots,p \}$, where $\mathbf{u_i}$ are the standard basis vectors in $\mathbb{R}^p$. Because we consider a linear projection, the image can be written as the Minkowski sum
\beq
    \mathcal{Z}(\hat{B})=\sum_{i=1}^p [-\hat{B}\mathbf{u_i},\hat{B}\mathbf{u_i}]=\sum_{i=1}^p [-\mathbf{b_i},\mathbf{b_i}].
\eeq
This is the initial definition we gave for the centrally symmetric zonotope as Minkowski sum of line segments, the vectors $\mathbf{b_i}$ being its generators.

Because the graphical zonotope $\mathcal{Z}_\Gamma$ has $E$ generators, one defined by each edge, it can be defined from the projection of the hypercube $C_E$. The choice of sign for the generators implies a choice of reference orientation $\vec{\Gamma}$. The matrix of the vector configuration defining the projection is $W=(\mathbf{b_1},\dots,\mathbf{b_E})$ with generators
\beq
\mathbf{b}_e = \mathbf{x}_i - \mathbf{x}_j \quad \text{for each edge } \{j \to i\} \in E(\vec{\Gamma}).
\eeq
Therefore, $B$ is the incidence matrix of the reference orientation digraph $\vec{\Gamma}$.
 
\paragraph{Graphic arrangement} Combinatorially, there is a dual description of the graphical zonotope of a graph $\Gamma$ via a central hyperplane arrangement called the \textit{graphic arrangement}\footnote{This is true for all zonotopes, but we restrict this discussion to graphical zonotopes}. We denote it by $\mathcal{A}_\Gamma$. It lives in $\mathbb{R}^N$, with coordinates $t_i$, $i\in \{1,\dots,N\}$ and its hyperplanes are defined by the equations
\beq
    t_i=t_j, \, \, \forall \,\{ij\} \in E(\Gamma).
    \label{eq:graphicarrangement}
\eeq
Since all hyperplanes contain the line spanned by $(1,\dots,1)$, it is often convenient to restrict the arrangement to the hyperplane $\sum_i t_i=0$.

The decomposition of $\mathbb{R}^N$ into polyhedral cones induced by $\mathcal{A}_\Gamma$ forms a fan, called the \textit{arrangement fan}. The fan of this hyperplane arrangement $\mathcal{A}_\Gamma$ is the normal fan of the graphical zonotope $\mathcal{Z}_\Gamma$~\cite{z-lop-95}. There is therefore a natural bijection between non-empty faces of $\mathcal{Z}_\Gamma$ and cones of $\mathcal{A}_\Gamma$. These are also in correspondence with covectors of the graphic oriented matroid. Namely, acyclic orientations of the graph can be put in correspondence both to vertices of $\mathcal{Z}_\Gamma$ and top-dimensional cones of the fan of~$\mathcal{A}_\Gamma$.

We physically interpret acyclic orientations of graphs as a (partial) time ordering of local interactions, where the $t_i$ variables can be interpreted as a time coordinate in a certain inertial reference frame. The hyperplane $t_i=t_j$ delimits the two half-spaces $t_i>t_j$, where interaction at vertex $i$ happens before interaction at vertex $j$, and $t_i<t_j$ where the time-ordering is reversed. The inequalities defining top-dimensional cones correspond to orienting edges of the graph without forming oriented cycles, otherwise the inequalities would be inconsistent. The absence of an edge between two vertices reflects the absence of time ordering between these two interactions for some causal configurations. In the Coleman-Norton picture, this makes the cones of the graphic arrangement fan Lorentz invariant, meaning that two points related by a Lorentz transformation belong to the same cone.

\begin{figure}[h]
\centering

\begin{minipage}{0.48\textwidth}
\centering
\begin{tikzpicture}[
    scale=2.2,
    >=Latex,
    every node/.style={font=\small},
    edge/.style={thick},
]


\coordinate (v1) at (1,0);
\coordinate (v2) at (0.5,0.866);
\coordinate (v3) at (-0.5,0.866);
\coordinate (v4) at (-1,0);
\coordinate (v5) at (-0.5,-0.866);
\coordinate (v6) at (0.5,-0.866);

\fill[blue!10] (v1)--(v2)--(v3)--(v4)--(v5)--(v6)--cycle;
\draw[thick] (v1)--(v2)--(v3)--(v4)--(v5)--(v6)--cycle;


\coordinate (base) at (-0.5,-0.866);

\draw[->,very thick,violet] (base) -- ++(1,0)
node[font=\scriptsize,pos=0.57,above] {$[e_1-e_2,e_2-e_1]$};

\draw[->,very thick,violet] (base) -- ++(0.5,0.866)
node[font=\scriptsize,right,pos=0.7,align=center] {$\,\,\, [e_1-e_3,$\\$e_3-e_1]$};

\draw[->,very thick,violet] (base) -- ++(-0.5,0.866)
node[font=\scriptsize,pos=0.9,right,align=center] {$[e_2-e_3,$\\$\,\,e_3-e_2]$};


\TriGraph{(1.3,0)}{0.55}{
\draw[-->-=0.8, thick] (A)--(B);
\draw[-->-=0.8, thick] (A)--(C);
\draw[-->-=0.8, thick] (C)--(B);
}{}

\TriGraph{(0.7,1.1)}{0.55}{
\draw[-->-=0.8, thick] (A) to (B);
\draw[-->-=0.8, thick] (A) to (C);
\draw[-->-=0.8, thick] (B) to (C);
}{}

\TriGraph{(-0.7,1.1)}{0.55}{
\draw[-->-=0.8, thick] (B)--(A);
\draw[-->-=0.8, thick] (B)--(C);
\draw[-->-=0.8, thick] (A)--(C);
}{}

\TriGraph{(-1.3,0)}{0.55}{
\draw[-->-=0.8, thick] (B)--(A);
\draw[-->-=0.8, thick] (C)--(A);
\draw[-->-=0.8, thick] (B)--(C);
}{}

\TriGraph{(-0.7,-1.1)}{0.55}{
\draw[-->-=0.8, thick] (B)--(A);
\draw[-->-=0.8, thick] (C)--(B);
\draw[-->-=0.8, thick] (C)--(A);
}{}

\TriGraph{(0.7,-1.1)}{0.55}{
\draw[-->-=0.8, thick] (A)--(B);
\draw[-->-=0.8, thick] (C)--(A);
\draw[-->-=0.8, thick] (C)--(B);
}{}



\TriGraph{(1.1,0.55)}{0.45}{
\draw [-->-=0.8, thick] (A)--(B);
\draw [-->-=0.8, thick] (A)--(C);
\draw [thick] (B)--(C);
\Tube[red]{(A)};
}{}

\TriGraph{(0,1.15)}{0.45}{
\draw [-->-=0.8, thick] (A)--(C);
\draw [-->-=0.8, thick] (B)--(C);
\draw[thick] (A)--(B);
\Tube[red,rounded corners=8pt, inner sep=4pt]{(A) (B)};
}{}

\TriGraph{(-1.1,0.55)}{0.45}{
\draw [-->-=0.8, thick] (B)--(A);
\draw [-->-=0.8, thick] (B)--(C);
\draw[thick] (A)--(C);
\Tube[red]{(B)};
}{}

\TriGraph{(-1.1,-0.55)}{0.45}{
\draw [-->-=0.8, thick] (B)--(A);
\draw [-->-=0.8, thick] (C)--(A);
\draw[thick] (C)--(B);
\Tube[red,rounded corners=8pt, inner sep=4pt]{(B) (C)};
}{}

\TriGraph{(0,-1.15)}{0.45}{
\draw [-->-=0.8, thick] (C)--(A);
\draw [-->-=0.8, thick] (C)--(B);
\draw[thick] (B)--(A);
\Tube[red]{(C)};
}{}

\TriGraph{(1.1,-0.55)}{0.45}{
\draw [-->-=0.8, thick] (A)--(B);
\draw [-->-=0.8, thick] (C)--(B);
\draw[thick] (C)--(A);
\Tube[red]{(A) (C)};
}{}

\end{tikzpicture}
\end{minipage}
\hfill
\begin{minipage}{0.48\textwidth}
\centering
\begin{tikzpicture}[scale=3]

\coordinate (O) at (0,0);

\draw (0,-1) -- (0,1);
\node[font=\small,above right] at (0,0.85) {$t_1=t_2$};

\draw (-0.866,-0.5) -- (0.866,0.5);
\node[font=\small] at (0.75,0.55) {$t_2=t_3$};

\draw (-0.866,0.5) -- (0.866,-0.5);
\node[font=\small] at (0.75,-0.55) {$t_1=t_3$};


\def\r{0.55}
\def\s{0.55}

\path (0:\r) coordinate (P1);
\TriGraph{P1}{\s}{
\draw[->-=0.8, thick] (A) to (B);
\draw[->-=0.8, thick] (A) to (C);
\draw[->-=0.8, thick] (C) to (B);
}{}

\path (60:\r) coordinate (P2);
\TriGraph{P2}{\s}{
\draw[->-=0.8, thick] (A) to (B);
\draw[->-=0.8, thick] (A) to (C);
\draw[->-=0.8, thick] (B) to (C);
}{}

\path (120:\r) coordinate (P3);
\TriGraph{P3}{\s}{
\draw[->-=0.8, thick] (B) to (A);
\draw[->-=0.8, thick] (A) to (C);
\draw[->-=0.8, thick] (B) to (C);
}{}

\path (180:\r) coordinate (P4);
\TriGraph{P4}{\s}{
\draw[->-=0.8, thick] (B) to (A);
\draw[->-=0.8, thick] (C) to (A);
\draw[->-=0.8, thick] (B) to (C);
}{}

\path (240:\r) coordinate (P5);
\TriGraph{P5}{\s}{
\draw[->-=0.8, thick] (B) to (A);
\draw[->-=0.8, thick] (C) to (A);
\draw[->-=0.8, thick] (C) to (B);
}{}

\path (300:\r) coordinate (P6);
\TriGraph{P6}{\s}{
\draw[->-=0.8, thick] (A) to (B);
\draw[->-=0.8, thick] (C) to (B);
\draw[->-=0.8, thick] (C) to (A);
}{}

\end{tikzpicture}
\end{minipage}
\caption{Graphical zonotope of the triangle graph $K_3$, whose facets correspond to ordered minimal cuts, and whose vertices are labeled by acyclic orientations of the graph (left) and graphic hyperplane arrangement dual to $\mathcal{Z}_{K_3}$ (right).} 
\label{fig:GZK3}
\end{figure}
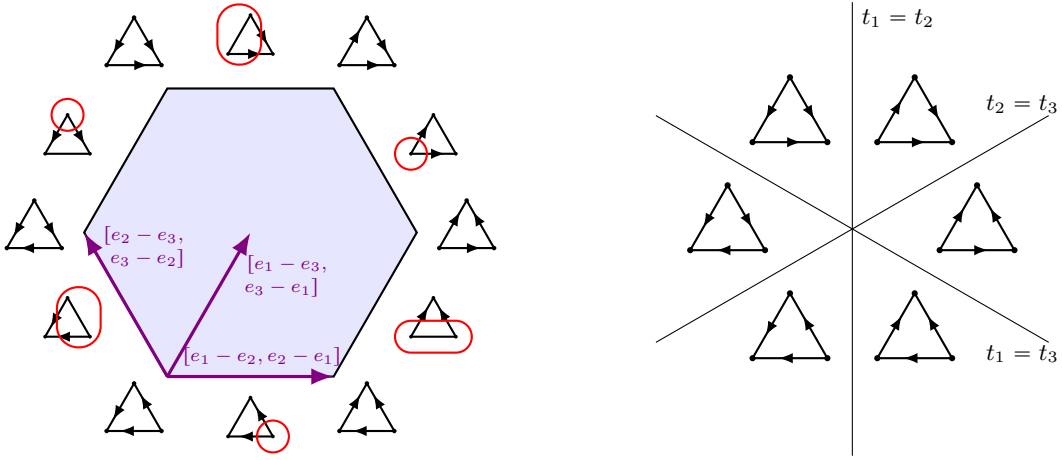

Fig.~\ref{fig:GZK3} shows the graphical zonotope of the triangle graph $K_3$, also called the \textit{3-permutohedron}, along with its dual hyperplane arrangement made of three lines. One can see the correspondence between minimal cuts and facets and between acyclic orientations and vertices of the graphical zonotope. A parity transformation of all coordinates, $x_i \rightarrow -x_i$ (i.e. $t_i \rightarrow -t_i$) for all $i$, corresponds to reversing the orientation of the graph, i.e. time reversal of the physical process. The hyperplane arrangement separates the space according to the ordering of the values of the $t_i$ variables, so that the top-dimensional regions that cannot be further decomposed into the union of smaller ones correspond to acyclic orientations. Here, the region characterized by the intersection of half-spaces $\{t_1<t_2\}\cap\{t_2<t_3\}$ corresponds to the acyclic orientation with vertex ordering $1 \rightarrow 2 \rightarrow 3$.

\subsection{Generalized permutohedra and deformed graphical zonotopes \label{sec:GeneralizedPermutohedra}}

The $N$-\textit{permutohedron} $P_N$ is a ($N-1$)-dimensional polytope whose vertices are in bijection with the permutations of $\{1,\dots,N\}$. The $N$-permutohedron is the graphical zonotope of the complete graph with $N$ vertices, i.e. the graph where all vertices are pairwise connected. In fact, acyclic orientations of a complete graph are in correspondence with all the possible total orders of its vertices. A realization of the $N$-permutohedron that we use in this article is 
\beq
    P_N=\sum_{1 \leq i < j \leq N}\left[\mathbf{\mathbf{x_j}}-\mathbf{\mathbf{x_i}}, \mathbf{\mathbf{x_i}}-\mathbf{\mathbf{x_j}} \right],
    \label{eq:PermutohedronDef}
\eeq
with $\mathbf{x_i}$ the standard basis vectors in $\mathbb{R}^N$. This differs from the standard realization by a scaling and translation, but has the same combinatorics.

\textit{Generalized permutohedra} (also called \textit{deformed permutohedra})~\cite{postnikov2005} are deformations of the $N$-permutohedron obtained by parallel translating the facets of the permutohedron without passing through a vertex. Because any graph induced by a vertex subset of a complete graph is connected, the minimal cuts of a complete graph correspond to any binary partition of the vertex set of the graph. Therefore, following the combinatorics of graphical zonotopes explained in the previous section, the facets of the $N$-permutohedron are in one-to-one correspondence with all non-empty proper subsets of vertices.

For the centrally symmetric realization of the $N$-permutohedron in Eq. (\ref{eq:PermutohedronDef}), its facet-defining inequalities are 
\beq
    \left|\sum_{i \in S} x_i\right| \le |S|(N-|S|) \equiv \alpha_S \;\;\forall\,\emptyset \neq S \subsetneq \{1,\dots,N\}.
\eeq
Along with the condition $\sum_{i=1}^N x_i = 0$, they define $P_N$. Then one can replace the offset of the facets $\alpha_S$ by some other $\beta_S$. However, in order for this to define a generalized permutohedron, the facets cannot be translated past any vertex. The set of allowed offsets forms the \textit{deformation cone}. It is known that the allowed deformations are given by the $\beta$ that are submodular functions. 

For a set $[N]\equiv\{1,\dots,N\}$, a submodular function is the map $\beta : 2^{[N]} \rightarrow \mathbb{R}$ satisfying, for every pair of subsets $A,B \subseteq [N]$ :
\beq
    \beta(A)+\beta(B) \geq \beta(A \cup B) + \beta(A \cap B).
    \label{eq:submodular}
\eeq
All submodular functions $\beta$ characterize all possible generalized permutohedra of a given dimension up to translation. Graphical zonotopes $\mathcal{Z}_\Gamma$ for any graph $\Gamma$, as well as their deformations, are in fact generalized permutohedra. But let us easily show this using another defining property of generalized permutohedra proven in Ref.~\cite{Postnikov2008}. 

A fan $\mathcal{F}$ is said to \textit{coarsen} another fan $\mathcal{F'}$ if every cone in $\mathcal{F}$ is a union of cones in $\mathcal{F'}$. Then, a polytope is a generalized permutohedron if and only if its normal fan coarsens the normal fan of the permutohedron, called the \textit{braid fan}. We know from Sec.~\ref{sec:DefCombinatoricsGZ} that the braid fan is the graphic arrangement of the complete graph, given by 
\beq
    t_i=t_j, \, \, \forall \,i<j \in [N].
\eeq
From the braid arrangement, one recovers the graphic arrangement of an arbitrary graph $\Gamma$ given in Eq. (\ref{eq:graphicarrangement}) by removing all hyperplanes $t_i=t_j$ such that $\{ij\} \notin E(\Gamma)$. This merges regions of the arrangement, so the graphic arrangement coarsens the braid fan, and the graphical zonotope is a generalized permutohedron.

In fact, deforming the permutohedron can give a polytope that is not combinatorially equivalent to it. This happens when there is linear dependence of facet normals of facets meeting at a vertex, and thus several vertices may merge forming a non-simple vertex, i.e. strictly more than $d$ facets meet at that vertex of the $d$-dimensional polytope. This is characterized by a saturation of an inequality in Def. (\ref{eq:submodular}) : 
\beq
\beta(A)+\beta(B) = \beta(A \cup B) + \beta(A \cap B).
\eeq
Here $A$ and $B$ are any subsets of vertices inducing connected subgraphs, as well as its complements. 

Now, we are interested in deformations of the graphical zonotope in which one starts with $\mathcal{Z}_\Gamma$ as defined in Sec.~\ref{sec:DefCombinatoricsGZ} and translates its facets remaining in the deformation cone. This was studied in Ref.~\cite{padrol:hal-03788987}, where it was shown that the deformation cone of the graphical zonotope, living in $\mathbb{R}^{2^N}$ is defined by the submodular function $\beta$ satisfying the following three conditions:
\begin{align}
    & \beta(\emptyset)=-\beta(V) \\
    &\beta(S\cup \{u\} )+\beta(S\cup \{v\})=\beta(S)+\beta(S\cup \{u,v\} ) \, \forall u,v \in V \text{ non-adjacent in } \Gamma \text{, and } \forall S \subseteq V \backslash\{u,v\} \label{eq:submodularityGZ1} \\
    &\beta(S\cup \{u\} )+\beta(S\cup \{v\} ) \geq \beta(S)+\beta(S\cup \{u,v\} )\, \forall u,v \in V \text{ adjacent in } \Gamma \text{, and } \forall S \subseteq V \backslash\{u,v\}. 
    \label{eq:submodularityGZ2}
\end{align}

Motivated by LTD, we are interested in a certain subset of this deformation cone of deformations of the graphical zonotope. Define $\mathcal{C}_S$ to be the subset of edges of $\Gamma$ with one endpoint in $S$ and the other in its complement $V\backslash S$, i.e. the edges cut by the partition $S\,|\,V\backslash S$. Consider the space $\mathbb{R}_{\geq 0} ^{E}$ with coordinates $y_e\geq0$ for all $e\in E(\Gamma)$. We will often denote it by $y_{ij}$ (typically $i<j$) for the edge connecting vertices $i$ and $j$. Then
\beq
    \beta(S)=\sum_{e\in \mathcal{C}_S} y_e,
    \label{eq:SubmodCuts}
\eeq
with $\beta(\emptyset)=\beta(V)=0$. For this choice of submodular function $\beta$, note that $\beta(S\cup \{u\} )+\beta(S\cup \{v\} )-(\beta(S)+\beta(S\cup \{u,v\} ))=2y_{uv}$. In the case of more than one edge connecting vertices $u$ and $v$, a sum over their on-shell energies $y_e$ should appear on the RHS.

Therefore, the condition in \Eq{eq:submodularityGZ1} only appears when $u$ and $v$ are non-adjacent. So, this $\beta$ function satisfies the submodularity conditions. Then, degeneracies of facets, i.e. linear dependences between facet normals, occur when some edges of the graph are removed. Because the function $\beta$ only depends on the variables $y_e$, the deformed graphical zonotope characterized by the function $\beta$ can be denoted $\mathcal{Z}_\Gamma^Y$, where $Y=\{y_1,\dots,y_E\}$ is a set of $y$ variables.

The cone defined by these submodular functions is $E$-dimensional in the case of a simple graph, and is therefore a subset of the deformation cone of the graphical zonotope, which is known to be of higher dimension in general. For triangle-free graphs, this cone given by the cut submodular function $\beta$ equals the full deformation cone of its graphical zonotope.

To sum up, the deformed graphical zonotopes we are interested in live in the space of external energies $x_i$ and are parametrized by on-shell energies of internal lines $y_e$. The facet description of a deformed graphical zonotope is 
\beq
    \left|\sum_{i \in S} x_i\right| \le \sum_{e\in \mathcal{C}_S} y_e 
\eeq
for all proper vertex subsets $S$ corresponding to a bond. By subtracting the left-hand side, we get
\beq
     \lambda_S   \geq 0,
     \label{eq:CSinequalities}
\eeq
where we have flipped the sign of $\sum_{i \in S} x_i$ in some of these inequalities by switching from $S$ to the complementary subset $V\backslash S$. $\lambda_S$ was defined in (\ref{eq:causalthreshold}). Note that $\lambda_S$ was defined in relation to lowest order singularities of the Feynman energy integrals. We showed that these singularities are in correspondence with facets of the graphical zonotope. $\mathcal{Z}_\Gamma^Y$ can also be written as a Minkowski sum of segments weighted by the $y$ variables:
\beq
    \mathcal{Z}_\Gamma^Y=\sum_{\{ij\}\in E} y_e [\mathbf{x_j}-\mathbf{x_i},\mathbf{x_i}-\mathbf{x_j}].
    \label{eq:DeformedZonotpeWeightedSum}
\eeq
The deformed graphical zonotope can be obtained from the affine projection of a hyperrectangle, whose dimensions are weighted by the $y_e$ variables controlling the deformation of the zonotope.

\begin{figure}[h]
\centering

\begin{minipage}{0.48\textwidth}
\centering
\begin{tikzpicture}[scale=1.3,line join=round,line cap=round]
\coordinate (v1) at (-1.48301,-1.70377);
\coordinate (v2) at (-1.64904,1.97489);
\coordinate (v3) at (-2.75318,0.79999);
\coordinate (v4) at (0.75318,-2.43298);
\coordinate (v5) at (-0.89737,-3.48834);
\coordinate (v6) at (-0.35096,-3.60788);
\coordinate (v7) at (-1.0634,0.19032);
\coordinate (v8) at (-0.51699,0.07077);
\coordinate (v9) at (-2.16754,-0.98458);
\coordinate (v10) at (2.16754,0.98458);
\coordinate (v11) at (0.51699,-0.07077);
\coordinate (v12) at (1.0634,-0.19032);
\coordinate (v13) at (0.35096,3.60788);
\coordinate (v14) at (0.89737,3.48834);
\coordinate (v15) at (-0.75318,2.43298);
\coordinate (v16) at (2.75318,-0.79999);
\coordinate (v17) at (1.64904,-1.97489);
\coordinate (v18) at (1.48301,1.70377);
\fill[blue!30,opacity=.18] (v9)--(v3)--(v2)--(v7)--cycle;
\fill[blue!30,opacity=.18] (v18)--(v14)--(v10)--(v16)--cycle;
\fill[blue!30,opacity=.18] (v4)--(v8)--(v18)--(v16)--cycle;
\fill[blue!30,opacity=.18] (v17)--(v6)--(v4)--(v16)--cycle;
\fill[blue!30,opacity=.18] (v8)--(v7)--(v2)--(v13)--(v14)--(v18)--cycle;
\fill[blue!30,opacity=.18] (v5)--(v9)--(v7)--(v8)--(v4)--(v6)--cycle;
\fill[blue!60,opacity=.32] (v5)--(v1)--(v3)--(v9)--cycle;
\fill[blue!60,opacity=.32] (v1)--(v11)--(v15)--(v3)--cycle;
\fill[blue!60,opacity=.32] (v15)--(v13)--(v2)--(v3)--cycle;
\fill[blue!60,opacity=.32] (v12)--(v17)--(v16)--(v10)--cycle;
\fill[blue!60,opacity=.32] (v11)--(v1)--(v5)--(v6)--(v17)--(v12)--cycle;
\fill[blue!60,opacity=.32] (v14)--(v13)--(v15)--(v11)--(v12)--(v10)--cycle;
\draw[dash pattern=on 3pt off 2pt,black!60,line width=.45pt] (v2)--(v7);
\draw[dash pattern=on 3pt off 2pt,black!60,line width=.45pt] (v7)--(v9);
\draw[dash pattern=on 3pt off 2pt,black!60,line width=.45pt] (v18)--(v14);
\draw[dash pattern=on 3pt off 2pt,black!60,line width=.45pt] (v16)--(v18);
\draw[dash pattern=on 3pt off 2pt,black!60,line width=.45pt] (v4)--(v8);
\draw[dash pattern=on 3pt off 2pt,black!60,line width=.45pt] (v8)--(v18);
\draw[dash pattern=on 3pt off 2pt,black!60,line width=.45pt] (v16)--(v4);
\draw[dash pattern=on 3pt off 2pt,black!60,line width=.45pt] (v6)--(v4);
\draw[dash pattern=on 3pt off 2pt,black!60,line width=.45pt] (v8)--(v7);
\draw[black,line width=.9pt] (v5)--(v1);
\draw[black,line width=.9pt] (v1)--(v3);
\draw[black,line width=.9pt] (v3)--(v9);
\draw[black,line width=.9pt] (v9)--(v5);
\draw[black,line width=.9pt] (v1)--(v11);
\draw[black,line width=.9pt] (v11)--(v15);
\draw[black,line width=.9pt] (v15)--(v3);
\draw[black,line width=.9pt] (v3)--(v2);
\draw[black,line width=.9pt] (v15)--(v13);
\draw[black,line width=.9pt] (v13)--(v2);
\draw[black,line width=.9pt] (v14)--(v10);
\draw[black,line width=.9pt] (v10)--(v16);
\draw[black,line width=.9pt] (v12)--(v17);
\draw[black,line width=.9pt] (v17)--(v16);
\draw[black,line width=.9pt] (v10)--(v12);
\draw[black,line width=.9pt] (v17)--(v6);
\draw[black,line width=.9pt] (v13)--(v14);
\draw[black,line width=.9pt] (v6)--(v5);
\draw[gray,line width=2pt] (v12)--(v11);

\fill[blue] (v11) circle (2.5pt);
\fill[brown] (v12) circle (2.5pt);

\begin{scope}[shift={($0.25*(v1)+0.3*(v3)+0.25*(v15)+0.22*(v11)$)}, scale=0.5]
  \BoxGraph{3}{}{
  \draw[thick] (A)--(B);
    \draw[thick] (D)--(C);
\draw [-->-=1, thick] (A)--(D);
\draw [-->-=1, thick] (B)--(D);
\draw [-->-=1, thick] (B)--(C);
\Tube[rounded corners=20pt, inner sep=12pt]{(A) (B)};
}{}
\end{scope}
\begin{scope}[shift={($0.2*(v12)+0.3*(v10)+0.3*(v16)+0.2*(v17)$)}, scale=0.45]
  \BoxGraph{3}{}{
    \draw[thick] (A)--(D);
    \draw[thick] (B)--(C);
\draw [-->-=1, thick] (A)--(B);
\draw [-->-=1, thick] (D)--(B);
\draw [-->-=1, thick] (D)--(C);
\Tube[rounded corners=17pt, inner sep=9pt]{(A) (D)};
}{}
\end{scope}
\begin{scope}[shift={($0.14*(v11)+0.14*(v12)+0.17*(v10)+0.17*(v15)+0.2*(v13)+0.2*(v14)$)}, scale=0.5]
  \BoxGraph{3}{}{
    \draw[thick] (B)--(C);
    \draw[thick] (D)--(C);
\draw [-->-=1, thick] (A)--(B);
\draw [-->-=1, thick] (A)--(D);
\draw [thick] (B)--(D);
\Tube[]{(A)};
}{}
\end{scope}
\begin{scope}[shift={($0.14*(v11)+0.14*(v12)+0.17*(v1)+0.17*(v17)+0.2*(v5)+0.2*(v6)$)}, scale=0.5]
  \BoxGraph{3}{}{
  \draw[thick] (A)--(B);
    \draw[thick] (A)--(D);
\draw [-->-=1, thick] (B)--(C);
\draw [-->-=1, thick] (D)--(C);
\draw [thick] (B)--(D);
\Tube[rounded corners=17pt, inner sep=9pt]{(A) (B)(D)};
}{}
\end{scope}
\begin{scope}[shift={($0.14*(v11)+0.14*(v12)+0.17*(v1)+0.17*(v17)+0.2*(v5)+0.2*(v6)$)}, scale=0.5]
  \BoxGraph{3}{}{
  \draw[thick] (A)--(B);
    \draw[thick] (A)--(D);
\draw [-->-=1, thick] (B)--(C);
\draw [-->-=1, thick] (D)--(C);
\draw [thick] (B)--(D);
\Tube[rounded corners=17pt, inner sep=9pt]{(A) (B)(D)};
}{}
\end{scope}
\begin{scope}[shift={($0.85*(v11)+0.15*(v3)$)}, scale=0.4]
  \BoxGraph[blue]{3}{}{
\draw[blue] [-->-=0.8, thick] (A)--(B);
\draw[blue] [-->-=0.8, thick] (A)--(D);
\draw[blue] [-->-=0.8, thick] (B)--(C);
\draw[blue] [-->-=0.8, thick] (B)--(D);
\draw[blue] [-->-=0.8, thick] (D)--(C);
}{}
\end{scope}
\begin{scope}[shift={($0.85*(v12)+0.15*(v5)$)}, scale=0.4]
  \BoxGraph[brown]{3}{}{
\draw[brown] [-->-=0.8, thick] (A)--(B);
\draw[brown] [-->-=0.8, thick] (A)--(D);
\draw[brown] [-->-=0.8, thick] (B)--(C);
\draw[brown] [-->-=0.8, thick] (D)--(B);
\draw[brown] [-->-=0.8, thick] (D)--(C);
}{}
\end{scope}
\begin{scope}[shift={($0.68*(v12)+0.26*(v11)+0.18*(v14)$)}, scale=0.4]
  \BoxGraph[gray]{3}{}{
\draw[gray] [-->-=0.8, thick] (A)--(B);
\draw[gray] [-->-=0.8, thick] (A)--(D);
\draw[gray] [-->-=0.8, thick] (B)--(C);
\draw[gray] [thick] (D)--(B);
\draw[gray] [-->-=0.8, thick] (D)--(C);
\Tube[gray]{(A)};
\Tube[gray,rounded corners=14pt, inner sep=6pt]{(A) (B) (D)};
}{}
\end{scope}
\end{tikzpicture}

\end{minipage}
\hfill
\begin{minipage}{0.48\textwidth}
\centering

\begin{tikzpicture}[scale=1.3,line join=round,line cap=round]
\coordinate (v1) at (-1.20981,-1.76354);
\coordinate (v2) at (-1.37583,1.91511);
\coordinate (v3) at (-2.47998,0.74022);
\coordinate (v4) at (0.47998,-2.37321);
\coordinate (v5) at (-0.62417,-3.54811);
\coordinate (v6) at (-0.79019,0.13055);
\coordinate (v7) at (-1.89434,-1.04435);
\coordinate (v8) at (1.89434,1.04435);
\coordinate (v9) at (0.79019,-0.13055);
\coordinate (v10) at (0.62417,3.54811);
\coordinate (v11) at (-0.47998,2.37321);
\coordinate (v12) at (2.47998,-0.74022);
\coordinate (v13) at (1.37583,-1.91511);
\coordinate (v14) at (1.20981,1.76354);
\fill[blue!30,opacity=.18] (v14)--(v10)--(v8)--(v12)--cycle;
\fill[blue!30,opacity=.18] (v14)--(v12)--(v4)--(v6)--cycle;
\fill[blue!30,opacity=.18] (v6)--(v4)--(v5)--(v7)--cycle;
\fill[blue!30,opacity=.18] (v7)--(v3)--(v2)--(v6)--cycle;
\fill[blue!30,opacity=.18] (v10)--(v14)--(v6)--(v2)--cycle;
\fill[blue!30,opacity=.18] (v4)--(v12)--(v13)--(v5)--cycle;
\fill[blue!60,opacity=.32] (v1)--(v3)--(v7)--(v5)--cycle;
\fill[blue!60,opacity=.32] (v3)--(v1)--(v9)--(v11)--cycle;
\fill[blue!60,opacity=.32] (v8)--(v10)--(v11)--(v9)--cycle;
\fill[blue!60,opacity=.32] (v3)--(v11)--(v10)--(v2)--cycle;
\fill[blue!60,opacity=.32] (v1)--(v5)--(v13)--(v9)--cycle;
\fill[blue!60,opacity=.32] (v12)--(v8)--(v9)--(v13)--cycle;
\draw[dash pattern=on 3pt off 2pt,black!60,line width=.45pt] (v14)--(v10);
\draw[dash pattern=on 3pt off 2pt,black!60,line width=.45pt] (v12)--(v14);
\draw[dash pattern=on 3pt off 2pt,black!60,line width=.45pt] (v12)--(v4);
\draw[dash pattern=on 3pt off 2pt,black!60,line width=.45pt] (v4)--(v6);
\draw[dash pattern=on 3pt off 2pt,black!60,line width=.45pt] (v6)--(v14);
\draw[dash pattern=on 3pt off 2pt,black!60,line width=.45pt] (v4)--(v5);
\draw[dash pattern=on 3pt off 2pt,black!60,line width=.45pt] (v7)--(v6);
\draw[dash pattern=on 3pt off 2pt,black!60,line width=.45pt] (v2)--(v6);
\draw[black,line width=.9pt] (v10)--(v8);
\draw[black,line width=.9pt] (v8)--(v12);
\draw[black,line width=.9pt] (v1)--(v3);
\draw[black,line width=.9pt] (v3)--(v7);
\draw[black,line width=.9pt] (v7)--(v5);
\draw[black,line width=.9pt] (v5)--(v1);
\draw[black,line width=.9pt] (v1)--(v9);
\draw[black,line width=.9pt] (v9)--(v11);
\draw[black,line width=.9pt] (v11)--(v3);
\draw[black,line width=.9pt] (v10)--(v11);
\draw[black,line width=.9pt] (v9)--(v8);
\draw[black,line width=.9pt] (v3)--(v2);
\draw[black,line width=.9pt] (v10)--(v2);
\draw[black,line width=.9pt] (v5)--(v13);
\draw[black,line width=.9pt] (v13)--(v9);
\draw[black,line width=.9pt] (v12)--(v13);

\fill[gray] (v9) circle (2.5pt);

\begin{scope}[shift={($0.25*(v11)+0.3*(v3)+0.25*(v1)+0.22*(v9)$)}, scale=0.5]
  \BoxGraph{3}{}{
  \draw[thick] (A)--(B);
    \draw[thick] (D)--(C);
\draw [-->-=1, thick] (A)--(D);
\draw [-->-=1, thick] (B)--(C);
\Tube[rounded corners=19pt, inner sep=10pt]{(A) (B)};
}{}
\end{scope}
\begin{scope}[shift={($0.2*(v9)+0.3*(v12)+0.3*(v8)+0.2*(v13)$)}, scale=0.45]
  \BoxGraph{3}{}{
  \draw[thick] (B)--(C);
    \draw[thick] (A)--(D);
\draw [-->-=1, thick] (A)--(B);
\draw [-->-=1, thick] (D)--(C);
\Tube[rounded corners=17pt, inner sep=9pt]{(A) (D)};
}{}
\end{scope}
\begin{scope}[shift={($0.27*(v10)+0.23*(v9)+0.3*(v8)+0.2*(v11)$)}, scale=0.5]
  \BoxGraph{3}{}{
    \draw[thick] (B)--(C);
    \draw[thick] (D)--(C);
\draw [-->-=1, thick] (A)--(B);
\draw [-->-=1, thick] (A)--(D);
\Tube[]{(A)};
}{}
\end{scope}
\begin{scope}[shift={($0.25*(v5)+0.25*(v9)+0.3*(v13)+0.2*(v1)$)}, scale=0.5]
  \BoxGraph{3}{}{
  \draw[thick] (A)--(B);
    \draw[thick] (A)--(D);
\draw [-->-=1, thick] (B)--(C);
\draw [-->-=1, thick] (D)--(C);
\Tube[rounded corners=17pt, inner sep=9pt]{(A) (B)(D)};
}{}
\end{scope}

\begin{scope}[shift={($0.75*(v9)+0.15*(v10)+0.1*(v8)$)}, scale=0.4]
  \BoxGraph[gray]{3}{}{
\draw[gray] [-->-=0.8, thick] (A)--(B);
\draw[gray] [-->-=0.8, thick] (A)--(D);
\draw[gray] [-->-=0.8, thick] (B)--(C);
\draw[gray] [-->-=0.8, thick] (D)--(C);
}{}
\end{scope}
\end{tikzpicture}

\end{minipage}

\caption[Deformed graphical zonotope]{Deformed graphical zonotope $\mathcal{Z}^Y_
{\begin{tikzpicture}
    \BoxGraph[black][1.2]{0.3}{}{
    \draw (A)--(B);
    \draw (A)--(D);
    \draw (B)--(C);
    \draw (D)--(C);
    \draw (D)--(B);
    }{}
\end{tikzpicture}}$ of the square graph with a chord, for a small value of the edge variable $y_c$ of the chord (left). Setting $y_c=0$ results in degeneration of faces and yields the deformed graphical zonotope of the square graph (right).}
\label{fig:DeformedGZSquareChord}
\end{figure}
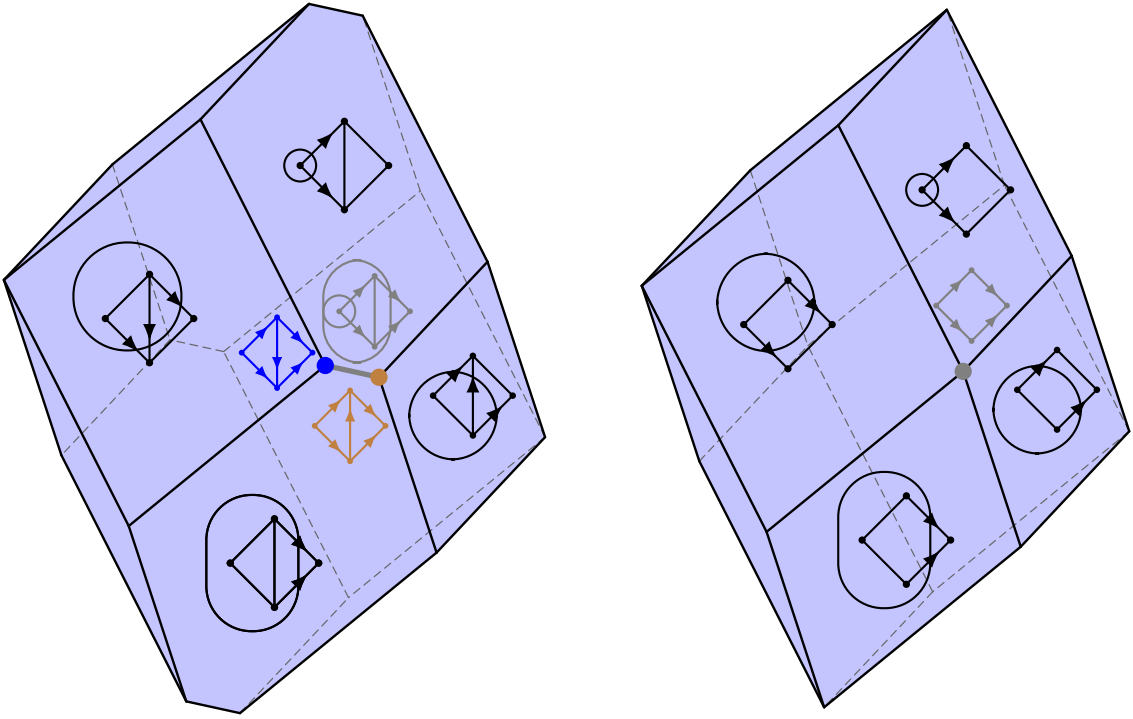

Let us now illustrate deformation of graphical zonotopes with the square graph with a chord $\begin{tikzpicture}
    \BoxGraph[black][1.2]{0.6}{}{
    \draw (A)--(B);
    \draw (A)--(D);
    \draw (B)--(C);
    \draw (D)--(C);
    \draw (D)--(B);
    }{}
\end{tikzpicture}$. A deformed graphical zonotope for this graph, $\mathcal{Z}^Y_
{\begin{tikzpicture}
    \BoxGraph[black][1.2]{0.3}{}{
    \draw (A)--(B);
    \draw (A)--(D);
    \draw (B)--(C);
    \draw (D)--(C);
    \draw (D)--(B);
    }{}
\end{tikzpicture}}$, is represented in Fig. \ref{fig:DeformedGZSquareChord} (left). The variable $y$ associated to the chord, let us call it $y_c$, is small compared to the other $y$ variables. This deformation makes four pairs of vertices very close (a pair is shown in blue and orange). Deforming further until $y_c=0$ merges these pairs into one single vertex (in gray) and the edge connecting both of them disappears, as shown on the right side of the figure. This deformation, characterized by $Y^{[y_c=0]}$ corresponds to a deformation of the graphical zonotope of the square graph : $\mathcal{Z}^{Y^{[y_c=0]}}_
{\begin{tikzpicture}
    \BoxGraph[black][1.2]{0.3}{}{
    \draw (A)--(B);
    \draw (A)--(D);
    \draw (B)--(C);
    \draw (D)--(C);
    \draw (D)--(B);
    }{}
\end{tikzpicture}}=\mathcal{Z}^{Y \backslash y_c}_{\begin{tikzpicture}
    \BoxGraph[black][1.2]{0.3}{}{
    \draw (A)--(B);
    \draw (A)--(D);
    \draw (B)--(C);
    \draw (D)--(C);
    }{}
\end{tikzpicture}}$.

For these deformed graphical zonotopes, setting some $y_e=0$ corresponds to removing the edge $e$ from the graph. Then, one passes from a strict inequality in the submodularity condition (\ref{eq:submodularityGZ2}) to the equality in condition (\ref{eq:submodularityGZ1}). As explained, it translates into a degeneration of facets, with the linear dependence of causal thresholds 
\begin{equation}
    \lambda_{(1,2)}+\lambda_{(1,3)}=\lambda_{(1)}+\lambda_{(1,2,3)},
    \label{eq:DegeneracySquareGraph}
\end{equation}
that we can pictorially represent as
\begin{equation}
    \begin{tikzpicture}
        \BoxGraph[black][1.2]{1.1}{
        \draw [-->-=0.8, thick] (A)--(B);
        \draw [-->-=0.8, thick] (A)--(D);
        \draw [-->-=0.8, thick] (B)--(C);
        \draw [-->-=0.8, thick] (D)--(C);
        \Tube[rounded corners=10pt, inner sep=5pt]{(A) (B)};
        }{}
    \end{tikzpicture} \quad
     \scalebox{3}{+} \quad
     \begin{tikzpicture}
        \BoxGraph[black][1.2]{1.1}{
        \draw [-->-=0.8, thick] (A)--(B);
        \draw [-->-=0.8, thick] (A)--(D);
        \draw [-->-=0.8, thick] (B)--(C);
        \draw [-->-=0.8, thick] (D)--(C);
        \Tube[rounded corners=10pt, inner sep=5pt]{(A) (D)};
        }{}
    \end{tikzpicture}
    \quad \scalebox{3}{=} \quad
    \begin{tikzpicture}
        \BoxGraph[black][1.2]{1.1}{
        \draw [-->-=0.8, thick] (A)--(B);
        \draw [-->-=0.8, thick] (A)--(D);
        \draw [-->-=0.8, thick] (B)--(C);
        \draw [-->-=0.8, thick] (D)--(C);
        \Tube{(A)};
        }{}
    \end{tikzpicture}
    \quad \scalebox{3}{+} \quad
    \begin{tikzpicture}
        \BoxGraph[black][1.2]{1.1}{
        \draw [-->-=0.8, thick] (A)--(B);
        \draw [-->-=0.8, thick] (A)--(D);
        \draw [-->-=0.8, thick] (B)--(C);
        \draw [-->-=0.8, thick] (D)--(C);
        \Tube[rounded corners=10pt, inner sep=5pt]{(A) (B) (D)};
        }{}
    \end{tikzpicture}
    \label{eq:DegeneracySquareGraph2}
\end{equation}

This linear dependence reflects the non-simplicity of the gray vertex (adjacent to four facets), obtained from merging two simple vertices. The vanishing of one or several $y$ variables combinatorially changes the graphical zonotope, possibly making faces of any codimension disappear. In fact, in some cases, facets can disappear by taking some $y_i \rightarrow 0$. It is however not the case here, because the facets of the square graph and the square with a chord graph are in correspondence, since both graphs have the same minimal cuts.

\subsection{Physics from the graphical zonotope \label{sec:Steinmann}}

The graphical zonotope $\mathcal{Z}_\Gamma$ provides valuable information about the physical properties of the corresponding Feynman integral. Although these properties were discussed in Ref.~\cite{Benincasa:2021qcb} through an analysis of the full SF, the graphical zonotope leads to the same conclusions in a more direct and transparent manner. 

\paragraph{Channel factorization}  Factorization of residues in causal channels translates into the fact that faces of $\mathcal{Z}_\Gamma^Y$ factorize into products of graphical zonotopes of subgraphs of $\Gamma$. This is very similar to the factorization of the faces of the ABHY associahedron~\cite{Arkani-Hamed:2017mur}. The simultaneous vanishing of $k$ orientation-compatible causal thresholds $\lambda_1,\dots,\lambda_k$, i.e., which do not cut any edge in opposite directions, gives a face of $\mathcal{Z}_\Gamma^Y$ of some codimension~$m$. We will see that $\mathcal{Z}_\Gamma^Y$ is a subset of $\mathcal{S}_\Gamma$, i.e. a linear section of it. So, the faces of $\mathcal{Z}_\Gamma^Y$ can be written as $\mathcal{Z}_\Gamma^Y\cap L_1 \cap \dots \cap L_k $ (the facets $L_S$ were defined in \ref{eq:DefLS}). We call $m$ the order of the cut induced by these faces, i.e. $\Gamma - (\bigcup_{i=1}^k \mathcal{C}_i)$ has $m+1$ connected components,
where $\mathcal{C}_i$ is the subset of edges intersected by the minimal cut corresponding to $\lambda_i$. Then we have
\begin{equation}
    \mathcal{Z}_\Gamma^Y\cap L_1 \cap \dots \cap L_k \cong \mathcal{Z}_{\Gamma_1}^Y \times \dots \times \mathcal{Z}_{\Gamma_{m+1}}^Y.
\end{equation}
This product decomposition is the graphical-zonotope counterpart of amplitude factorization, since the canonical form factorizes into $\Omega(\mathcal{Z}_{\Gamma_1}^Y) \wedge \dots \wedge \Omega(\mathcal{Z}_{\Gamma_{m+1}}^Y)$. The exact relation between the canonical form of $\mathcal{Z}_{\Gamma}^Y$ will become clear in the next section. 

Whenever $k=m$ and the $L_1,\dots,L_k$ are linearly independent hyperplanes, this translates algebraically into 
\begin{equation}
    \mathrm{Res}_{\lambda_k=0}\dots \mathrm{Res}_{\lambda_1=0}\,\, \Omega(\mathcal{Z}_{\Gamma}^Y)=\Omega(\mathcal{Z}_{\Gamma_1}^Y) \wedge \dots \wedge \Omega(\mathcal{Z}_{\Gamma_{k+1}}^Y)\footnote{Here the order of the causal thresholds in the nested residues should be taken such that $\bigcup_{i=1}^a \mathcal{C}_i$ is a cut of order $a$ for any $a$ such that $1\leq a\leq k$, otherwise the nested residues vanish. This is explained in more detail in Sec.~\ref{sec:RefinementFan}.}.
\end{equation}
We give an argument based on oriented matroids to explain why the faces of $\mathcal{Z}_\Gamma^Y$ factorize in this way. The face $\mathcal{Z}_\Gamma^Y \cap L_1 \cap \dots \cap L_k $ considered above is encoded by a signed covector $X$ of the graphic oriented matroid, and its zero set is $X^0=E(\Gamma_1 \cup \dots \cup \Gamma_{m+1})$. In fact, only the edges intersected by the oriented cut defining the face, corresponding to edges flowing between the partition subsets, are assigned a "+" or "-". Applying Eq. (\ref{eq:FaceGZ}), we get
\begin{equation}
    \mathcal{Z}_\Gamma^Y \cap L_1 \cap \dots \cap L_k = \sum_{s=1}^{m+1}\left(\sum_{e=\{ij\}\in \Gamma_s} y_e \left[\mathbf{x_j}-\mathbf{x_i}, \mathbf{x_i}-\mathbf{\mathbf{x_j}} \right] \right)+ \sum_{e=\{ij\}\in X^+}y_e \left( \mathbf{x_i}-\mathbf{x_j} \right)+ \sum_{e=\{ij\}\in X^-} y_e \left( \mathbf{x_j}-\mathbf{x_i} \right).
     \label{eq:FaceGZ2}
\end{equation}

The last two terms in \Eq{eq:FaceGZ2} only contribute to a translation of the face, indicating the position of the face inside $\mathcal{Z}_\Gamma^Y$. We are interested in the first term, which defines its geometry. Each term in the sum over $s$ is the graphical zonotope $\mathcal{Z}_{\Gamma_s}^Y$, so that, forgetting translations, 
\begin{equation}
    \mathcal{Z}_\Gamma^Y \cap L_1 \cap \dots \cap L_k \cong\sum_{i=1}^{m+1} \mathcal{Z}_{\Gamma_s}^Y.
\end{equation}
Because all subgraphs $\Gamma_s$ are disjoint, their vertex sets are, and the graphical zonotopes $\mathcal{Z}_{\Gamma_s}^Y$ are linearly independent (their spans are independent). Then, the Minkowski sum becomes a product, i.e.
\begin{equation}
    \mathcal{Z}_\Gamma^Y \cap L_1 \cap \dots \cap L_k \cong \mathcal{Z}_{\Gamma_1}^Y \times \dots \times \mathcal{Z}_{\Gamma_{m+1}}^Y.
    \label{eq:FactorizationGZ}
\end{equation}

\paragraph{Generalized Steinmann relations}

The relation between generalized Steinmann relations, and the CP and the SF was studied in~\cite{Benincasa:2021qcb,Benincasa:2020aoj}. In the physical region, Steinmann relations state that double discontinuities vanish in partially overlapping channels. Consider any two vertex subsets $S_1,S_2\subset V(\Gamma)$ inducing connected subgraphs and their complements $\bar{S}_1,\bar{S}_2$. Call $\lambda_1$ and $\lambda_2$ the denominators whose vanishing is associated to the ordered partitions $S_1 \mid \bar{S}_1$ and $S_2 \mid \bar{S}_2$ respectively. Then the Steinmann relations are
\begin{equation}
      \left\{\begin{aligned}
    & S_1 \cap S_2 \neq \emptyset ,\\
    & \bar{S}_1 \cap S_2 \neq \emptyset, \\
    & S_1 \cap \bar{S}_2 \neq \emptyset, \\
    & \bar{S}_1 \cap \bar{S}_2 \neq \emptyset,
    \end{aligned}\right. \quad \Rightarrow  \quad \mathrm{Disc}_{S_1}\mathrm{Disc}_{S_2} M=0 ,
    \label{eq:Steinmann}
\end{equation}
where $M$ is a scattering amplitude. Satisfying all conditions on the left-hand side of \Eq{eq:Steinmann} defines a pair of overlapping channels. These Steinmann relations have a counterpart in the Feynman energy integral $\mathcal{I}_\Gamma$, they correspond to the vanishing of iterated residues in the causal thresholds. However, there are other pairs of subsets $S_1, S_2$ that do not satisfy these conditions above but for which the corresponding double residue of $\mathcal{I}_\Gamma$ vanishes. 

Recall from the Landau analysis in Sec.~\ref{sec:Landau} that the SF must have two kinds of facets: the first kind of facets are the $\{y_e=0\}\cap \mathcal{S}_\Gamma$, while the second kind are the $L_i$, where $i$ labels the oriented bonds/causal thresholds. We therefore would like to know which facets of the second kind intersect in the SF.

Consider $Y$ with all $y_e>0$. Then, it turns out that $k$ of these facets, such that they correspond to orientation-compatible bonds, have a common intersection in $\mathcal{S}_\Gamma$ in codimension $k$, if and only if they do so in $\mathcal{Z}_\Gamma^Y$: 
\begin{equation}
    \mathrm{codim}_{\mathcal{S}_\Gamma} (L_1 \cap \dots \cap L_k)=k \Leftrightarrow\mathrm{codim}_{\mathcal{Z}_\Gamma^Y}(\mathcal{Z}_\Gamma^Y \cap L_1 \cap \dots \cap L_k)=k.
    \label{eq:relCodim}
\end{equation}
This should become clear when we study the relation between both polytopes. Therefore, the face structure of the graphical zonotope encodes all the information on whether iterated residues of the integral vanish. Considering more than two channels gives us the generalized Steinmann relations.

We know that the dimension of $\mathcal{Z}_{\Gamma_i}^Y$ for a connected graph $\Gamma_i$ with $N_{\Gamma_i}$ vertices is $N_{\Gamma_i}-1$. Therefore, from the factorization property (\ref{eq:FactorizationGZ}),
\begin{equation}
    \mathrm{dim}(\mathcal{Z}_\Gamma^Y\cap L_1 \cap \dots \cap L_k)=\sum_{i=1}^{m+1}\mathrm{dim}(\mathcal{Z}_{\Gamma_i}^Y)=\sum_{i=1}^{m+1} N_{\Gamma_i}-m-1=\mathrm{dim}(\mathcal{Z}_\Gamma^Y)-m,
\end{equation}
recalling that $L_1,\dots,L_k$ must correspond to orientation-compatible causal thresholds. Therefore, the rule is the following: $k$ facets of $\mathcal{Z}_\Gamma^Y$ intersect in codimension $k$ if and only if the corresponding $k$ orientation-compatible cuts partition the graph into $k+1$ connected subgraphs. Call $\mathcal{C}_i$ the minimal cut associated to causal threshold $\lambda_i$. Then, in order for $\mathrm{Res}_{\lambda_k=0} \dots \mathrm{Res}_{\lambda_1=0} \, \, \Omega(\mathcal{Z}_\Gamma^Y)\neq 0$ we need the cut $\bigcup_{i=1}^k \mathcal{C}_i$ to be of order~$k$. 

For $k=2$, one can see that if $S_1$ and $S_2$ are partially overlapping channels, then the union of their associated cuts $\mathcal{C}_1\cup \mathcal{C}_2$ partitions $\Gamma$ into at least four connected subgraphs, so $\mathrm{Res}_{\lambda_{S_2}=0}  \mathrm{Res}_{\lambda_{S_1}=0} \, \, \Omega(\mathcal{Z}_\Gamma^Y)= 0$. Recalling property (\ref{eq:relCodim}) and relation (\ref{eq:IntegralCanonicalForm}) this becomes
\begin{equation}
    \mathrm{Res}_{\lambda_{S_2}=0}  \mathrm{Res}_{\lambda_{S_1}=0} \, \, \mathcal{I}_\Gamma= 0.
\end{equation}
This is the residue counterpart of the usual Steinmann relation. However, there may be pairs of channels which are not partially overlapping but nevertheless give a vanishing double residue of $\mathcal{I}_\Gamma$.

For $k>2$, it is not enough to check whether the $k$ facets intersect on codimension $k$ to conclude that the $k$-fold iterated residues is non-vanishing. Obviously, we need $\mathrm{Res}_{\lambda_a=0} \dots \mathrm{Res}_{\lambda_1=0} \, \, \Omega(\mathcal{Z}_\Gamma^Y)\neq0$ for all $a=1,\dots,k$. Consequently, the rule is
\begin{equation}
     \mathrm{Res}_{\lambda_k=0} \dots \mathrm{Res}_{\lambda_1=0} \, \, \mathcal{I}_\Gamma\neq 0 \quad \Leftrightarrow \quad \bigcup_{i=1}^a \mathcal{C}_i \text{ is of order } a \text{ for all }a=1,\dots,k,
     \label{eq:RuleRes}
\end{equation}
where the cuts $\mathcal{C}_i$ are assumed to be pairwise orientation-compatible. Note that rule (\ref{eq:RuleRes}) shows how in many cases the order in which the residues are taken determines whether the iterated residue vanishes.
  
Let us show a couple of examples, starting with a case of a vanishing double residue that does not correspond to partially overlapping channels as defined by the conditions in (\ref{eq:Steinmann}). The two following minimal cuts $\mathcal{C}_1$ and $\mathcal{C}_2$ in the square graph 
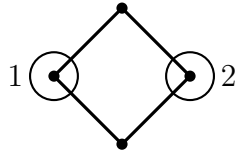
\begin{equation}
    \tikz[baseline=-0.5ex]{
\BoxGraph{3}{
  \node[font=\large] at ($(A)+(-0.17,0)$) {1};
  \node[font=\large] at ($(C)+(0.17,0)$) {2};
  \draw[very thick] (A)--(D);
  \draw[very thick] (A)--(B);
  \draw[very thick] (B)--(C);
  \draw[very thick] (C)--(D);
  \Tube[rounded corners=9pt, inner sep=9pt]{(A)};
  \Tube[rounded corners=9pt, inner sep=9pt]{(C)};
}
} \notag
\end{equation}
cut the four edges of the graph, thus partitioning the graph into $4$ connected components, where each connected component is a vertex. This means $\mathcal{C}_1\cup\mathcal{C}_2 $ is a cut of order $3$, and so corresponds to a codimension $3$ face of $\mathcal{Z}_{\begin{tikzpicture}
    \BoxGraph[black][1.2]{0.3}{}{
    \draw (A)--(B);
    \draw (A)--(D);
    \draw (B)--(C);
    \draw (D)--(C);
    }{}
\end{tikzpicture}}$, i.e. a vertex. A non-vanishing double residue would correspond to a codimension $2$ face of $\mathcal{Z}_{\begin{tikzpicture}
    \BoxGraph[black][1.2]{0.3}{}{
    \draw (A)--(B);
    \draw (A)--(D);
    \draw (B)--(C);
    \draw (D)--(C);
    }{}
\end{tikzpicture}}$, 
so 
\begin{equation}
    \mathrm{Res}_{\lambda_2=0}\mathrm{Res}_{\lambda_1=0}\,\, \mathcal{I}_{\, \begin{tikzpicture}
    \BoxGraph[black][1.2]{0.3}{}{
    \draw (A)--(B);
    \draw (A)--(D);
    \draw (B)--(C);
    \draw (D)--(C);
    }{}
\end{tikzpicture}}=0.
\end{equation}

Let us consider a case with three minimal cuts, two of them overlapping:
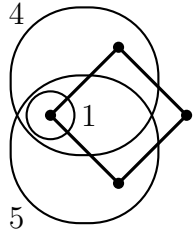
\begin{equation}
    \tikz[baseline=-0.5ex]{
\BoxGraph{3}{
  \node[font=\large] at ($(A)+(0.17,0)$) {1};
  \node[font=\large] at ($0.5*(A)+0.5*(B)+(-0.3,0.3)$) {4};
  \node[font=\large] at ($0.5*(A)+0.5*(D)+(-0.3,-0.3)$) {5};
  \draw[very thick] (A)--(D);
  \draw[very thick] (A)--(B);
  \draw[very thick] (B)--(C);
  \draw[very thick] (C)--(D);
  \Tube[rounded corners=9pt, inner sep=9pt]{(A)};
  \Tube[rounded corners=26pt, inner sep=15pt]{(A) (B)};
  \Tube[rounded corners=26pt, inner sep=15pt]{(A) (D)};
}
} \notag
\end{equation}
The cut $\mathcal{C}_1\cup\mathcal{C}_4 \cup \mathcal{C}_5$ is also of order $3$ since it partitions the graph into its four vertices. The intersection of $L_1$, $L_4$ and $L_5$ is therefore of codimension $3$ in $\mathcal{S}_{\begin{tikzpicture}
    \BoxGraph[black][1.2]{0.3}{}{
    \draw (A)--(B);
    \draw (A)--(D);
    \draw (B)--(C);
    \draw (D)--(C);
    }{}
\end{tikzpicture}}$, and in the same way we see that the intersection of $L_1$ and $L_4$ is of codimension 2. Then, the nested residue is non-vanishing:
\begin{equation}
    \mathrm{Res}_{\lambda_5=0}\mathrm{Res}_{\lambda_4=0}\mathrm{Res}_{\lambda_1=0}\,\, \mathcal{I}_{\, \begin{tikzpicture}
    \BoxGraph[black][1.2]{0.3}{}{
    \draw (A)--(B);
    \draw (A)--(D);
    \draw (B)--(C);
    \draw (D)--(C);
    }{}
\end{tikzpicture}}\neq 0.
\end{equation}
Note that other orderings for the nested residues give a vanishing result, like $\mathrm{Res}_{\lambda_1=0}\mathrm{Res}_{\lambda_4=0}\mathrm{Res}_{\lambda_5=0}\,\, \mathcal{I}_{\, \begin{tikzpicture}
    \BoxGraph[black][1.2]{0.3}{}{
    \draw (A)--(B);
    \draw (A)--(D);
    \draw (B)--(C);
    \draw (D)--(C);
    }{}
\end{tikzpicture}}$. This is because $\mathcal{C}_4 \cup \mathcal{C}_5$ partitions $\Gamma$ into $4$ subgraphs, so the intersection of $L_4$ and $L_5$ is of codimension $3$ : $\mathrm{Res}_{\lambda_4=0}\mathrm{Res}_{\lambda_5=0}\,\, \mathcal{I}_{\, \begin{tikzpicture}
    \BoxGraph[black][1.2]{0.3}{}{
    \draw (A)--(B);
    \draw (A)--(D);
    \draw (B)--(C);
    \draw (D)--(C);
    }{}
\end{tikzpicture}}=0$. The channels $4$ and $5$ are in fact partially overlapping channels.

\section{The scattering facet as a Cayley polytope \label{sec:CayleyPolytope}}

From Sec.~\ref{sec:GraphicalZonotope}, we can discern a clear relation between the graphical zonotope and the SF. The relation is made transparent by treating the SF $\mathcal{S}_\Gamma$ as a \textit{Cayley polytope}, which is obtained through an embedding of segments in edge space $Y$. Then, $\mathcal{S}_\Gamma$ admits a base-fiber description, in which the fibers are deformed graphical zonotopes. This geometric picture will be helpful to study the face structure of $\mathcal{S}_\Gamma$ and, especially, its triangulations. It is important not to confuse these Cayley polytopes, studied in the mathematical literature, with another family of polytopes known under the same name in the context of positive geometries, such as in Ref.~\cite{He:2018pue}, which constitute a generalization of the associahedron. As far as we know, there is no relation between them.

\subsection{The Cayley embedding \label{sec:CayleyEmbedding}}

We follow the presentation in Ref.~\cite{ubt_eref1116}. Let $A_1,\dots,A_n$ be polytopes in $\mathbb{R}^m$, a space with coordinates $x_i$, and $\mathbf{y_1}, \dots, \mathbf{y_n}$ the standard basis vectors in a different space $\mathbb{R}^n$ and coordinates $y_i$. Define the embedding $\mu_i:\mathbb{R}^m \rightarrow \mathbb{R}^m \times  \mathbb{R}^{n} $, $\mu_i(\mathbf{x})=(\mathbf{x},\mathbf{y_i})$. Then, the \textit{Cayley embedding} $\mathcal{C}(A_1,\dots,A_n) \subset \mathbb{R}^m \times  \left\{ y_i : \sum_i  y_i=1 \right\} \cong \mathbb{R}^m \times \mathbb{R}^{n-1}$ is 
\begin{equation}
    \mathcal{C}(A_1,\dots,A_n)\equiv \bigcup_{i=1}^n  \mu_i(A_i).
\end{equation}
The convex hull of this union is then called a \textit{Cayley polytope}:
\begin{equation}
    \rm Cay(A_1,\dots,A_n)  \equiv \rm conv \big(\mathcal{C}(A_1,\dots,A_n)\big)
\end{equation}
Actually, the Cayley polytope can simply be obtained by taking the convex hull of the union of vertices of the embedded polytopes, $\rm Cay(A_1,\dots,A_n)=\rm conv \big(  \bigcup_{i=1}^n \mu_i(V(A_i)) \big)$, $V(A_i)$ being the set of vertices of the polytope $A_i$. 

Cayley polytopes can be seen as fibers (polytopes themselves) over a base space, which is the simplex $\Delta^{n-1}=\left\{y_i \geq 0, \sum_i y_i=1\right\}$. Then, there is a projection $\pi :\rm Cay(A_1,\dots,A_n) \rightarrow \Delta^{n-1}$, $\pi(x,y)=y$. Take $y=(y_1,\dots,y_n) \in \Delta^{n-1}$. The fiber over $y$ is 

\begin{equation}
    \pi^{-1}(y)=  \left(\sum_{i=1}^n y_i A_i,y\right),
    \label{eq:CayleyFiber}
\end{equation}
where the summation is a Minkowski sum. 

Let us now go back to the SF. Let $\mathbf{y_e}$, $e\in E(\Gamma)$ and $\mathbf{x_i}$, $i=1,\dots,N$ be the standard basis vectors in kinematic space as in the previous sections. Define the segments in $X$ space $w_{e}\equiv [\mathbf{x_j}-\mathbf{x_i},\mathbf{x_i}-\mathbf{x_j}]\subset \mathbb{R}^{N-1}$, restricting to the $\sum x_i=0$ hyperplane in which the segments live, such that $i$ and $j$ label the vertices that are the endpoints of $e\in E(\Gamma)$. Then, for a graph $\Gamma$, $A_e=w_{e}$. The embedding is done in $Y$ space, $\mu_e(w_{e})=(w_{e},\mathbf{y_e})$. We get 
\begin{equation}
    \mathcal{C}(A_1,\dots,A_n)\equiv \bigcup_{e \in E(\Gamma)}  (w_{e},\mathbf{y_e})  \subset \mathbb{R}^{N-1} \times \mathbb{R}^{E-1} .
\end{equation}
Then its associated Cayley polytope is the convex hull of $\bigcup_{e=ij\in E(\Gamma)} \{ \mathbf{x_i}-\mathbf{x_j} +\mathbf{y_e}, \mathbf{x_j}-\mathbf{x_i} +\mathbf{y_e} \}$. This is precisely the vertex definition of the SF $\mathcal{S}_\Gamma$, as shown in Eq. (\ref{eq:VerticesSF}). 

The fiber over a point $y=(y_1,\dots,y_E)\in \Delta^{E-1}$ is $\sum_{e \in E(\Gamma)} y_e w_e$ according to Eq. (\ref{eq:CayleyFiber}), so it is in correspondence with a deformed graphical zonotope living in $X$ space (see \Eq{eq:DeformedZonotpeWeightedSum}). Then, we can view the simplex in edge space $\Delta^{E-1}$ as parametrizing the deformations of the graphical zonotope. These deformations are the family of deformations characterized by the submodular functions defined in (\ref{eq:SubmodCuts}). They lead to the inequalities (\ref{eq:CSinequalities}) defining the deformed graphical zonotopes. From our Landau analysis in Sec.~\ref{sec:Landau}, we know them to be some of the inequalities from the facet description of the SF. The fiber over the boundary of the simplex characterized by $y_e=0$ gives Minkowski sums with the vanishing weight $y_e=0$. This corresponds to a deformed graphical zonotope of the graph $\Gamma$ with the edge $e$ removed. If there is no other edge with the same endpoints as $e$, this leads to degenerations of faces of the graphical zonotope as discussed in Sec.~\ref{sec:GeneralizedPermutohedra}.

\begin{figure}[h]
\centering

\begin{minipage}{0.48\textwidth}
\centering
\begin{tikzpicture}[scale=2,vertex/.style={circle,fill=black,inner sep=1.2pt},
        every node/.style={font=\small}]

        \coordinate (v1) at (-0.78446,1.27811);
    \coordinate (v2) at (0.78446,0.64914);
    \coordinate (v3) at (1.41128,0.02433);
    \coordinate (v4) at (-1.41128,-0.02433);
    
    \coordinate (d1) at ($1.3*(v1)-0.3*(v2)$);   
    \coordinate (d2) at ($1.15*(v2)-0.15*(v1)$);  
    \draw [dash pattern=on 4pt off 4pt,line width=1pt,black] (d1)--(d2);
    \coordinate (d3) at ($1.15*(v3)-0.15*(v4)$);   
    \coordinate (d4) at ($1.15*(v4)-0.15*(v3)$);  
    \draw [dash pattern=on 4pt off 4pt,line width=1pt,black] (d3)--(d4);
    \node[above]  at (d1) {$y_{12}=1$};
    \node[below]  at (d4) {$y_{12}=0$};
    
    
    \coordinate (z1) at (0.752558, 0.400461);
    \coordinate (z2) at (-1.22324, 0.366405);
    \coordinate (z3) at (1.22324, 0.211769);
    \coordinate (z4) at (-0.752558, 0.177714);
    
    \coordinate (w1) at ($1.3*(z1)-0.3*(z4)$);   
    \coordinate (w2) at ($1.3*(z2)-0.3*(z3)$);  
    \coordinate (w3) at ($1.3*(z3)-0.3*(z2)$);   
    \coordinate (w4) at ($1.3*(z4)-0.3*(z1)$);  
    \draw[thick] (w1)--(w3);
    \draw[thick] (w3)--(w4);
    \draw[thick] (w4)--(w2);
    
    \node[black,above right] at (w2) {$y_{12}=0.3$};

    \fill[blue!10,opacity=.5] (v4)--(v1)--(v2)--cycle;
    \fill[blue!10,opacity=.5] (v3)--(v4)--(v2)--cycle;
    \fill[blue!10,opacity=.5] (v1)--(v3)--(v2)--cycle;
    \fill[blue!10,opacity=.5] (v4)--(v3)--(v1)--cycle;
    \draw[dash pattern=on 3pt off 2pt,black!60,line width=.9pt] (v2)--(v4);
    \draw[black,line width=0.9pt] (v4)--(v1);
    \draw[blue,line width=1.5pt] (v1)--(v2);
    
    \draw[black,line width=.9pt] (v3)--(v2);
    \draw[red,line width=1.5pt] (v3)--(v4);
    
    \fill[gray!45,opacity=1] (z1)--(z3)--(z4)--(z2)--cycle;
    
    \draw[black,line width=.9pt] (v1)--(v3);
    
    \node[blue,above] at ($(v1)!0.5!(v2)$) {$w_{12}$};
    \node[red,below] at ($(v3)!0.5!(v4)$) {$w_{23}$};

    \draw[thick] (w1)--(w2);
    
    \foreach \P in {v1,v2,v3,v4}
       \node[vertex] at (\P) {};
    
    \node[above right]  at (v1) {$\mathbf y_{12}+\mathbf x_1-\mathbf x_2$};
    \node[above right] at (v2) {$\mathbf y_{12}-\mathbf x_1+\mathbf x_2$};
    \node[below]  at (v3) {$\mathbf y_{23}+\mathbf x_2-\mathbf x_3$};
    \node[below right] at (v4) {$\mathbf y_{23}-\mathbf x_2+\mathbf x_3$};

\end{tikzpicture}
\end{minipage}
\hfill
\begin{minipage}{0.48\textwidth}
\centering
\begin{tikzpicture}[scale=2]

    
    \coordinate (v1) at (-0.0707107, 0.857321);
    \coordinate (v2) at (0.919239, -0.857321);
    \coordinate (v3) at (-0.919239, 0.857321);
    \coordinate (v4) at (0.0707107, -0.857321);
    
    \fill[gray!25] (v1)--(v3)--(v4)--(v2)--cycle;
    \draw[thick] (v1)--(v3)--(v4)--(v2)--cycle;

    \draw[very thick,blue] (v4)--(v2)
    node[above right] {$0.3\,w_{12}$};
    \draw[very thick,red] (v4)--(v3)
    node[pos=0.8,left] {$0.7\,w_{23}$};

    \coordinate (d1) at ($1.05*(v1)-0.05*(v4)$);
    \coordinate (d2) at ($1.05*(v2)-0.05*(v3)$);
    \coordinate (d3) at ($1.05*(v3)-0.05*(v2)$);
    \coordinate (d4) at ($1.05*(v4)-0.05*(v1)$);
    
    \ThreeTree{d1}{0.55}{
    \draw[-->-=0.8, thick] (A)--(B);
    \draw[-->-=0.8, thick] (B)--(C);
    }{}     
    \ThreeTree{d2}{0.55}{
    \draw[-->-=0.8, thick] (A)--(B);
    \draw[-->-=0.8, thick] (C)--(B);
    }{}  
    \ThreeTree{d3}{0.55}{
    \draw[-->-=0.8, thick] (B)--(A);
    \draw[-->-=0.8, thick] (B)--(C);
    }{}  
    \ThreeTree{d4}{0.55}{
    \draw[-->-=0.8, thick] (B)--(A);
    \draw[-->-=0.8, thick] (C)--(B);
    }{}  

    \coordinate (e12) at ($0.75*(v1)+0.75*(v2)-0.25*(v4)-0.25*(v3)$);
    \coordinate (e34) at ($0.75*(v3)+0.75*(v4)-0.25*(v1)-0.25*(v2)$);
    \coordinate (e13) at ($0.45*(v3)+0.45*(v1)+0.05*(v2)+0.05*(v4)$);
    \coordinate (e24) at ($0.45*(v2)+0.45*(v4)+0.05*(v1)+0.05*(v3)$);
    
    \ThreeTree{e12}{0.45}{
    \draw [-->-=0.8, thick] (A)--(B);
    \Tube[rounded corners=7pt,inner sep=7pt]{(A)};
    }{}
    \ThreeTree{e34}{0.45}{
    \draw [-->-=0.8, thick] (B)--(A);
    \Tube[rounded corners=8pt, inner sep=7pt]{(B) (C)};
    }{}
    \ThreeTree{e13}{0.45}{
    \draw [-->-=0.8, thick] (B)--(C);
    \Tube[rounded corners=8pt,inner sep=7pt]{(A) (B)};
    }{}
    \ThreeTree{e24}{0.45}{
    \draw [-->-=0.8, thick] (C)--(B);
    \Tube[rounded corners=7pt,inner sep=7pt]{(C)};
    }{}

\end{tikzpicture}
\end{minipage}

\caption{SF of the tree graph $T_3$ (left). The section obtained by intersecting it with the plane $\{ y_{12}=0.3 \}$ is the deformed graphical zonotope of $T_3$, with generators $0.3 w_{12}$ and $0.7 w_{23}$ (right).}
\label{fig:Tetrahedron}
\end{figure}
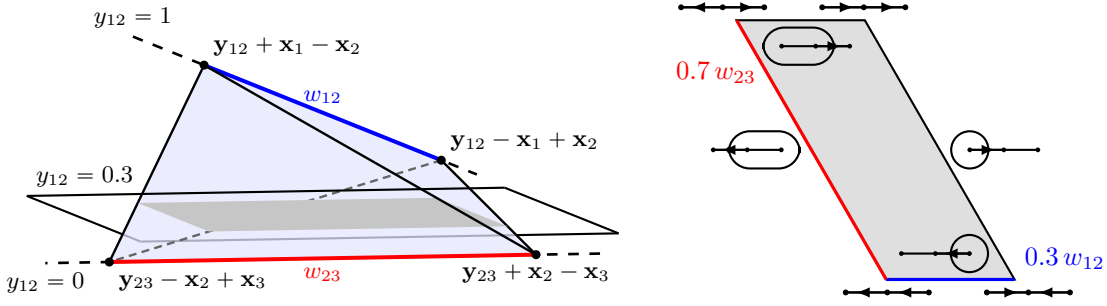

An example of the base-fiber structure of the SF is given by the tree graph with three vertices 
$T_3 \equiv \begin{tikzpicture}
    \ThreeTree{(0,0)}{1}{}{}
\end{tikzpicture}$ in Fig. \ref{fig:Tetrahedron}. Because the graph is a tree, the SF is a simplex of dimension $E+N-2=3$: it is a tetrahedron, as depicted in Fig.~\ref{fig:Tetrahedron} (left). It can be decomposed into the base space simplex $\Delta^1$, which ranges from $y_{12}=0$ (i.e. $y_{23}=1$) to $y_{12}=1$, and the fiber over each value of $y_{12}$, which is a deformed graphical zonotope of $T_3$. The intersection of $\mathcal{S}_{T_3}$ with the plane $\{ y_{12}=0.3 \}$ gives a zonotope with generating segments $0.3 w_{12}$ and $0.7 w_{23}$, as seen in Fig.~\ref{fig:Tetrahedron} (right). Looking at the boundaries of the base space $\Delta^1$, $y_{12}=0$ and $y_{23}=0$, one of the generators vanishes, $y_{12}w_{12}=0$ and $y_{23}w_{23}=0$ respectively, and the fibers over them degenerate and become a line segment, corresponding to the graphical zonotope of the tree graph with two vertices and one edge $T_2$.

\subsection{Face structure of the scattering facet \label{sec:FaceStructure}}

From our Landau analysis in Sec.~\ref{sec:Landau}, we know that facets of $\mathcal{S}_\Gamma$ are given by
\begin{itemize}
    \item $\lambda_S=0$ for every oriented bond labeled by $S$.
    \item $y_e=0$ for every edge $e$ that is not a bridge.
\end{itemize}
The fact that the latter is not a facet for bridge edges is illustrated in Fig.~\ref{fig:Tetrahedron}. Each facet corresponds to one of the four minimal cuts of the $T_3$ graph, but faces $\mathcal{S}_\Gamma\cap\{y_{12}=0\}$ and $\mathcal{S}_\Gamma\cap\{y_{23}=0\}$ are of codimension two, the red and blue segments respectively. 

In fact, let $S_e$ be the subset corresponding to a bond cutting a bridge $e$. Then
\begin{equation}
    2y_e=\lambda_{S_e}+\lambda_{V\backslash S_e},
    \label{eq:BridgeDegenration}
\end{equation}
so  
\begin{equation}
    y_e=0  \implies \lambda_{S_e}=0 \text{ and } \lambda_{V\backslash S_e}=0
    \label{eq:BridgeDegenration2}
\end{equation}
because of positivity of the causal thresholds. In this case the extra relation obtained when taking $y_e=0$ means that the corresponding face lies at codimension $2$. It only happens for bridges, since for an edge $e'$ that is not a bridge, any partition of the graph cutting $e'$ would also cut at least another edge. Then the degeneration \Eq{eq:BridgeDegenration} does not happen.

In Sec.~\ref{sec:Landau}, we showed that solutions to the Landau equations could be seen as combinations of minimal cuts and edge deletions. In Sec.~\ref{sec:DefCombinatoricsGZ} we defined a cut of order $n$, i.e. a partition dividing the graph into $n+1$ connected subgraphs, together with an acyclic orientation between partition subgraphs. Such a cut can be decomposed into $n$ independent causal thresholds, i.e. gives $n$ independent relations on the external parameters, of the form $\lambda_{\dots}=0$. 

The deletion of an edge $e$ adds one relation, $y_e=0$, except when it disconnects the graph, in which case it gives another extra relation $\lambda_{S_e}=0$, as we showed above. Therefore, any face of $\mathcal{S}_\Gamma$ can be represented as a set of $n$ independent minimal cuts on any subgraph, with at least one edge~\footnote{Otherwise we get $\sum_{e \in E(\Gamma)} y_e=0$.}, of $\Gamma$ obtained by removing edges from it. We call \textit{decorated graph of the face $\mathcal{F}$} this subgraph with cuts used to represent a face $\mathcal{F} \subseteq \mathcal{S}_\Gamma$, and we denote it by $G_\mathcal{F}$. The codimension of a face $\mathcal{F}$ is then
\begin{equation}
    \rm codim(\mathcal{F})= \# \text{(deleted edges)}+\# \text{(connected components)}-1+(\text{order of the cut}).
    \label{eq:codim}
\end{equation}

Let us illustrate this with the square graph $\begin{tikzpicture}
    \BoxGraph[black][1.2]{0.6}{}{
    \draw (A)--(B);
    \draw (A)--(D);
    \draw (B)--(C);
    \draw (D)--(C);
    }{}
\end{tikzpicture}$. Let $\{12\}$ and $\{34\} $ correspond to two opposite edges in the graph. Consider the face $\mathcal{F}_{
\begin{tikzpicture}
\begin{scope}[shift={(0,0)},scale=0.3]
\def\boxcolor{black}

\coordinate (A) at (-0.3,0);
\coordinate (B) at (0,0.3);
\coordinate (C) at (0.3,0);
\coordinate (D) at (0,-0.3);

\fill[\boxcolor] (A) circle (1.2pt);
\fill[\boxcolor] (B) circle (1.2pt);
\fill[\boxcolor] (C) circle (1.2pt);
\fill[\boxcolor] (D) circle (1.2pt);

\draw[\boxcolor] (A)--(B);
\draw[\boxcolor] (C)--(D);

\end{scope}
\end{tikzpicture}
} = \mathcal{S}_{\begin{tikzpicture}
    \BoxGraph[black][1.2]{0.3}{}{
    \draw (A)--(B);
    \draw (A)--(D);
    \draw (B)--(C);
    \draw (D)--(C);
    }{}
\end{tikzpicture}}\cap \{ y_{12}=0 \}\cap \{ y_{34}=0 \}$, whose decorated graph is $\begin{tikzpicture}
\begin{scope}[shift={(0,0)},scale=0.6]
\def\boxcolor{black}

\coordinate (A) at (-0.3,0);
\coordinate (B) at (0,0.3);
\coordinate (C) at (0.3,0);
\coordinate (D) at (0,-0.3);

\fill[\boxcolor] (A) circle (1.2pt);
\fill[\boxcolor] (B) circle (1.2pt);
\fill[\boxcolor] (C) circle (1.2pt);
\fill[\boxcolor] (D) circle (1.2pt);

\draw[\boxcolor] (A)--(B);
\draw[\boxcolor] (C)--(D);

\end{scope}
\end{tikzpicture}$. It has codimension three according to \Eq{eq:codim}, because it implies the extra relation $\lambda_{(1,3)}=0$ (or $\lambda_{(2,4)}=0$). On the other hand, consider the face $\mathcal{F}_{
\begin{tikzpicture}
\begin{scope}[shift={(0,0)},scale=0.4]
\def\boxcolor{black}

\coordinate (A) at (-0.3,0);
\coordinate (B) at (0,0.3);
\coordinate (C) at (0.3,0);
\coordinate (D) at (0,-0.3);

\fill[\boxcolor] (A) circle (1.2pt);
\fill[\boxcolor] (B) circle (1.2pt);
\fill[\boxcolor] (C) circle (1.2pt);
\fill[\boxcolor] (D) circle (1.2pt);

\draw[\boxcolor] (A)--(B);
\draw[\boxcolor] (C)--(D);
\draw[\boxcolor] (B)--(C);
\Tube[rounded corners=3pt,inner sep=1.7pt]{(A) (B)};

\end{scope}
\end{tikzpicture}
} =  \{ y_{12}=0 \}\cap L_{(1,3)}$, whose decorated graph is $\begin{tikzpicture}
\begin{scope}[shift={(0,0)},scale=0.6]
\def\boxcolor{black}

\coordinate (A) at (-0.3,0);
\coordinate (B) at (0,0.3);
\coordinate (C) at (0.3,0);
\coordinate (D) at (0,-0.3);

\fill[\boxcolor] (A) circle (1.2pt);
\fill[\boxcolor] (B) circle (1.2pt);
\fill[\boxcolor] (C) circle (1.2pt);
\fill[\boxcolor] (D) circle (1.2pt);

\draw[\boxcolor] (A)--(B);
\draw[\boxcolor] (C)--(D);
\draw[\boxcolor] (B)--(C);
\Tube[rounded corners=5pt,inner sep=2.5pt]{(A) (B)};

\end{scope}
\end{tikzpicture}$. It has codimension two. 

This is useful to know which iterated residues of the canonical form of the SF could vanish. We know that a residue of order $n$ of the canonical form can be non-zero only if the corresponding $n$ facets intersect in codimension $n$. Therefore $\rm Res_{y_{12}} \rm Res_{y_{34}}  \, \Omega(\mathcal{S}_{\begin{tikzpicture}
    \BoxGraph[black][1.2]{0.3}{}{
    \draw (A)--(B);
    \draw (A)--(D);
    \draw (B)--(C);
    \draw (D)--(C);
    }{}
\end{tikzpicture}})=0$, while $\rm Res_{y_{12}} \rm Res_{\lambda_{(1,3)}}  \, \Omega(\mathcal{S}_{\begin{tikzpicture}
    \BoxGraph[black][1.2]{0.3}{}{
    \draw (A)--(B);
    \draw (A)--(D);
    \draw (B)--(C);
    \draw (D)--(C);
    }{}
\end{tikzpicture}})\neq 0$. For an ordered iterated residue, this condition must hold at every intermediate step, as explained in Sec.~\ref{sec:Steinmann}. In fact, one has
\begin{align}
    &\rm Res_{y_{34}} \rm Res_{y_{12}} \rm Res_{\lambda_{(1,3)}}  \, \Omega(\mathcal{S}_{\begin{tikzpicture}
    \BoxGraph[black][1.2]{0.3}{}{
    \draw (A)--(B);
    \draw (A)--(D);
    \draw (B)--(C);
    \draw (D)--(C);
    }{}
\end{tikzpicture}})=\rm Res_{y_{34}} \rm Res_{\lambda_{(1,3)}} \rm Res_{y_{12}}   \, \Omega(\mathcal{S}_{\begin{tikzpicture}
    \BoxGraph[black][1.2]{0.3}{}{
    \draw (A)--(B);
    \draw (A)--(D);
    \draw (B)--(C);
    \draw (D)--(C);
    }{}
\end{tikzpicture}})\neq 0, \notag \\
&\rm Res_{\lambda_{(1,3)}} \rm Res_{y_{34}} \rm Res_{y_{12}}   \, \Omega(\mathcal{S}_{\begin{tikzpicture}
    \BoxGraph[black][1.2]{0.3}{}{
    \draw (A)--(B);
    \draw (A)--(D);
    \draw (B)--(C);
    \draw (D)--(C);
    }{}
\end{tikzpicture}})= 0,
\end{align}
where this mirrors the work done previously on generalized Steinmann relations, but applied to the whole SF, which includes the $y_e=0$ poles.

Now let us examine the incidence of faces. The projective polytope $\mathcal{S}_\Gamma$ is defined from the intersection of a cone $\mathcal{C}$ and the hyperplane $\{ \sum_{e\in E(\Gamma)}  y_e=1 \}$. All $\{y_e=0\}$ and $\{\lambda_S=0\}$ facets contain the origin $O$ of the cone $\mathcal{C}$. If any set of faces intersects only at $O$, then they do not intersect in $\mathcal{S}_\Gamma$. If the intersection is not only at $O$, it must also happen at the hyperplane $\{ \sum_{e \in E(\Gamma)}  y_e=1 \}$, so in $\mathcal{S}_\Gamma$. This is so because there is no point $O \neq z \in \mathcal{C}$ such that $a.z=0$, since that would imply that some $y$ coordinate is strictly negative. Then, by rescaling by some $\lambda>0$, $\lambda z \in \mathcal{S}_\Gamma$.

Therefore, a subset of faces $\mathcal{F}_i$, $i=1,\dots,n$ has a common intersection in $\mathcal{S}_\Gamma$ if and only if their intersection does not lie entirely in $\bigcap_{e \in E(\Gamma)} \{ y_e=0 \}$. Remember from the analysis of Landau equations that $y_e=0$ is imposed by a solution whenever we take $\alpha_e^{(+)}=\alpha_e^{(-)}>0$, which are the parameters corresponding to the two orientations of edge $e$. Schematically, we have $\alpha_e^{(+)}>0$ whenever there is a cut of edge $e$ following the reference orientation of the graph, or whenever the edge $e$ is deleted. 

Starting from a subset of faces $\mathcal{F}^{(n)}\equiv \{ \mathcal{F}_1,\dots, \mathcal{F}_n\}$, we draw their decorated graphs $G_{\mathcal{F}_1}, \dots, G_{\mathcal{F}_n}$. They intersect in $\mathcal{S}_\Gamma$ if and only if there exists at least one edge $e$ which is not deleted in any of the decorated graphs, and such that it is not cut in both directions (in different graphs). This means that the intersection does not belong to the hyperplane $\{y_e=0\}$ and happens at the SF.

One can find the decorated graph $G_{\mathcal{F}^{(n)}}\equiv G_{\mathcal{F}_1 \cap\dots\cap \mathcal{F}_n}$ for an intersection of the subset of faces $\mathcal{F}^{(n)}$ in the following way. If the edge $e$ does not appear in all decorated graphs $G_{\mathcal{F}_i}$, it does not appear in $G_{\mathcal{F}^{(n)}}$. If, among all decorated graphs, the edge $e$ is cut following its two orientations, it is also absent from $G_{\mathcal{F}^{(n)}}$. Then add all cuts appearing in all $G_{\mathcal{F}_i}$ to $G_{\mathcal{F}^{(n)}}$, following the vertex subset they define. Do not add "cuts" that do not cut any edge. Then, all the faces in $\mathcal{F}^{(n)}$ intersect in $\mathcal{S}_\Gamma$ if and only if $G_{\mathcal{F}^{(n)}}$ has at least one edge. The codimension of $\mathcal{F}^{(n)}$, $\rm codim(\mathcal{F}^{(n)})$, is obtained by applying \Eq{eq:codim}. 

\textit{Example 1} : Consider the intersection of faces $\mathcal{F}_{\begin{tikzpicture}
        \BoxGraph[black][1.2]{0.4}{
        \draw (A)--(B);
        \draw (A)--(D);
        \draw (B)--(C);
        \draw (A)--(C);
        \Tube[rounded corners=2.5pt, inner sep=2.5pt]{(A)};
        }{}
    \end{tikzpicture}}$ and $\mathcal{F}_{\begin{tikzpicture}
        \BoxGraph[black][1.2]{0.4}{
        \draw (A)--(B);
        \draw (A)--(D);
        \draw (B)--(C);
        \draw (D)--(C);
        \draw (A)--(C);
        \Tube[rounded corners=2.5pt, inner sep=2.5pt]{(C)};
        }{}
    \end{tikzpicture}}$, of respective codimension two and one, where the subindices indicate their decorated graphs. Because one of the decorated graphs has a missing edge, it does not appear either in the decorated graph of the intersection. Moreover, the edge corresponding to the chord of the square is cut in opposite directions in both decorated graphs, so it is also removed in the intersection. In the remaining graph with three edges, one simply adds the two minimal cuts:
\begin{equation}
    \begin{tikzpicture}
        \BoxGraph[black][1.2]{1.1}{
        \draw[thick] (A)--(B);
        \draw[thick] (A)--(D);
        \draw[thick] (B)--(C);
        \draw[thick] (A)--(C);
        \Tube{(A)};
        }{}
    \end{tikzpicture} \quad
     \scalebox{2.5}{$\cap$} \quad
     \begin{tikzpicture}
        \BoxGraph[black][1.2]{1.1}{
        \draw[thick] (A)--(B);
        \draw[thick] (A)--(D);
        \draw[thick] (B)--(C);
        \draw[thick] (D)--(C);
        \draw[thick] (A)--(C);
        \Tube{(C)};
        }{}
    \end{tikzpicture}
    \quad \scalebox{2.5}{=} \quad
    \begin{tikzpicture}
        \BoxGraph[black][1.2]{1.1}{
        \draw[thick] (A)--(B);
        \draw[thick] (A)--(D);
        \draw[thick] (B)--(C);
        \Tube{(C)};
        \Tube{(A)};
        }{}
    \end{tikzpicture}.
\end{equation}
The resulting face has codimension five.

\textit{Example 2} : In this example for the triangle graph $K_3$, we look at the intersection of three faces of the SF of the triangle graph $K_3$. None of the three edges remains in the decorated graph of the intersection:
\begin{equation}
    \begin{tikzpicture}
        \TriGraph{(0,0)}{1.1}{
        \draw[thick] (A)--(C);
        \draw[thick] (C)--(B);
        \Tube{(A) (C)};
        \Tube[rounded corners=9pt, inner sep=4pt]{(B) (C)};
        }{} 
    \end{tikzpicture} \quad
     \scalebox{2.5}{$\cap$} \quad
     \begin{tikzpicture}
        \TriGraph{(0,0)}{1.1}{
        \draw[thick] (A)--(C);
        \draw[thick] (C)--(B);
        \Tube{(B)};
        }{} 
    \end{tikzpicture}
    \quad \scalebox{2.5}{$\cap$} \quad
    \begin{tikzpicture}
        \TriGraph{(0,0)}{1.1}{
        \draw[thick] (A)--(C);
        \draw[thick] (C)--(B);
        \draw[thick] (A)--(B);
        \Tube{(A)};
        }{} 
    \end{tikzpicture}
    \quad \scalebox{2.5}{=} \quad
    \begin{tikzpicture}
        \TriGraph{(0,0)}{1.1}{
        }{} 
    \end{tikzpicture} \, \, \, ,
\end{equation}
which means that these three faces do not have a common intersection. 

\subsection{Base-fiber representation of the canonical form}

In this section, we prove that the fiber structure of the SF explained in Sec.~\ref{sec:CayleyEmbedding} is reflected in the canonical form. In fact, it can be written as the canonical form of the deformed graphical zonotopes that constitute the fibers, parametrized by the $y_e$ variables, times the canonical form of the base simplex $\Delta^{E-1}$. This will lead us in Sec.~\ref{sec:CausalRepresentation} to the causal representation of the SF.

Our starting point is the contour integral representation of the canonical form of $\mathcal{S}_\Gamma$ in \Eq{eq:ContourIntegral}. We got rid of the delta function by solving the system of equations $\hat{p}=VC$, with $V$ the matrix given in \Eq{eq:VMatrix}, whose ordering of columns depends on a reference orientation $\vec{\Gamma}$. We will work directly with the canonical function instead of the canonical form. By performing the change of variables $\{C_{e}^+ ,C_{e}^-\}  \rightarrow \{C_e^+-C_e^-,C_e^++C_e^-\}$, $e =1,\dots, E$, we get
\begin{equation}
    VC= 
    \begin{pmatrix}
        \hat{B} & 0 \\
        0 & Id_{E}
    \end{pmatrix} 
    \begin{pmatrix}
        C_e^+-C_e^- \\
        C_e^++C_e^-
    \end{pmatrix}.
    \label{eq:UMatrix}
\end{equation}
Then, from \Eq{eq:ContourIntegral}, the canonical function becomes 
\begin{align}
    \tilde{\Omega}(\mathcal{S}_\Gamma)(p) &=\frac{1}{2^E(2 \pi i)^{L}} \int_{\mathbb{R}^{2E}} \frac{d^{E} C^-d^{E} C^+ \delta^{E}\left(C^++C^--y\right)\delta^{N-1}\left(\hat{B}C^+-\hat{B}C^--\hat{x}\right)}{\prod_{e=1}^E  (C_e^--i\epsilon_e^-/2)(C_e^+-i\epsilon_e^+/2)}  
    \label{eq:ContourIntegralChangeVariables}
\end{align}
We first integrate out the first delta function by integrating over $C^-$, and we then define the variable $u_e\equiv 2C_e^+-y_e$. The Jacobian of this change of variables gives a factor $(1/2)^E$. This yields
\begin{equation}
    \tilde{\Omega}(\mathcal{S}_\Gamma)(p) =\frac{1}{\prod_{e=1}^E (2y_e)} \left[\frac{1}{(2 \pi i)^{L}} \int_{\mathbb{R}^{E}} d^{E} u \prod_{e=1}^E \left(\frac{ 2y_e}{ (q_{e,0}^{(+)}-u_e)(q_{e,0}^{(+)}+u_e)}\right) \delta^{N-1}\left(\hat{B}u-\hat{x}\right)\right],
    \label{eq:ContourIntegralHypercubeProjection}
\end{equation}
where we factored out $\frac{1}{\prod_{e=1}^E (2y_e)}$ for convenience. This prefactor is proportional to $\tilde{\Omega}(\Delta^{E-1})(Y)$, i.e. the canonical function of the base simplex. One can easily check that, dropping the imaginary prescription, the product over edge variables inside the integral is the canonical function of the hyperrectangle 
\begin{equation}
    R_Y \equiv [-y_1,y_1]\times \dots \times [-y_E,y_E].
    \label{eq:Hyperrectangle}
\end{equation}

Recall from Sec.~\ref{sec:GraphicalZonotope} that deformed graphical zonotopes can be obtained by affine projection $\pi$ of the hyperrectangle in \Eq{eq:Hyperrectangle}, whose segments are weighted by the $y_e$ variables, and the projection is given by a reduced incidence matrix $\hat{B}$. Canonical forms are compatible with this linear projection through integration along the fibers. Indeed, the faces of the image are projections of exposed faces of $R_Y$, and the residue recursion of $\Omega(R_Y)$ is therefore inherited by the projected form. Then, up to a global orientation sign,
\begin{equation}
    \Omega(\mathcal{Z}_\Gamma^Y)=\pi_\ast\Omega(R_Y),
\end{equation}
which is the expression between square brackets in \Eq{eq:ContourIntegralHypercubeProjection}, since the projection is carried out by the delta function. Consequently, we have
\begin{equation}
    \tilde{\Omega}(\mathcal{S}_\Gamma)=\frac{1}{\prod_{e=1}^E (2y_e)}\tilde{\Omega}(\mathcal{Z}_\Gamma^Y), 
    \label{eq:CanonicalFormFactorization}
\end{equation}
so the canonical form of the SF decomposes into the canonical forms of the base simplex and the fibers.

\section{The causal representation \label{sec:CausalRepresentation}}

Starting from the base-fiber structure of the canonical form in \Eq{eq:CanonicalFormFactorization}, we derive the causal representation, which captures the time structure of the physical process underlying the Feynman diagram. It is also interesting for numerical loop integration, since we can write it in a form that is free of spurious poles. We will finally show how to bootstrap the causal representation without performing any integration or taking any residues.

By performing partial fraction decomposition, expanding and discarding the infinitesimal complex prescription that sits outside the integral, the canonical function becomes 
\begin{equation}
    \tilde{\Omega}(\mathcal{S}_\Gamma)(p) =\frac{1}{\prod_{e=1}^E (2y_e)} \left[\sum_{\kappa \in \{+,-\}^E} \frac{1}{(2 \pi i)^{L}} \int_{\mathbb{R}^E} d^{E} u \prod_{e=1}^E \frac{ 1}{ q_{e,0}^{(+)}+\kappa_e u_e} \delta^{N-1}\left(\hat{B}u-\hat{x}\right)\right],
    \label{eq:ContourIntegralCausal}
\end{equation}
where $R_Y=[-y_1,y_1]\times\dots\times[-y_E,y_E]$ is a hyperrectangle. This expression corresponds to a sum over all possible orientations of $\Gamma$, where $\kappa_e=+1$ (respectively $\kappa_e=-1$) corresponds to edge $e$ in $\Gamma$ having the same (respectively opposite) orientation as the reference orientation defining the incidence matrix~$B$. Equivalently, 
 \begin{equation}
     \tilde{\Omega}(\mathcal{S}_\Gamma)(p) =\frac{1}{\prod_{e=1}^E (2y_e)}\frac{1}{(2 \pi i)^{L}} \sum_{\kappa \in \{+,-\}^E} \int_{\mathbb{R}^L} \frac{\prod_{i=1}^L d \ell_{i,0}}{\prod_{e \in E(\Gamma)} [q_{e,0}^{(+)}-\kappa_e q_{e,0}]},
     \label{eq:ContourIntegralCausal2}
 \end{equation}
 which can also be easily obtained from \Eq{eq:ContourFeynmanIntegral} via partial fraction decomposition.

Now, consider the charge configuration $\mathcal{Q}_\kappa$ generated by vectors $\tilde{\beta}_e^\kappa \in \{ -1,0,1 \}^L$, such that $\tilde{\beta}_e^\kappa=\kappa_e\beta_e \in \mathbb{R}^L$, where the $\beta$ vectors were defined in (\ref{eq:MomentaAssig}). There is one such charge configuration for each term in \Eq{eq:ContourIntegralCausal2}. For each edge $e$, the component $(\tilde{\beta}_e)_i$, tells us whether $\ell_{i,0}$ and $i \epsilon_e^\kappa$ are preceded by the same sign (and $(\tilde{\beta}_e)_i=0$ if $q_{e,0}$ does not depend on $\ell_{i,0}$) in the denominator factor corresponding to edge $e$. We can imagine this $\mathbb{R}^L$ space as the "imaginary loop space" in which the integration contour, initially at the origin, can be deformed to avoid poles of the integrand. Call $\eta$ the vector in imaginary loop space indicating the deformation direction of the integration contour. Then, for example, if the contour is deformed along $\eta=\tilde{\beta}_e$ vector, it encounters the pole given by the denominator $G_{\rm F}^{(\kappa_e)}(q_{e})$. For the integration contour to go past it and be deformed to infinity, the residue at this pole is taken. In general, the integration contour encounters poles whose corresponding vectors in the fan generate a cone containing $\eta$ (we will explain this in more detail in Sec.~\ref{sec:JeffreyKirwan} when we explain Jeffrey-Kirwan residues). 

Every orientation of $\Gamma$, and thus every term in \Eq{eq:ContourIntegralCausal2}, is labeled by a sign vector $\kappa\in\{+,-\}^E$ with respect to the reference orientation. The vectors $\beta_e$ realize the cographic oriented matroid of $\Gamma$. By Farkas' lemma, the orientation labeled by $\kappa$ is cyclic if and only if there exists a nonzero vector $\eta\in\mathbb{R}^L$ such that
\begin{equation}
    \widetilde{\beta}_e^\kappa\cdot\eta\leq 0
    \qquad \text{for every }e,
\end{equation}
up to an overall reversal of the inequality depending on the contour convention. Indeed, the vector with components $\widetilde{\beta}_e^\kappa\cdot\eta$ is a sign-compatible circulation, and its nonzero support is a union of directed cycles.

Therefore, for a cyclic orientation one can deform the loop-energy contour along the corresponding direction $\eta$ without crossing any pole and push it to infinity. The corresponding contour is null-homologous and the term vanishes. Consequently, only acyclic orientations contribute. This is analogous to the IBP relations for Feynman integrals. Homology kills cyclic orientations, which has a nice physical interpretation as only allowing causal orderings consistent with the linear evolution of time. We can consequently write  $\mathcal{S}_\Gamma$
 \begin{equation}
     \tilde{\Omega}(\mathcal{S}_\Gamma)(p) =\frac{1}{\prod_{e=1}^E (2y_e)}\frac{1}{(2 \pi i)^{L}} \sum_{\vec{\Gamma} \text{ acyclic}} \int_{\mathbb{R}^L} \frac{\prod_{i=1}^L d \ell_{i,0}}{\prod_{e \in E(\Gamma)} [q_{e,0}^{(+)}-\kappa_e^{\vec{\Gamma}} q_{e,0} ]},
     \label{eq:CausalRepresentationAcyclic}
 \end{equation}
 where we sum over all acyclic orientations $\vec{\Gamma}$ of $\Gamma$ and $\kappa_e^{\vec{\Gamma}}$ denotes the tope given by $\vec{\Gamma}$. Upon integration, this sum over acyclic orientations is the LTD causal representation.

 We see that each term of the causal representation corresponds to a vertex of $\mathcal{Z}_\Gamma$. Applying a Landau analysis to the terms of the causal representation as we did in Sec.~\ref{sec:Landau}, one sees that the poles appearing in the term corresponding to orientation $\vec{\Gamma}$ are the minimal cuts compatible with that orientation.

 \subsection{The cone decomposition}

The causal representation corresponds to applying a \textit{cone decomposition} to $\Omega(\mathcal{Z}_\Gamma^Y)$ after decomposing the canonical form of the SF as in \Eq{eq:CanonicalFormFactorization}. The cone decomposition arises by localizing the geometry near each vertex $v$. At a vertex, the facets incident on $v$ determine the normal cone $N_vP$ generated by the normal vectors to the facets incident to $v$. Since the canonical form is determined by its logarithmic singularities on boundaries, one may reconstruct its canonical form by summing the contributions of these local cones in the way explained below.
 
The cone decomposition for a simple polytope $P$ in $d$ dimensions takes a very simple form. Each such vertex is the intersection of only $d$ facets, which appear as simple poles in the canonical form. Since the canonical function of the graphical zonotope is a homogeneous function of degree $-(N-1)$ in the $x$ and $y$ variables, we expect the term related to each vertex to be proportional to the inverse of the product of the denominators related to each facet, as shown in Ref.~\cite{Salvatori:2019phs}.
 
The graphical zonotope $\mathcal{Z}_\Gamma^Y$ is in general not simple, but some of its vertices may be simple. Let us call $\tilde{\Omega}(\mathcal{Z}_\Gamma^Y)_{v}$ the term in $\tilde{\Omega}(\mathcal{Z}_\Gamma^Y)$ corresponding to the cone $N_vP$ normal to a simple vertex $v$. Then,
 \begin{equation}
     \tilde{\Omega}(\mathcal{Z}_\Gamma^Y)_{v}=\prod_{\substack{\text{Minimal cuts} \\ \text{compatible with $\vec{\Gamma}_v$}}} \frac{1}{\lambda_S}
 \end{equation}
is the term in the canonical function associated with the acyclic orientation $\vec{\Gamma}_v$.

\paragraph{The permutohedron.} It is particularly easy to obtain the canonical form of simple graphical zonotopes. The only simple graphical zonotopes are $(N-1)$-dimensional permutohedra $P_{N}$ and products of permutohedra, which respectively come from complete graphs $K_{N}$ and complete graphs by blocks, i.e. complete graphs joined by only a vertex, or disconnected. The acyclic orientations of complete graphs give total orders on the vertices, so that each vertex of the permutohedron $P_{N}$ corresponds to a permutation $\sigma \in S_N$ on the $N$ vertices. Its minimal cuts are of the form $\sigma(1),\dots,\sigma(i)|\sigma(i+1),\dots, \sigma(N)$, with $i=1,\dots,N-1$. Therefore, the causal thresholds compatible with the orientation given by the total order $\sigma$ are $\lambda_{(\sigma(1),\dots,\sigma(i))}$. Then, its canonical function is 
 \begin{equation}
     \tilde{\Omega}(P_{N})=\sum_{\sigma \in S_N} \prod_{i=1}^{N-1} \frac{1}{\lambda_{(\sigma(1),\dots,\sigma(i))}}.
 \end{equation}

\paragraph{Simplicial refinements of the cones.} The cone decomposition of the canonical form of $P$ admits a natural geometrical interpretation in terms of subdivisions of the dual polytope $P^\ast$. It is obtained by subdividing $P^\ast$ into cells obtained by joining a distinguished point $p_\infty$ to each facet of $P^\ast$. Each cell in the subdivision contains exactly one facet $F_v^\ast$ of the dual polytope, each dual to a vertex $v \in P$. $p_\infty$ is the point dual to the hyperplane at infinity. More  precisely, one considers the decomposition
\begin{equation}
     P^\ast = \bigcup_{v\in \mathrm{Vert}(\mathcal P)} \mathrm{Conv}(p_\infty, F_v^\ast),
\end{equation}
where each cell is a pyramid with base $F_v^\ast$ and apex $p_\infty$. One can easily see that their union reconstructs $P^\ast$. By duality, each such pyramid is associated with the normal cone at the corresponding vertex $v$.

If $P$ is simple, then each facet $F_v^\ast$ is a simplex, and the corresponding pyramid is simplicial. In this case, no further subdivision is required, and each vertex contributes a single term in the cone decomposition, as we saw above. For a general polytope, the facets $F_v^\ast$ need not be simplicial. In this case, one chooses a triangulation of each facet,
\begin{equation}
    F_v^\ast = \bigcup_{\alpha} \Delta_{v,\alpha},
\end{equation}
into simplices $\Delta_{v,\alpha}$. This induces a subdivision of the corresponding pyramid into simplices
\begin{equation}
    \mathrm{Conv}(p_\infty, F_v^\ast)=\bigcup_{\alpha} \mathrm{Conv}(p_\infty, \Delta_{v,\alpha}).
    \label{eq:TriangPyramid}
\end{equation}

Because vertices of $P^\ast$ are dual to facets of $P$, the extremal rays of this cone are normal to the facets of $P$. Therefore, the triangulations of all pyramids in \Eq{eq:TriangPyramid} correspond to a simplicial refinement of the normal fan $N_P$ of $P$. 
\begin{equation}
    N_v P = \bigcup_{\alpha} C_{v,\alpha},
\end{equation}
where we called $N_vP$ the top-dimensional cone of $N_P$ corresponding to vertex $v$, with each simplicial cone $C_{v,\alpha}$ corresponding to a simplex $\mathrm{Conv}(p_\infty, \Delta_{v,\alpha})$ in $ P^\ast$. Globally, the cone decomposition corresponds to a subdivision of $ P^\ast$ into simplices obtained by triangulating each facet and adjoining the point $p_\infty$.

This perspective explains why the canonical form of $P$ can be subdivided according to the normal cones $N_vP$ at its vertices. Since the canonical form can be identified with the volume form of $ P^\ast$, and since volume is additive under subdivision, the canonical form decomposes as a sum over the simplices in this subdivision. Each simplex corresponds to a simplicial cone obtained from triangulating $N_vP$, and contributes one term to the canonical function. Independence of the final expression from the choice of triangulation follows from the fact that different triangulations yield subdivisions of $\mathcal P^\ast$ with the same total volume.

An example is shown in Fig. \ref{fig:CausalSubdivision} (left), which represents the dual polytope $\mathcal{Z}^{Y\ast}_{\begin{tikzpicture}
    \BoxGraph[black][1.2]{0.3}{}{
    \draw (A)--(B);
    \draw (A)--(D);
    \draw (B)--(C);
    \draw (D)--(C);
    }{}
    \end{tikzpicture}}$ of the generalized graphical zonotope $\mathcal{Z}^Y_{\begin{tikzpicture}
    \BoxGraph[black][1.2]{0.3}{}{
    \draw (A)--(B);
    \draw (A)--(D);
    \draw (B)--(C);
    \draw (D)--(C);
    }{}
    \end{tikzpicture}}$ (see left panel of Fig. \ref{fig:DeformedGZSquareChord}). One cell in the causal subdivision of $\mathcal{Z}^{Y\ast}_{\begin{tikzpicture}
    \BoxGraph[black][1.2]{0.3}{}{
    \draw (A)--(B);
    \draw (A)--(D);
    \draw (B)--(C);
    \draw (D)--(C);
    }{}
    \end{tikzpicture}}$ is depicted in salmon color. It is $\mathrm{Conv}(p_\infty, F_v^\ast)$, where the facet $F_v^\ast \subset \mathcal{Z}^{Y\ast}_{\begin{tikzpicture}
    \BoxGraph[black][1.2]{0.3}{}{
    \draw (A)--(B);
    \draw (A)--(D);
    \draw (B)--(C);
    \draw (D)--(C);
    }{}
    \end{tikzpicture}}$ corresponds to acyclic orientation $\begin{tikzpicture}
        \BoxGraph[black][1]{0.6}{
        \draw [->-=0.5] (A)--(B);
        \draw [->-=0.5] (A)--(D);
        \draw [->-=0.5] (B)--(C);
        \draw [->-=0.5] (D)--(C);
        }{}
\end{tikzpicture}$, and dual in $\mathcal{Z}^{Y}_{\begin{tikzpicture}
    \BoxGraph[black][1.2]{0.3}{}{
    \draw (A)--(B);
    \draw (A)--(D);
    \draw (B)--(C);
    \draw (D)--(C);
    }{}
    \end{tikzpicture}}$ to a vertex we call $v$. Doing this for all facets yields the causal subdivision. The edges going from $p_\infty$ to the vertices of the facet can be seen as extremal rays of a cone with origin $p_\infty$. It corresponds to the top-dimensional cone $N_v\mathcal{Z}^Y_{\begin{tikzpicture}
    \BoxGraph[black][1.2]{0.3}{}{
    \draw (A)--(B);
    \draw (A)--(D);
    \draw (B)--(C);
    \draw (D)--(C);
    }{}
    \end{tikzpicture}}$.
    Then, in order to compute the term $\tilde{\Omega}(\mathcal{Z}^Y_{\begin{tikzpicture}
    \BoxGraph[black][1.2]{0.3}{}{
    \draw (A)--(B);
    \draw (A)--(D);
    \draw (B)--(C);
    \draw (D)--(C);
    }{}
    \end{tikzpicture}})_{v}$, we do a triangulation of $\mathrm{Conv}(p_\infty, F_v^\ast)$, or equivalently a simplicial refinement of $N_v\mathcal{Z}^Y_{\begin{tikzpicture}
    \BoxGraph[black][1.2]{0.3}{}{
    \draw (A)--(B);
    \draw (A)--(D);
    \draw (B)--(C);
    \draw (D)--(C);
    }{}
    \end{tikzpicture}}$. In Fig.~\ref{fig:CausalSubdivision} (right) one of the two possible triangulations is represented. 

\begin{figure}[h]
\centering

\begin{minipage}{0.48\textwidth}
\centering

\begin{tikzpicture}[scale=8,line join=round,line cap=round]
\coordinate (p) at (0,0);
\coordinate (v1) at (0.46393,0.21042);
\coordinate (v2) at (0.3849,0.09205);
\coordinate (v3) at (-0.46393,-0.21042);
\coordinate (v4) at (-0.1174,-0.12202);
\coordinate (v5) at (-0.3849,-0.09205);
\coordinate (v6) at (0.1174,0.12202);
\coordinate (v7) at (-0.37556,0.16638);
\coordinate (v8) at (-0.30387,0.29949);
\coordinate (v9) at (0,0.39284);
\coordinate (v10) at (0,-0.39284);
\coordinate (v11) at (0.30387,-0.29949);
\coordinate (v12) at (0.37556,-0.16638);
\fill[blue!30,opacity=.18] (v11)--(v4)--(v2)--cycle;
\fill[blue!30,opacity=.18] (v2)--(v9)--(v1)--cycle;
\fill[blue!30,opacity=.18] (v7)--(v4)--(v3)--cycle;
\fill[blue!30,opacity=.18] (v7)--(v8)--(v9)--cycle;
\fill[blue!30,opacity=.18] (v3)--(v4)--(v11)--(v10)--cycle;
\fill[blue!30,opacity=.18] (v2)--(v1)--(v12)--(v11)--cycle;
\fill[blue!60,opacity=.32] (v10)--(v5)--(v3)--cycle;
\fill[blue!60,opacity=.32] (v11)--(v12)--(v10)--cycle;
\fill[blue!60,opacity=.32] (v6)--(v12)--(v1)--cycle;
\fill[blue!60,opacity=.32] (v6)--(v8)--(v5)--cycle;
\fill[blue!60,opacity=.32] (v10)--(v12)--(v6)--(v5)--cycle;
\fill[blue!60,opacity=.32] (v6)--(v1)--(v9)--(v8)--cycle;
\fill[blue!60,opacity=.32] (v5)--(v8)--(v7)--(v3)--cycle;
\draw[dash pattern=on 3pt off 2pt,black!60,line width=.45pt] (v11)--(v4);
\draw[dash pattern=on 3pt off 2pt,black!60,line width=1pt] (v4)--(v2);
\draw[dash pattern=on 3pt off 2pt,black!60,line width=.45pt] (v2)--(v11);
\draw[dash pattern=on 3pt off 2pt,black!60,line width=1pt] (v2)--(v9);
\draw[dash pattern=on 3pt off 2pt,black!60,line width=.45pt] (v1)--(v2);
\draw[dash pattern=on 3pt off 2pt,black!60,line width=1pt] (v7)--(v4);
\draw[dash pattern=on 3pt off 2pt,black!60,line width=.45pt] (v4)--(v3);
\draw[dash pattern=on 3pt off 2pt,black!60,line width=1pt] (v9)--(v7);
\draw[black,line width=0.9pt] (v10)--(v5);
\draw[black,line width=.9pt] (v5)--(v3);
\draw[black,line width=.9pt] (v3)--(v10);
\draw[black,line width=.9pt] (v11)--(v12);
\draw[black,line width=0.9pt] (v12)--(v10);
\draw[black,line width=.9pt] (v10)--(v11);
\draw[black,line width=.9pt] (v9)--(v1);
\draw[black,line width=0.9pt] (v6)--(v12);
\draw[black,line width=.9pt] (v12)--(v1);
\draw[black,line width=.9pt] (v1)--(v6);
\draw[black,line width=.9pt] (v6)--(v8);
\draw[black,line width=.9pt] (v8)--(v5);
\draw[black,line width=.9pt] (v5)--(v6);
\draw[black,line width=.9pt] (v3)--(v7);
\draw[black,line width=.9pt] (v7)--(v8);
\draw[black,line width=.9pt] (v8)--(v9);
\draw[dash pattern=on 3pt off 2pt,red,line width=1.2pt] (p)--(v2);
\draw[dash pattern=on 3pt off 2pt,red,line width=1.2pt] (p)--(v7);
\draw[dash pattern=on 3pt off 2pt,red,line width=1.2pt] (p)--(v4);
\draw[dash pattern=on 3pt off 2pt,red,line width=1.2pt] (p)--(v9);
\fill[red!40,opacity=.18] (v9)--(v2)--(v4)--(v7)--cycle;

\fill (p) circle (0.25pt) [color=red];
\fill (v2) circle (0.2pt);
\fill (v7) circle (0.2pt);
\fill (v4) circle (0.2pt);
\fill (v9) circle (0.2pt);

\node[above right]  at (p) {$p_\infty$};

\begin{scope}[shift={($1.15*(v7)-0.15*(p)$)}, scale=0.15]
\BoxGraph{1}{}{
    \draw[thick] (A)--(B);
    \draw[thick] (A)--(D);
    \draw[thick] (B)--(C);
    \draw[thick] (D)--(C);
    \Tube[rounded corners=5pt, inner sep=5pt]{(A)};
}{}
\end{scope}
\begin{scope}[shift={($1.15*(v9)-0.15*(p)$)}, scale=0.15]
\BoxGraph{1}{}{
    \draw[thick] (A)--(B);
    \draw[thick] (A)--(D);
    \draw[thick] (B)--(C);
    \draw[thick] (D)--(C);
    \Tube[rounded corners=11pt, inner sep=5pt]{(A) (B)};
}{}
\end{scope}
\begin{scope}[shift={($1.2*(v2)-0.2*(v11)$)}, scale=0.15]
\BoxGraph{1}{}{
    \draw[thick] (A)--(B);
    \draw[thick] (A)--(D);
    \draw[thick] (B)--(C);
    \draw[thick] (D)--(C);
    \Tube[rounded corners=10.5pt, inner sep=5pt]{(A) (B) (D)};
}{}
\end{scope}
\begin{scope}[shift={($0.8*(v4)+0.2*(v10)$)}, scale=0.15]
\BoxGraph{1}{}{
    \draw[thick] (A)--(B);
    \draw[thick] (A)--(D);
    \draw[thick] (B)--(C);
    \draw[thick] (D)--(C);
    \Tube[rounded corners=11pt, inner sep=5pt]{(A) (D)};
}{}
\end{scope}

\coordinate (digraph) at ($0.7*(v9)+0.7*(v2)-0.2*(v7)-0.2*(v4)$);      
\coordinate (facet) at ($0.4*(v9)+0.4*(v2)+0.1*(v7)+0.1*(v4)$);   
\draw[->, dashed, thick]
  (digraph) to[out=0, in=0] (facet);

\fill (digraph) circle (1pt) [color=white];

\begin{scope}[shift={($1.06*(digraph)-0.06*(v7)-0*(v4)$)}, scale=0.2]
\BoxGraph[black][1]{0.9}{
    \draw [-->-=0.8, thick] (A)--(B);
    \draw [-->-=0.8, thick] (A)--(D);
    \draw [-->-=0.8, thick] (B)--(C);
    \draw [-->-=0.8, thick] (D)--(C);
}{}
\end{scope}

\end{tikzpicture}

\end{minipage}
\hfill
\begin{minipage}{0.48\textwidth}
\centering

\begin{tikzpicture}[scale=8,line join=round,line cap=round]
\coordinate (p) at (0,0);
\coordinate (v2) at (0.3849,0.09205);
\coordinate (v4) at (-0.1174,-0.12202);
\coordinate (v7) at (-0.37556,0.16638);
\coordinate (v9) at (0,0.39284);
\draw[black!60,line width=1pt] (v4)--(v2);
\draw[black!60,line width=1pt] (v2)--(v9);
\draw[black!60,line width=1pt] (v9)--(v7);
\draw[black,line width=1pt] (v4)--(v7);
\draw[red,line width=1.2pt] (p)--(v2);
\draw[red,line width=1.2pt] (p)--(v7);
\draw[line width=1pt] (p)--(v4);
\draw[line width=1pt] (p)--(v9);
\draw[dash pattern=on 3pt off 2pt,red,line width=1.2pt] (v2)--(v7);
\fill[orange!70,opacity=.5] (v4)--(v2)--(v7)--cycle;
\fill[red!70,opacity=.25] (v9)--(v2)--(p)--(v7)--cycle;

\fill (p) circle (0.25pt) [color=red];
\fill (v2) circle (0.2pt);
\fill (v7) circle (0.2pt);
\fill (v4) circle (0.2pt);
\fill (v9) circle (0.2pt);

\node[above right]  at (p) {$p_\infty$};

\begin{scope}[shift={($1.15*(v7)-0.15*(p)$)}, scale=0.15]
\BoxGraph{1}{}{
    \draw[thick] (A)--(B);
    \draw[thick] (A)--(D);
    \draw[thick] (B)--(C);
    \draw[thick] (D)--(C);
    \Tube[rounded corners=5pt, inner sep=5pt]{(A)};
}{}
\end{scope}
\begin{scope}[shift={($1.15*(v9)-0.15*(p)$)}, scale=0.15]
\BoxGraph{1}{}{
    \draw[thick] (A)--(B);
    \draw[thick] (A)--(D);
    \draw[thick] (B)--(C);
    \draw[thick] (D)--(C);
    \Tube[rounded corners=11pt, inner sep=5pt]{(A) (B)};
}{}
\end{scope}
\begin{scope}[shift={($1.25*(v2)-0.25*(v11)$)}, scale=0.15]
\BoxGraph{1}{}{
    \draw[thick] (A)--(B);
    \draw[thick] (A)--(D);
    \draw[thick] (B)--(C);
    \draw[thick] (D)--(C);
    \Tube[rounded corners=10.5pt, inner sep=5pt]{(A) (B) (D)};
}{}
\end{scope}
\begin{scope}[shift={($0.8*(v4)+0.2*(v10)$)}, scale=0.15]
\BoxGraph{1}{}{
    \draw[thick] (A)--(B);
    \draw[thick] (A)--(D);
    \draw[thick] (B)--(C);
    \draw[thick] (D)--(C);
    \Tube[rounded corners=11pt, inner sep=5pt]{(A) (D)};
}{}
\end{scope}

\coordinate (digraph) at ($0.7*(v9)+0.7*(v2)-0.2*(v7)-0.2*(v4)$);      
\coordinate (facet) at ($0.4*(v9)+0.4*(v2)+0.1*(v7)+0.1*(v4)$);   
\draw[->, dashed, thick]
  (digraph) to[out=0, in=0] (facet);

\fill (digraph) circle (1pt) [color=white];

\begin{scope}[shift={($1.06*(digraph)-0.06*(v7)-0*(v4)$)}, scale=0.2]
\BoxGraph[black][1]{0.9}{
    \draw[thick] [-->-=0.8] (A)--(B);
    \draw[thick] [-->-=0.8] (A)--(D);
    \draw[thick] [-->-=0.8] (B)--(C);
    \draw[thick] [-->-=0.8] (D)--(C);
}{}
\end{scope}

\end{tikzpicture}

\end{minipage}
\caption[Dual polytope]{Dual polytope $\mathcal{Z}^{Y\ast}_{\begin{tikzpicture}
    \BoxGraph[black][1.2]{0.3}{}{
    \draw (A)--(B);
    \draw (A)--(D);
    \draw (B)--(C);
    \draw (D)--(C);
    }{}
    \end{tikzpicture}}$ (left) and one of its cells in the causal subdivision, corresponding to acyclic orientation $\begin{tikzpicture}
        \BoxGraph[black][1]{0.6}{
        \draw [->-=0.5] (A)--(B);
        \draw [->-=0.5] (A)--(D);
        \draw [->-=0.5] (B)--(C);
        \draw [->-=0.5] (D)--(C);
        }{}
\end{tikzpicture}$ (right). It is triangulated into two simplices, corresponding to a simplicial refinement of a cone in the normal fan $N_v\mathcal{Z}^Y_{\begin{tikzpicture}
    \BoxGraph[black][1.2]{0.3}{}{
    \draw (A)--(B);
    \draw (A)--(D);
    \draw (B)--(C);
    \draw (D)--(C);
    }{}
    \end{tikzpicture}}$.}
\label{fig:CausalSubdivision}
\end{figure}

One could wonder why only denominators corresponding to the facets intersecting at vertex $v$ appear in $\tilde{\Omega}(P)_{v}$, and not a factor $\lambda_\infty$ for the hyperplane at infinity, dual to $p_\infty$, which appears in every simplex of the triangulation. The projective formula for the volume of a simplex obtained from the causal triangulation of $P^\ast$ is 
\begin{equation}
    \operatorname{Vol}_A\left(
        \operatorname{Conv}(p_\infty,\Delta_{v,\alpha})
    \right)
    =
    \frac{
        \left|\left\langle Z_0 Z_1 \cdots Z_{N-1}\right\rangle\right|
    }{
        \prod_{i=0}^{N-1}(A\cdot Z_i)
    },
\end{equation}
where $Z_0=p_\infty$ and $A\cdot Z_i=\lambda_i(A)$.
Choosing $p_\infty=(1,0,\dots,0)$ and the affine chart
$A=(1,x)$ gives $A\cdot p_\infty=1$, so only the denominators
associated with the facets incident at $v$ remain.

Let us comment briefly on why we chose this causal representation method to obtain $\Omega(\mathcal{Z}_\Gamma^Y)$ instead of resorting to triangulations of $\mathcal{Z}_\Gamma^Y$. The reason is that triangulations introduce internal facets, translating into undesired spurious poles in the canonical form. Another possibility would have been to triangulate the dual $\mathcal{Z}_\Gamma^{Y\ast}$ and sum over the volume of each simplex, as in the causal representation, but without introducing the central point $p_\infty$. The simple poles appearing in this representation are causal thresholds, so that there are no spurious logarithmic singularities. However, each term in this representation mixes orientation-incompatible causal thresholds, so that the intersection of the corresponding facets in $\mathcal{Z}_\Gamma^Y$ is empty. This translates into a spurious pole of higher order, which can be seen by taking iterated residues, that are nonzero for some individual terms but vanish when applied to $\Omega(\mathcal{Z}_\Gamma^Y)$. For the causal representation, all causal thresholds appearing in a term have a common intersection, and in Sec.~\ref{sec:NumInt} we show how to obtain a version of the causal representation free of spurious poles of any order.

\subsection{Refinement of the normal fan \label{sec:RefinementFan}} 

We just showed that the complexity of obtaining the causal representation of the canonical form of a polytope, in our case the deformed graphical zonotope $\mathcal{Z}_\Gamma^Y$, lies in finding a simplicial refinement of the normal fan $N_vP$. 

From Sec.~\ref{sec:GeneralizedPermutohedra}, we know that, since $\mathcal{Z}_\Gamma^Y$ is a generalized permutohedron, its normal fan is refined by the braid fan of the same dimension, i.e. the normal fan of the permutohedron. Because the permutohedron is a simple polytope, its normal fan is simplicial. Therefore the braid fan is a simplicial refinement of $N_v\mathcal{Z}_\Gamma^Y$. In the braid fan there is one top-dimensional simplicial cone per ordering of vertices $\sigma$, and, for each cut $\sigma(1),\dots,\sigma(i)|\sigma(i+1),\dots, \sigma(N)$, one ray. The span of the $(N-1)$ rays of orientation-compatible cuts gives the cone of the corresponding permutation. 

However, these cuts may not correspond to bonds anymore, since the subgraph induced by the subset of vertices $\{ \sigma(1),\dots,\sigma(i)\}$ (or $\{ \sigma(i+1),\dots,\sigma(N)\}$) of a general graph $\Gamma$ may not be connected for some $\sigma$ and some $i$. In that case, the resulting denominator $\lambda_{(\sigma(1),\dots,\sigma(i))}$ corresponds to an extremal ray of the braid fan that is not an extremal ray of the normal fan of $\mathcal{Z}_\Gamma^Y$. Dually, from the point of view of the graphical zonotope, we can see it as coming from an internal facet which appears in the triangulation of the cone but is not a real boundary of $\mathcal{Z}_\Gamma^Y$. In fact, the deformation of generalized permutohedra described in Sec.~\ref{sec:GeneralizedPermutohedra}, that gets $\mathcal{Z}_\Gamma^Y$ from $P_N$ by setting some $y_e$ to zero, makes some facets disappear. Still, we can write
 \begin{equation}
     \tilde{\Omega}(\mathcal{Z}_\Gamma^Y)=\sum_{\sigma \in S_N} \prod_{i=1}^{N-1} \frac{1}{\lambda_{(\sigma(1),\dots,\sigma(i))}}, 
     \label{eq:PermutohedronCF}
 \end{equation}
knowing that some of the denominators may generate spurious poles, which are eliminated in the sum over all terms. 

Note that this expression can be represented graphically as all possible collections of $(N-1)$ nested subsets ($I$ and $J$ are called nested subsets if $I\subset J$ or $J\subset I$). For a permutation $\sigma$, this collection is $\{\sigma(1)\}\subset\{\sigma(1), \sigma (2)\}\subset \dots \subset \{\sigma(1), \sigma (2), \sigma (N-1) \}$. Each of these subsets corresponds to a causal threshold, and each collection of $N-1$ nested subsets gives a term in Eq. (\ref{eq:PermutohedronCF}). However, \Eq{eq:PermutohedronCF} contains spurious poles that may undermine the numerical integration over the loop three-momenta.

\subsubsection{Bond decomposition method \label{sec:BondDecomposition}}

We now present a method to get rid of the denominators in \Eq{eq:PermutohedronCF} that do not correspond to bonds, and are thus at the origin of spurious poles. For a vertex subset $S$, let
\begin{equation}
    S=S^{(1)}\sqcup\cdots\sqcup S^{(c_S)}
\end{equation}
be the decomposition of the subgraph induced by $S$ into its $c_S$ connected components. If $\lambda_S$ is a spurious pole we have three cases:
\begin{enumerate}
    \item $S$ does not induce a connected subgraph and its complement $\bar{S}$ does: $\lambda_S=\sum_{a=1}^{c_S} \lambda_{S^{(a)}}$.
    \item $S$ induces a connected subgraph and its complement $\bar{S}$ does not: $\lambda_S=\sum_{a=1}^{c_{\bar{S}}} \lambda_{\overline{\bar{S}^{(a)}}}$.
    \item Both $S$ and $\bar{S}$ induce disconnected subgraphs, so one of the two bond decompositions above must be chosen. In this case 3, we will choose from now on the decomposition $\lambda_S=\sum_{a=1}^{c_S} \lambda_{S^{(a)}}$. 
\end{enumerate}

Let us explain how this decomposition arises directly from the sum over permutations in \Eq{eq:PermutohedronCF}. Consider the above cases 1 and 3, in which $S$ induces a disconnected subgraph. Fix an ordering $ \sigma_a=(v_{a,1},\ldots,v_{a,n_a}) $ of the vertices of every connected component $S^{(a)}$, and define the corresponding nested subsets $ S^{(a)}_r=\{v_{a,1},\ldots,v_{a,r}\}$, with $ r=1,\ldots,n_a$, and $ S^{(a)}_0=\emptyset$.
The permutations of the vertices of $S$ which preserve the internal ordering $\sigma_a$ of every component are precisely the shuffles $ \sigma\in\operatorname{Sh}(\sigma_1,\ldots,\sigma_{c_S})$.  If $r^\sigma_a(k)$ denotes the number of vertices of $S^{(a)}$ appearing among the first $k$ entries of a given shuffle $\sigma$, its $k$-th initial subset is 
 \begin{equation}
     S^\sigma_k \equiv \{\sigma(1),\dots,\sigma(k)\}= \bigsqcup_{a=1}^{c_S}S^{(a)}_{r^\sigma_a(k)}.
 \end{equation} 
 
We can perform the bond decomposition on the denominators appearing in the term in \Eq{eq:PermutohedronCF} corresponding to $\sigma$:
\begin{equation}
    \lambda_{S^\sigma_k} = \sum_{a=1}^{c_S}\lambda_{S^{(a)}_{r_a^\sigma(k)}}.
\end{equation} 
The sum over shuffles then satisfies the identity 
\begin{equation}
    \sum_{\sigma\in\operatorname{Sh}(\sigma_1,\ldots,\sigma_{c_S})} \prod_{k=1}^{|S|} \frac{1}{\lambda_{S^\sigma_k}} = \prod_{a=1}^{c_S} \prod_{r=1}^{n_a} \frac{1}{\lambda_{S^{(a)}_r}}
    \label{eq:ShuffleBond}
\end{equation}
This is the multi-component generalization of    
   \begin{equation}
       \frac{1}{a(a+b)}+\frac{1}{b(a+b)} = \frac{1}{ab}.
   \end{equation} 

Denote by $F(n_1,\ldots,n_c)$ the left-hand side of \Eq{eq:ShuffleBond}. We have
\begin{equation}
    F(i_1,\ldots,i_c)=\frac{1}{\sum_{a:i_a>0} \lambda_{S_{i_a}^{(a)}}} \sum_{a:i_a>0} F(i_1,\dots,i_a-1,\dots,i_c).
\end{equation}
Now, we can prove \Eq{eq:ShuffleBond} by induction. By assuming it to be true for $F(i_1,\dots,i_a-1,\dots,i_c)$ for all $a=1,\dots,c$, we get
\begin{align}
    F(i_1,\ldots,i_c)&=\frac{1}{\sum_{b:i_b>0} \lambda_{S_{i_b}^{(b)}}}\sum_{j:i_j>0} \left(\lambda_{S_{i_j}^{(j)}} \prod_{a=1}^{c_S} \prod_{r=1}^{i_{a}} \frac{1}{\lambda_{S^{(a)}_r}} \right) \notag \\
    &=\frac{1}{\sum_{b:i_b>0} \lambda_{S_{i_b}^{(b)}}} \prod_{a=1}^{c_S} \prod_{r=1}^{i_{a}} \left( \frac{1}{\lambda_{S^{(a)}_r}}\right) \sum_{j:i_j>0} \left(\lambda_{S_{i_j}^{(j)}}  \right) \notag\\
    &=\prod_{a=1}^{c_S} \prod_{r=1}^{i_{a}} \left( \frac{1}{\lambda_{S^{(a)}_r}}\right),
\end{align}
as desired.

In a similar fashion, in case 2, i.e. when $S$ induces a connected subgraph and $\bar{S}$ does not, we find the decomposition identity
\begin{equation}
    \sum_{\bar{\sigma}\in\operatorname{Sh}(\bar{\sigma}_1,\dots,\bar{\sigma}_{c_{\bar S}})} \prod_{k=1}^{|\bar{S}|} \frac{1}{\lambda_{\overline{\bar{S}^{\bar \sigma}_k}}} = \prod_{a=1}^{c_{\bar S}} \prod_{r=1}^{n_a} \frac{1}{\lambda_{\overline{{\bar S}^{(a)}_r}}},
    \label{eq:ShuffleBondComplement}
\end{equation}
$n_a=|\bar{S}^{(a)}|$. We also define $\bar{\sigma}_a$ as an ordering of the vertices of every connected component $\bar{S}^{(a)}$, and $\bar{S}^{\bar \sigma}_k =\{{\bar \sigma}(1),\dots,{\bar \sigma}(k)\}$ is the subset made of the first $k$ elements of ${\bar S}$ given by the shuffle ordering ${\bar \sigma}$.

After applying Eqs. (\ref{eq:ShuffleBond}) and (\ref{eq:ShuffleBondComplement}), some of the resulting subsets $S_r^{(a)}$ or $\overline{{\bar S}^{(a)}_r}$ may still fail to define oriented bonds, either because they or their complements induce disconnected subgraphs. Whenever this happens, we apply the same procedure recursively, taking $S \equiv S_r^{(a)}$, until the resulting expression contains no spurious poles.

Let us call \textit{compatible} the vertex subsets such that their corresponding causal thresholds appear in the same term in the causal representation after the simplification obtained from identities (\ref{eq:ShuffleBond}) and (\ref{eq:ShuffleBondComplement}). Note that these identities were obtained from an arbitrary choice of bond decomposition for case 3, and thus give a certain causal representation, i.e. simplicial refinement, of many possible. The notion of compatibility here is therefore dependent on our chosen representation.
We implicitly assume that the subsets correspond to oriented bonds. 

In cases 1 and 3, $S_u^{(a)}$ and $S_v^{(b)}$ are compatible for any $u=1,\dots,n_a$, $v=1,\dots,n_b$ and any connected components of $S$ labeled by $a$ and $b$. This is due to the bond decomposition $\lambda_S=\sum_{a=1}^{c_S} \lambda_{S^{(a)}}$. Similarly, in case 2, $\overline{\bar{S}_u^{(a)}}$ and $\overline{\bar{S}_v^{(b)}}$ are compatible. On the other hand, nested sets are always mutually compatible, since their corresponding causal thresholds already appeared together in some terms in \Eq{eq:PermutohedronCF}, before the removal of spurious thresholds.

We call two disjoint subsets adjacent if there exists an edge with one endpoint in each subset. Then, two subsets $S_1$ and $S_2$ are compatible if and only if one of the following three conditions is satisfied.
\begin{itemize}
    \item $S_1$ and $S_2$ are nested.
    \item $S_1$ and $S_2$ are disjoint and non-adjacent.
    \item $\overline{S_1}$ and $\overline{S_2}$ are disjoint and non-adjacent, and $S_1\cap S_2$ induces a connected subgraph.
\end{itemize}
A collection of subsets containing a pair of incompatible subsets is incompatible. However, there may be a collection of subsets such that they are all pairwise compatible, but the collection is incompatible. In other words, there may exist 'higher-order incompatibilities'. 

A collection $\mathcal{C}$ of subsets is compatible if and only if they are all pairwise compatible and, for any group of subsets $S_1,\dots,S_m \subset C$ such that they are all pairwise compatible by the last condition, we have that $S_1 \cap \dots \cap S_m $ induces a connected subgraph.
If $S_1 \cap \dots \cap S_m $ induced a disconnected subgraph, it would correspond to case 3, so the decomposition would be $\lambda_{S_1 \cap \dots \cap S_m}= \sum_{a=1}^{c_S} \lambda_{(S_1 \cap \dots \cap S_m)^{(a)}}$, which does not make the $S_1 , \dots, S_m $ compatible.

We denote by \textit{maximal compatible set} a compatible collection of subsets such that no other subset can be added without making this collection incompatible. A remarkable property of these compatibility conditions is that the resulting maximal compatible subsets are all of size $N-1$. This happens because each maximal compatible set corresponds to a term in a refined causal expression, and all the subsets in a maximal compatible set give the causal thresholds appearing in each term. The final expression must have $N-1$ causal thresholds in each term, since it is obtained from \Eq{eq:PermutohedronCF} and the number of causal thresholds does not vary in the bond decomposition procedure.

The complete bipartite graph $K_{2,3}$ is the simplest example of a graph for which compatibilities are not determined only by pairwise compatibilities of subsets. Consider the following collection of subsets, where the dashed tube represents the complement of the encircled vertex, i.e. the subset $\{1,2,4,5\}$.
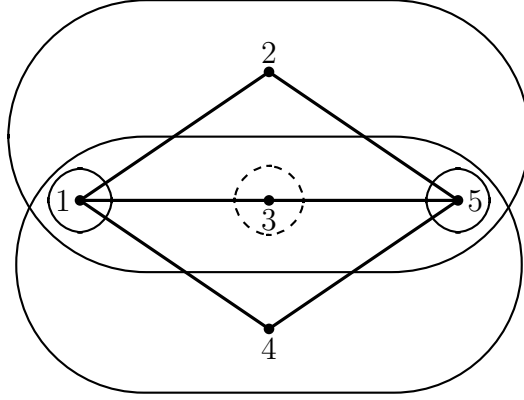
\begin{equation}
     \begin{tikzpicture}
    \coordinate (A) at (-2.5,0);
    \coordinate (B) at (0,1.7);
    \coordinate (C) at (0,0);
    \coordinate (D) at (0,-1.7);
    \coordinate (E) at (2.5,0);
    
    \fill (A) circle[radius=2pt];
    \fill (B) circle[radius=2pt];
    \fill (C) circle[radius=2pt];
    \fill (D) circle[radius=2pt];
    \fill (E) circle[radius=2pt];
    \node[font=\large, anchor=east]  at (A) {1};
    \node[font=\large, anchor=south] at (B) {2};
    \node[font=\large, anchor=north]  at (C) {3};
    \node[font=\large, anchor=north] at (D) {4};
    \node[font=\large, anchor=west] at (E) {5};

    \draw[font=\large,very thick] (A)--(C);
    \draw[font=\large-,very thick] (A)--(B);
    \draw[font=\large,very thick] (A)--(D);
    \draw[font=\large,very thick] (E)--(C);
    \draw[font=\large,very thick] (B)--(E);
    \draw[font=\large,very thick] (E)--(D);

    \Tube[rounded corners=13pt,inner sep=12pt]{(A)};
    \Tube[dashed,rounded corners=14pt,inner sep=13pt]{(C)};
    \Tube[rounded corners=13pt,inner sep=12pt]{(E)};
    \Tube[rounded corners=52pt, inner sep=27pt]{(A) (B) (C) (E)};
    \Tube[rounded corners=48pt, inner sep=24pt]{(A) (D) (C) (E)};
    
    \end{tikzpicture} \notag
\end{equation}
All subsets are pairwise compatible: $\{1\}$ and $\{5\}$ are disjoint and non-adjacent, and they are both nested in $\{1,2,3,5\}$, in $\{1,2,4,5\}$ and in $\{1,3,4,5\}$. These three subsets are pairwise compatible, since their complements are disjoint and non-adjacent, and $\{1,2,3,5\}\cap\{1,2,4,5\}=\{1,2,5\}$, $\{1,2,3,5\}\cap\{1,3,4,5\}=\{1,3,5\}$ and $\{1,2,4,5\}\cap\{1,3,4,5\}=\{1,4,5\}$, all three inducing connected subgraphs.
However, $\{1,2,3,5\}$, $\{1,2,4,5\}$ and $\{1,3,4,5\}$ are pairwise compatible by the last condition, but the three subsets are not mutually compatible, since $\{1,2,3,5\}\cap\{1,2,4,5\}\cap \{1,3,4,5\}=\{1,5\}$ does not induce a connected subgraph. Therefore this collection of five subsets of $K_{2,3}$ is not compatible. 
By removing either $\{1,2,3,5\}$, $\{1,2,4,5\}$ or $\{1,3,4,5\}$, one gets a maximal compatible subset of size $4$, as expected. Therefore, the following three terms appear in the causal expression of the graph $K_{2,3}$:

\begin{equation}
    \frac{1}{\lambda_{(1)}\lambda_{(5)}}\left( \frac{1}{\lambda_{(1,2,3,5)}\lambda_{(1,2,4,5)}} +\frac{1}{\lambda_{(1,2,3,5)}\lambda_{(1,3,4,5)}}+\frac{1}{\lambda_{(1,2,4,5)}\lambda_{(1,3,4,5)}}  \right).
\end{equation}

\subsubsection{Orlik-Solomon relations} Before presenting an alternative method to obtain the causal expression, it is useful to understand the linear dependences among the hyperplanes intersecting at vertex $v$. Let us denote these hyperplanes by $H_i=\{\lambda_i=0\}$ for $i=1,\dots,m$. Each of them has an associated one-form $\omega_i=d\log(\lambda_i)$. The linear dependences among the equations of these hyperplane imply that the forms $\omega_i$ satisfy the so-called \textit{Orlik-Solomon (OS) relations}~\cite{Orlik1980}. 

If we encode these dependencies in a matroid $\mathcal M$ whose ground set is the set of hyperplanes, then the relations are generated by
\begin{equation}
    \partial \omega_S \equiv \sum_{i=1}^{s} (-1)^i \,\omega_1 \wedge \dots \wedge \widehat{\omega_i} \wedge \dots \wedge \omega_s = 0,
    \label{eq:Orlik-Solomon}
\end{equation}
for any circuit $S=\{\omega_1,\dots,\omega_s\}$ of $\mathcal M$. In other words, there is one such generating relation for every inclusion-minimal subset of linearly dependent hyperplanes intersecting at $v$. These relations extend to any dependent set of $\mathcal M$, whose corresponding OS relation is generated by those in~\Eq{eq:Orlik-Solomon}.

To relate these identities to the canonical function, we express the forms in local coordinates $x=(x_1,\dots,x_d)$ on $P$. Since each $\lambda_i$ is linear near the vertex, we can write
\begin{equation}
d\log(\lambda_i)=\frac{d\lambda_i}{\lambda_i}=\frac{\gamma_i \cdot dx}{\lambda_i},
\end{equation}
where $\gamma_i\equiv \nabla \lambda_i$. Then, for any subset $\{i_1,\dots,i_d\}$,
\begin{equation}
d\log(\lambda_{i_1})\wedge \cdots \wedge d\log(\lambda_{i_d})=\frac{\det(\gamma_{i_1},\dots,\gamma_{i_d})}{\lambda_{i_1}\cdots \lambda_{i_d}}\, d^dx.
\end{equation}
Thus, each term in \Eq{eq:Orlik-Solomon} is proportional to the same volume form $d^dx$, and factoring it out yields an identity between rational functions. Denoting by $M_{\widehat{i}}$ the $d \times d$ minors of the matrix with columns $\gamma_{i_1},\dots, \gamma_{i_d}$, obtained by removing the $i$-th hyperplane from the circuit, we obtain
\begin{equation}
     \sum_{i=1}^{s} (-1)^i \, M_{\widehat{i}} \prod_{\substack{j=1 \\ j\neq i}}^s  \frac{1}{\lambda_j}   =0.
    \label{eq:Orlik-Solomon2}   
\end{equation}

To illustrate this, let us look at the vertex of the deformed graphical zonotope $\mathcal{Z}_{\begin{tikzpicture}
        \BoxGraph[black][1.2]{0.3}{
        \draw (A)--(B);
        \draw (A)--(D);
        \draw (B)--(C);
        \draw (D)--(C);
        }{}
\end{tikzpicture}}^Y$ of the square graph which corresponds to the acyclic orientation $\begin{tikzpicture}
        \BoxGraph[black][1]{0.6}{
        \draw [->-=0.5] (A)--(B);
        \draw [->-=0.5] (A)--(D);
        \draw [->-=0.5] (B)--(C);
        \draw [->-=0.5] (D)--(C);
        }{}
\end{tikzpicture}$. This vertex lies at the intersection of four hyperplanes $\{\lambda_{(1,2)}=0\},\{\lambda_{(1,3)}=0\},\{\lambda_{(1)}=0\}$ and $\{\lambda_{(1,2,3)}=0\}$, which define facets of $\mathcal{Z}_{\begin{tikzpicture}
        \BoxGraph[black][1.2]{0.3}{
        \draw (A)--(B);
        \draw (A)--(D);
        \draw (B)--(C);
        \draw (D)--(C);
        }{}
\end{tikzpicture}}^Y$ and correspond to the four minimal cuts compatible with that acyclic orientation. The linear dependence of the normals of these facets is given by \Eq{eq:DegeneracySquareGraph}, which is controlled by the submodular function defining the deformed graphical zonotope. Therefore these four hyperplanes form a circuit. We can choose $x_1$, $x_2$ and $x_3$ as the coordinates of $\mathcal{Z}_{\begin{tikzpicture}
        \BoxGraph[black][1.2]{0.3}{
        \draw (A)--(B);
        \draw (A)--(D);
        \draw (B)--(C);
        \draw (D)--(C);
        }{}
\end{tikzpicture}}^Y$, so that the $d\lambda \rightarrow dx$ change of coordinates matrix, with columns corresponding to $\lambda_{(1,2)}$, $\lambda_{(1,3)}$, $\lambda_{(1)}$ and $\lambda_{(1,2,3)}$ in that order, is 
\begin{equation}
        \frac{d \lambda}{dx}=
    \begin{pmatrix}
        1 & 1 & 1 & 1 \\
        1 & 0 & 0 & 1 \\
        0 & 1 & 0 & 1 \\
    \end{pmatrix}.
\end{equation}

By evaluating the minors of the matrix and applying \Eq{eq:Orlik-Solomon2}, we get
\begin{equation}
    \frac{1}{\lambda_{(1,2)}\lambda_{(1,3)}\lambda_{(1)}}+\frac{1}{\lambda_{(1,2)}\lambda_{(1,3)}\lambda_{(1,2,3)}}=\frac{1}{\lambda_{(1,2)}\lambda_{(1)}\lambda_{(1,2,3)}}+\frac{1}{\lambda_{(1,3)}\lambda_{(1)}\lambda_{(1,2,3)}},
    \label{eq:OSSquare}
\end{equation}
where each side of the equality corresponds to the two different simplicial refinements of the normal cone into simplicial cones. The right-hand side comes from the refinement given by the braid fan, whose simplicial cones correspond to the total orderings $1\rightarrow2 \rightarrow 3 \rightarrow 4$ and $1\rightarrow3 \rightarrow 2 \rightarrow 4$. The left-hand side does not correspond to the normal fan of any graphical zonotope, and can therefore not be obtained with the method mentioned previously. In fact, crossing cuts like $\lambda_{(1,2)}$, $\lambda_{(1,3)}$ do not come from linear extensions of the reachability poset of the graph. Then, the method described in what follows will allow us to obtain any simplicial refinement of the normal cones, and so the different ways of expressing the causal representation.

\subsubsection{NBC bases method}

Ref.~\cite{Brown:2025jjg} proposes the following combinatorial method to get the term of the canonical form associated to an ordered linearly dependent set of hyperplanes $\{H_1,\dots,H_m\}$, forming an arrangement of hyperplanes $\mathcal A$. Consider now in $\mathbb{P}^n$ an ordered subset of hyperplanes $\{H_1,\dots,H_s\} \in C(\mathcal A)$ which is a circuit of the hyperplane arrangement $\mathcal A$. Then the corresponding \textit{broken circuit} is obtained by removing the first element from the circuit, i.e. $\{H_2,\dots,H_s\}$. A \textit{no-broken circuit} (NBC) is a subset of hyperplanes $\{H_{i_1},\dots,H_{i_k}\}$ that does not contain any broken circuit. 

It was proven in Ref.~\cite{Orlik2009} that $\omega_{i_1}\wedge \dots \wedge \omega_{i_k}$ forms a basis of the $k$-th cohomology of $ \mathbb{P}^n \backslash \mathcal A $, where the $\omega_{i_j}$ are the logarithmic forms corresponding to each hyperplane in the NBC basis. This means the following: the terms in the canonical form corresponding to the cone $N_vP$ are written as a linear combination of the $\omega_{i_1}\wedge \dots \wedge \omega_{i_n}$ for all elements in the chosen NBC basis of the hyperplane arrangement $\mathcal{A}_v$ defined by the hyperplanes intersecting at $v$. The different choices of NBC bases, obtained from different orderings of hyperplanes, give different simplicial refinements of the cones. 

In order to get the coefficient of each element of the NBC basis in the canonical form, one can "project" these elements to the canonical form by taking iterated residues. In fact, the terms appearing in the canonical form must have the singularity structure of the polytope $P$ under study. As we know, an $n$-fold iterated residue must correspond to a boundary of $P$ of codimension $n$. The iterated residues are taken from last to first with respect to the chosen ordering of hyperplanes. We label by $I=\{{i_1}, \dots , {i_n}\}$ a subset of hyperplanes of the arrangement and we define $\rm Res _I\equiv Res _{\{H_{i_1}=0\}}\dots Res _{\{H_{i_n}=0\}}$ the iterated residues and $\omega_I \equiv \omega_{i_1}\wedge \dots \wedge \omega_{i_n}$. Then 
\beq  \Omega(P) =\sum_{\text{Vertices }v  \in P}\left( \sum_{\substack{I \in NBC(\mathcal{A}_v) \\ |I|=n}} c_I \, \omega_I  \right), \qquad
    c_I \equiv\mathrm{Res}_I \Omega(P)_v
\eeq
Because of the properties of canonical forms, we expect these coefficients to be either $1$ or $0$. The condition $\mathrm{Res}_I \left( d\rm log (\lambda)_I\right)=1$ geometrically requires the hyperplanes $I$ to define, in reverse order, a flag of faces of $P$.

We now apply this to the cone $N_v\mathcal{Z}_\Gamma^Y$, where the hyperplanes are now $\{\lambda_i=0\}$ such that the causal thresholds $\lambda_i$ are represented as minimal cuts $\mathcal{C}(\lambda_i)$ of $\Gamma$ compatible with the orientation $\vec{\Gamma}_v$. As explained in Sec.~\ref{eq:Steinmann}, in order for $\mathrm{Res}_I \left( d\rm log (\lambda)_I\right)=1$, with $I \in NBC(\mathcal{A}_v)$, we need $\mathcal{O}(\mathcal{C}(\lambda_{i_{n-a+1}})\cup \dots \cup \mathcal{C}(\lambda_{i_n}))=a$ for all $a=1,\dots,n$. In other words, $I$ must define a sequence of minimal cuts such that, for all $a$, the last $a$ minimal cuts partition $\Gamma$ into exactly $a+1$ connected subgraphs.

We summarize this method to obtain $\mathcal{I}_\Gamma$ for a graph $\Gamma$ with $N$ vertices:
\begin{enumerate}
\item Choose an acyclic orientation $\vec{\Gamma}$ and choose an ordering $\{\lambda_1,\dots,\lambda_m\}$ for all minimal cuts compatible with that orientation. This set defines a hyperplane arrangement.

\item Find all circuits and then the NBC basis of size $N-1$ for that hyperplane arrangement.

\item For the subsets of minimal cuts $\{\lambda_{i_1},\dots,\lambda_{i_{N-1}}\}$ corresponding to each element of the NBC basis, check if $\mathcal{O}(\mathcal{C}(\lambda_{i_{N-a}})\cup \dots \cup \mathcal{C}(\lambda_{i_{N-1}}))=a$ for all $a=1,\dots,N-1$. 

\item In that case, the term $\frac{(-1)^E}{\prod_{e=1}^E 2y_e}\frac{1}{\lambda_{i_1}\times \dots\times\lambda_{i_{N-1}}}$ appears in $\mathcal{I}_\Gamma$. Sum over all such terms.

\item Proceed in this way for all acyclic orientations of $\Gamma$ and sum over all of them.
\end{enumerate}

Let us apply this to find the term $\tilde{\Omega}(\mathcal{Z}^Y_{\begin{tikzpicture}
    \BoxGraph[black][1.2]{0.3}{}{
    \draw (A)--(B);
    \draw (A)--(D);
    \draw (B)--(C);
    \draw (D)--(C);
    }{}
\end{tikzpicture}})_v$, where $\vec{\Gamma}_v=\begin{tikzpicture}
        \BoxGraph[black][1]{0.6}{
        \draw [->-=0.5] (A)--(B);
        \draw [->-=0.5] (A)--(D);
        \draw [->-=0.5] (B)--(C);
        \draw [->-=0.5] (D)--(C);
        }{}
\end{tikzpicture}$.
The minimal cuts compatible with this orientation are $\begin{tikzpicture}
        \BoxGraph[black][1]{0.6}{
        \draw (A)--(B);
        \draw (A)--(D);
        \draw (B)--(C);
        \draw (D)--(C);
        \Tube[rounded corners=3pt, inner sep=3pt]{(A)};
        }{}
\end{tikzpicture}$, $\begin{tikzpicture}
        \BoxGraph[black][1]{0.6}{
        \draw (A)--(B);
        \draw (A)--(D);
        \draw (B)--(C);
        \draw (D)--(C);
        \Tube[rounded corners=6pt, inner sep=3pt]{(A) (B)};
        }{}
\end{tikzpicture}$, $\begin{tikzpicture}
        \BoxGraph[black][1]{0.6}{
        \draw (A)--(B);
        \draw (A)--(D);
        \draw (B)--(C);
        \draw (D)--(C);
        \Tube[rounded corners=6pt, inner sep=3pt]{(A) (D)};
        }{}
\end{tikzpicture}$ and $\begin{tikzpicture}
        \BoxGraph[black][1]{0.6}{
        \draw (A)--(B);
        \draw (A)--(D);
        \draw (B)--(C);
        \draw (D)--(C);
        \Tube[rounded corners=6.5pt, inner sep=3pt]{(A) (B) (D)};
        }{}
\end{tikzpicture}$ corresponding respectively to hyperplanes $\{\lambda_{(1)}=0\}$, $\{\lambda_{(1,2)}=0\}$, $\{\lambda_{(1,3)}=0\}$ and $\{\lambda_{(1,2,3)}=0\}$, that define the hyperplane arrangement $\mathcal A_v$, and we choose to order in this way.
The vertex $v$ lies at the intersection of these four hyperplanes. Because $\rm dim (\mathcal{Z}^Y_{\begin{tikzpicture}
    \BoxGraph[black][1.2]{0.3}{}{
    \draw (A)--(B);
    \draw (A)--(D);
    \draw (B)--(C);
    \draw (D)--(C);
    }{}
\end{tikzpicture}})=3$, the vertex $v$ is not simple. In fact, we know that this set of hyperplanes is itself a circuit, satisfying the relation (\ref{eq:DegeneracySquareGraph}). Then, the broken circuit is $\{ \{\lambda_{(1,2)}=0\}, \{\lambda_{(1,3)}=0\}, \{\lambda_{(1,2,3)}=0\}  \}$ and the NBCs are 
\begin{align}
    B_1&= \{ \{\lambda_{(1)}=0\}, \{\lambda_{(1,2)}=0\}, \{\lambda_{(1,3)}=0\}  \}, \quad B_2= \{ \{\lambda_{(1)}=0\}, \{\lambda_{(1,2)}=0\}, \{\lambda_{(1,2,3)}=0\}  \}, \notag \\
    B_3&= \{ \{\lambda_{(1)}=0\}, \{\lambda_{(1,3)}=0\}, \{\lambda_{(1,2,3)}=0\}  \}.
\end{align}
We see that $\mathcal{C}(\lambda_{(1,2)} \cup \lambda_{(1,3)})=\begin{tikzpicture}
        \BoxGraph[black][1]{0.6}{
        \draw (A)--(B);
        \draw (A)--(D);
        \draw (B)--(C);
        \draw (D)--(C);
        \Tube[rounded corners=6pt, inner sep=3pt]{(A) (B)};
        \Tube[rounded corners=6pt, inner sep=3pt]{(A) (D)}
        }{}
\end{tikzpicture}$ partitions the graph into four connected subgraphs, so it has order three, i.e. these two hyperplanes intersect only at vertex $v$. Therefore, the projection of basis element $B_1$ in the canonical form vanishes. This corresponds to 
\begin{align}
    \mathrm{Res}_{\lambda_{(1)}}\mathrm{Res}_{\lambda_{(1,2)}}\mathrm{Res}_{\lambda_{(1,3)}} \left( \tilde{\Omega}(\mathcal{Z}^Y_{\begin{tikzpicture}
    \BoxGraph[black][1.2]{0.3}{}{
    \draw (A)--(B);
    \draw (A)--(D);
    \draw (B)--(C);
    \draw (D)--(C);
    }{}
    \end{tikzpicture}})_v \right)=0
\end{align}

As for $B_2$ and $B_3$ , we see that each time a cut is added in the inverse order of the NBC, the order of the cut increases exactly by one. Algebraically, this translates into the fact that the iterated residue in the reverse order of the corresponding NBC basis element is non-zero. Consequently, the simplicial cones defined by these two basis elements form a simplicial refinement of $N_v\mathcal{Z}^Y_{\begin{tikzpicture}
    \BoxGraph[black][1.2]{0.3}{}{
    \draw (A)--(B);
    \draw (A)--(D);
    \draw (B)--(C);
    \draw (D)--(C);
    }{}
    \end{tikzpicture}}$ and appear in the canonical function. Then, we get 
\begin{equation}
    \tilde{\Omega}(\mathcal{Z}^Y_{\begin{tikzpicture}
    \BoxGraph[black][1.2]{0.3}{}{
    \draw (A)--(B);
    \draw (A)--(D);
    \draw (B)--(C);
    \draw (D)--(C);
    }{}
    \end{tikzpicture}})_v=\frac{1}{\lambda_{(1)}\lambda_{(1,2)}\lambda_{(1,2,3)}}+\frac{1}{\lambda_{(1)}\lambda_{(1,3)}\lambda_{(1,2,3)}}.
    \label{eq:NBCMethodSquare}
\end{equation}
Consider now starting with the ordering of hyperplanes $\{ \{\lambda_{(1,2)}=0\}, \{\lambda_{(1)}=0\},\{\lambda_{(1,3)}=0\}, \{\lambda_{(1,2,3)}=0\}   \}$. Then we find 
\begin{equation}
    \tilde{\Omega}(\mathcal{Z}^Y_{\begin{tikzpicture}
    \BoxGraph[black][1.2]{0.3}{}{
    \draw (A)--(B);
    \draw (A)--(D);
    \draw (B)--(C);
    \draw (D)--(C);
    }{}
    \end{tikzpicture}})_v=\frac{1}{\lambda_{(1)}\lambda_{(1,2)}\lambda_{(1,3)}}+\frac{1}{\lambda_{(1,2)}\lambda_{(1,3)}\lambda_{(1,2,3)}}.
    \label{eq:NBCMethodSquare2}
\end{equation}
Both results are equal, as we showed thanks to the Orlik-Solomon relations in \Eq{eq:OSSquare}. They correspond to the two possible simplicial refinements of $N_v\mathcal{Z}^Y_{\begin{tikzpicture}
    \BoxGraph[black][1.2]{0.3}{}{
    \draw (A)--(B);
    \draw (A)--(D);
    \draw (B)--(C);
    \draw (D)--(C);
    }{}
    \end{tikzpicture}}$ that do not introduce spurious simple poles. The simplicial refinement for the first expression obtained corresponds to the braid fan (restricted to the cone $N_v\mathcal{Z}^Y_{\begin{tikzpicture}
    \BoxGraph[black][1.2]{0.3}{}{
    \draw (A)--(B);
    \draw (A)--(D);
    \draw (B)--(C);
    \draw (D)--(C);
    }{}
    \end{tikzpicture}}$). It is precisely given by the triangulation illustrated in Fig. \ref{fig:CausalSubdivision} (right).

\subsection{Unrefined form \label{sec:NumInt}}

The method with the NBC basis described above does not introduce any spurious singularity of codimension one, since the simplicial refinement only uses hyperplanes that define boundaries of the polytope. This wasn't the case for the braid fan refinement, where some causal thresholds appearing in the amplitude correspond to facets existing in the permutohedron, but not the polytope itself.

Although no new linear denominators are introduced, the simplicial subdivision makes certain products of physical facet denominators appear in individual terms even when the corresponding collection of facets does not define a face of the expected codimension. These spurious higher-codimension singularities cancel after summing all simplicial cones refining the same normal cone. 
This means that there may exist iterated residues that vanish when applied to the causal representation of the canonical form as obtained with the NBC basis method, or by refinement of the normal fan, but do not vanish for each term individually. These residues, i.e. spurious poles, vanish when all terms corresponding to the same normal cone $N_vP$ are added. When a normal cone at a vertex is not simplicial and needs to be refined, we call the resulting terms in the expression of the canonical function refined terms.

We can illustrate with the previous example of the square graph, where we have
\begin{align}
    \mathrm{Res}_{\lambda_{(1)}}\mathrm{Res}_{\lambda_{(1,2,3)}} \left( \tilde{\Omega}(\mathcal{Z}^Y_{\begin{tikzpicture}
    \BoxGraph[black][1.2]{0.3}{}{
    \draw (A)--(B);
    \draw (A)--(D);
    \draw (B)--(C);
    \draw (D)--(C);
    }{}
    \end{tikzpicture}})_v \right)=0,
\end{align}
but 
\begin{align}
    \mathrm{Res}_{\lambda_{(1)}}\mathrm{Res}_{\lambda_{(1,2,3)}} \left(  \frac{1}{\lambda_{(1)}\lambda_{(1,2)}\lambda_{(1,2,3)}}\right)=-\mathrm{Res}_{\lambda_{(1)}}\mathrm{Res}_{\lambda_{(1,2,3)}} \left(  \frac{1}{\lambda_{(1)}\lambda_{(1,3)}\lambda_{(1,2,3)}}\right)\neq 0.
\end{align}
This is certainly not desirable if we want to integrate the expression obtained over the loop three-momenta on which the $y_e$ variables depend. With expressions like \Eq{eq:NBCMethodSquare} obtained from simplicial refinement, for some values of the loop three momenta, terms in the integrand diverge even if the full integrand doesn't, leading to big and undesirable cancellations.

There is a way to avoid this by gathering in a single term all the refined terms corresponding to the same acyclic orientation, i.e. the terms that come from the same top-dimensional cone in the normal fan. Because the simplicial refinement is done cone by cone for the top-dimensional cones in the normal fan, the cancellations of spurious poles only happen between simplicial cones coming from the simplicial refinement of the same cone of the normal fan. Therefore, having a single term for each top-dimensional cone of the normal guarantees that the expression is free of the spurious divergences introduced by the simplicial refinement. This expression does not correspond to a certain simplicial refinement of the normal fan, even if we obtained it by first refining the fan. This is why we call this way of expressing the canonical form as the \textit{unrefined causal representation}.

The example above gives us the term
\begin{equation}
    \tilde{\Omega}(\mathcal{Z}^Y_{\begin{tikzpicture}
    \BoxGraph[black][1.2]{0.3}{}{
    \draw (A)--(B);
    \draw (A)--(D);
    \draw (B)--(C);
    \draw (D)--(C);
    }{}
    \end{tikzpicture}})_v=\frac{\lambda_{(1,2)}+\lambda_{(1,3)}}{\lambda_{(1)}\lambda_{(1,2)}\lambda_{(1,3)}\lambda_{(1,2,3)}}
\end{equation}
for the canonical function of the square graph. Because of the relation in \Eq{eq:DegeneracySquareGraph}, we obtain the same numerator if we started from \Eq{eq:NBCMethodSquare} or from \Eq{eq:NBCMethodSquare2} to get the unrefined causal representation. It is independent of the simplicial refinement used.

Start from a refined causal representation, and let $v$ be a vertex of $\mathcal{Z}^Y_\Gamma$. Let $I$ label a subset of energy denominators such that they all appear together in the denominator of a refined term of $\tilde{\Omega}(\mathcal{Z}^Y_\Gamma)_v$. Let us call $K_v$ the set of all these subsets of energy denominators, so that each element $I \in K_v$ corresponds to a refined term in $\tilde{\Omega}(\mathcal{Z}^Y_\Gamma)_v$. Call $L_v$ the set labeling all energy denominators corresponding to the facets incident at $v$, i.e. compatible with acyclic orientation $\vec{\Gamma}_v$. Then, the general form of the term associated with $v$ in the unrefined causal representation is 
\begin{equation}
    \tilde{\Omega}(\mathcal{Z}^Y_\Gamma)_v=\frac{\sum_{I \in K_v}\prod_{i \in L_v \backslash I} \lambda_i}{\prod_{j\in L_v}\lambda_j}.
 \end{equation}
This formula corresponds to putting the refined terms over the common denominator $\prod_{j\in L_v}\lambda_j$.

\section{Triangulations and the spanning tree representation \label{sec:SpanningTree}}

\subsection{Zonotopal tilings and the Cayley trick \label{sec:zonotopaltilings}}

Consider the $d$-dimensional zonotope generated by a vector configuration $W=\{\mathbf{w}_1,\dots, \mathbf{w}_n\}$ : $\mathcal Z_{\mathbf{w}}=[-\mathbf{w}_1,\mathbf{w}_1]+\dots+[-\mathbf{w}_n,\mathbf{w}_n] \subset \mathbb{R}^d$. The vector configuration $W$ defines an oriented matroid $\mathcal M_W$. \textit{Zonotopal tilings} are subdivisions of zonotopes into smaller $d$-dimensional zonotopes, called \textit{tiles}. A tile can be labeled by a sign vector $\sigma \in\{+,-,0\}^n$ on the generators of the zonotope, such that the subset of vectors with sign $0$ is of rank $d$ in $\mathcal M_W$, i.e. it generates $\mathbb{R}^d$. We call $\sigma ^+$, $\sigma ^-$ and $\sigma ^0$ the subsets of vectors with signs $+$, $-$ and $0$, respectively, in $\sigma$. Then tiles are of the form
\begin{equation}
    \mathcal{Z}_{W}^{\sigma}=\sum_{i \in \sigma^0} [-\mathbf{w}_i,\mathbf{w}_i] +\sum_{j \in \sigma^+} \mathbf{w}_j-\sum_{k \in \sigma^-} \mathbf{w}_k,
    \label{eq:ZonotopalTiling}
\end{equation}
Thus, the subset $\sigma^0$ indexes the generator vectors in the tile, so it controls its shape, while the vectors in $\sigma^+$ and $\sigma^-$ translate it, i.e. they determine where the tile is placed inside the zonotope. A collection of such tiles that do not overlap and completely fill the zonotope yields a zonotopal tiling. We are interested in zonotopal tilings that are \textit{fine}, and we will also give particular attention to \textit{regular} zonotopal tilings.

A \textit{fine zonotopal tiling} is a zonotopal tiling such that all its tiles are generated by linearly independent vectors $\mathbf{w}_i$, $i \in I$. Such zonotopes are called parallelotopes. Therefore fine tilings are given by sign vectors such that $\sigma^0$ corresponds to a basis of $\mathcal M_W$. In a fine zonotopal tiling, there is a bijection between tiles and bases of $\mathcal M_W$.

\textit{Regular zonotopal tilings} are defined by lifting the generators of the zonotopes at certain heights $h_i$, $i=1,\dots,n$, and projecting the lower faces of the resulting zonotope, analogous to the way in which regular triangulations are defined.

In the case of the graphical zonotope of the graph $\Gamma$, the oriented matroid controlling zonotopal tilings is the graphic oriented matroid $\mathcal{M}_{\vec{\Gamma}}$, where $\vec{\Gamma}$ is a reference orientation. Then, generators  are of the form $\mathbf{w}_e=\mathbf{e}_j-\mathbf{e}_i$, where $e=1,\dots,E$ is the edge in $\vec{\Gamma}$ going from vertex $i$ to vertex $j$. So a vector configuration requires a choice of orientation for each edge, i.e. an orientation of $\Gamma$. Bases of $\mathcal{M}_{\vec{\Gamma}}$ are spanning trees of $\Gamma$. We call $\mathcal{T}(\Gamma)$ the set of spanning trees in $\Gamma$. 

Graphically, one can represent the fine tile of the graphical zonotope $\mathcal{Z}_\Gamma$ given by $\sigma$ in the following way.The edges indexed by $\sigma_0$ (edges in the spanning tree) are bioriented, while the edges $e$ such that $\sigma_e=+$ (resp. $\sigma_e=-$) are oriented in the same direction as $\vec{\Gamma}$ (resp. opposite direction). We call \textit{sign graphs} the resulting graphs with orientations or biorientations of edges.

For each spanning tree $\mathcal{T}$ in $\Gamma$, there is exactly one tile in the fine zonotopal tiling such that $\sigma^0=E(\mathcal{T})$. For a given zonotopal tiling, we label $\sigma_{\mathcal{T}}$ the sign vector of that tile. 
The most complex task consists of determining which combinations of signed sets give a fine zonotopal tiling. For each tile in a zonotopal tiling, the subsets $\sigma_{\mathcal{T}}^+$ and $\sigma_{\mathcal{T}}^-$ must translate the tile (see Eq. (\ref{eq:ZonotopalTiling})) such that they do not overlap each other, thus covering the zonotope. We must therefore find a way of determining, for every tile, the signs in $\sigma_{\mathcal{T}}$ of the edges not in the spanning tree, i.e. determining $\sigma_{\mathcal{T}}^+$ and $\sigma_{\mathcal{T}}^-$.

 One possible way would be using the Bohne-Dress theorem~\cite{bohne1992kombinatorische}, which establishes a bijection between any zonotopal tiling (not only regular) and one-element liftings of $\mathcal{M}_{\vec{\Gamma}}$. However, there are simpler ways, and this question has been covered in the mathematical literature. We will treat it in Sec.~\ref{eq:ExternalAtlas}. 

The importance of zonotopal tilings can be understood by applying the \textit{Cayley trick}~\cite{ubt_eref1116}. This theorem deals with a polytope $A=A_1+\dots+A_n$ obtained from a Minkowski sum, which can be subdivided into \textit{Minkowski cells} $B_1+\dots+B_n$, where $B_i$ is a face of $A_i$, yielding a \textit{mixed subdivision} of the polytope. \textit{Fine mixed subdivisions} are obtained from an affinely independent set of $B_i$, where the $B_i$ are simplices. The Cayley trick states that there is a bijection between mixed subdivisions of $A$ and polyhedral subdivisions of $\mathcal{C}(A_1,\dots,A_n)$. Recall that $\mathcal{C}$ denotes the Cayley polytope. Correspondingly, fine mixed subdivisions of $A$ are in bijection with triangulations of $\mathcal{C}(A_1,\dots,A_n)$.

Zonotopal tilings are therefore special cases of mixed subdivisions, where all $A_i$ are segments, and fine zonotopal tilings correspond to fine mixed subdivisions. In fact, from \Eq{eq:ZonotopalTiling}, we can see that the choice of faces $B_i$ for each segment corresponds to a choice of orientation or biorientation of each edge, i.e. a sign vector $\sigma \in \{+,-,0\}^E$. We then have the following bijections:
\begin{align}
    \left\{  \text{(Regular) zonotopal tilings of } \mathcal{Z}_\Gamma\right\} & \Leftrightarrow  \left\{ \text{(Regular) subdivisions of } \mathcal{S}_\Gamma \right\} \\
    \left\{  \text{(Regular) fine zonotopal tilings of } \mathcal{Z}_\Gamma\right\} & \Leftrightarrow  \left\{ \text{(Regular) triangulations of } \mathcal{S}_\Gamma \right\}.
\end{align}
Then, given a triangulation T of $\mathcal{S}_\Gamma$, we can call each simplex of the triangulation $\Delta^T_{\mathcal{T}}$, labeled by the corresponding tree $\mathcal{T}$, since we know that fine tiles in a zonotopal tiling are in bijection with both spanning trees of $\Gamma$ and simplices in the corresponding triangulation. We call $\tau_\mathcal{T}$ the corresponding fine tile in $\mathcal{Z}_\Gamma$.

The simplex $\mathcal{C}(B_1,\dots,B_n)$ in the Cayley polytope, where $B_1+\dots+B_n$ corresponds to a fine Minkowski cell, is the convex hull of the vertices of $B_1,\dots,B_n$, after Cayley embedding. From \Eq{eq:ZonotopalTiling}, each simplex is generated by $|\sigma^0|=N-1$ segments, since each spanning tree has $N-1$ edges, and $|\sigma^+|+|\sigma^-|=E-(N-1)=L$ points. Therefore, the simplex is the convex hull of $2(N-1)+E-(N-1)=E+N-1=\mathrm{dim}(\mathcal{S}_\Gamma)+1$ points, as expected. This is not surprising, since the SF of a tree graph is a simplex. The remarkable fact is that the simplices in a triangulation of the SF can be labeled by its spanning trees. 

Moreover, the sign vector $\sigma_\mathcal{T}$ of each tile $\tau_\mathcal{T}$ tells us about the vertices in the corresponding simplex, where $\sigma(e)=0$ means both $\mathbf{y_e}+\mathbf{x_j}-\mathbf{x_i} \in \Delta_\mathcal{T}$ and $\mathbf{y_e}+\mathbf{x_i}-\mathbf{x_j}\in \Delta_\mathcal{T}$, $\sigma(e)=+$ (resp. $\sigma(e)=-$) means $\mathbf{y_e}+\mathbf{x_j}-\mathbf{x_i} \in \Delta_\mathcal{T}$ and $\mathbf{y_e}+\mathbf{x_i}-\mathbf{x_j} \notin \Delta_\mathcal{T}$ (resp. $\mathbf{y_e}+\mathbf{x_i}-\mathbf{x_j} \in \Delta_\mathcal{T}$ and $\mathbf{y_e}+\mathbf{x_j}-\mathbf{x_i} \notin \Delta_\mathcal{T}$). Then, $\Delta_\mathcal{T}$ is simply the convex hull of these vertices of $\mathcal{S}_\Gamma$ that also belong to it.

\begin{figure}[h]
\centering

\begin{minipage}{0.48\textwidth}
\centering
\begin{tikzpicture}[
    scale=2.2,
    >=Latex,
    every node/.style={font=\small},
    edge/.style={thick},
]


\coordinate (v1) at (1,0);
\coordinate (v2) at (0.5,0.866);
\coordinate (v3) at (-0.5,0.866);
\coordinate (v4) at (-1,0);
\coordinate (v5) at (-0.5,-0.866);
\coordinate (v6) at (0.5,-0.866);
\coordinate (p) at (0,0);

\fill[blue!10] (v1)--(v2)--(v3)--(v4)--(v5)--(v6)--cycle;
\draw[thick] (v1)--(v2)--(v3)--(v4)--(v5)--(v6)--cycle;


\draw[thick] (v1)--(v2)--(v3)--(p)--cycle;
\draw[thick] (v3)--(v4)--(v5)--(p)--cycle;
\draw[thick] (v5)--(v6)--(v1)--(p)--cycle;


\begin{scope}[shift={($0.45*(p)+0.55*(v2)$)}, scale=0.15]
\TriGraph[
  \node at ($(A)!0.5!(B)+(-0.1,0.06)$) {$a$};
  \node at ($(A)!0.5!(C)+(0,-0.14)$) {$b$};
  \node at ($(B)!0.5!(C)+(0.1,0.06)$) {$c$};
]{(0,0)}{5}{
\draw[double arrow,red,thick] (A)--(B);
\draw[double arrow,red,thick] (B)--(C);
\draw[->-,thick] (A)--(C);
}{}
\end{scope}

\begin{scope}[shift={($0.5*(v3)+0.5*(v5)$)}, scale=0.15]
\TriGraph[
  \node at ($(A)!0.5!(B)+(-0.1,0.06)$) {$a$};
  \node at ($(A)!0.5!(C)+(0,-0.14)$) {$b$};
  \node at ($(B)!0.5!(C)+(0.1,0.06)$) {$c$};
]{(0,0)}{5}{
\draw[double arrow,red,thick] (A)--(C);
\draw[double arrow,red,thick] (B)--(C);
\draw[->-,thick] (B)--(A);
}{}
\end{scope}

\begin{scope}[shift={($0.5*(v1)+0.5*(v5)$)}, scale=0.15]
\TriGraph[
  \node at ($(A)!0.5!(B)+(-0.1,0.06)$) {$a$};
  \node at ($(A)!0.5!(C)+(0,-0.14)$) {$b$};
  \node at ($(B)!0.5!(C)+(0.1,0.06)$) {$c$};
]{(0,0)}{5}{
\draw[double arrow,red,thick] (A)--(B);
\draw[double arrow,red,thick] (A)--(C);
\draw[->-,thick] (C)--(B);
}{}
\end{scope}


\coordinate (base) at (-0.5,-0.866);

\draw[->,violet] (base) -- ++(1,0)
node[font=\scriptsize,pos=0.55,above] {$\mathbf{w}_a$};

\draw[->,violet] (base) -- ++(0.5,0.866)
node[font=\scriptsize,right,pos=0.5,align=center] {$\mathbf{w}_b$};

\draw[->,violet] (base) -- ++(-0.5,0.866)
node[font=\scriptsize,right,pos=0.5,align=center] {$\mathbf{w}_c$};


\TriGraph{(1.3,0)}{0.55}{
\draw[-->-, thick] (A)--(B);
\draw[-->-, thick] (A)--(C);
\draw[-->-, thick] (C)--(B);
}{}

\TriGraph{(0.7,1.1)}{0.55}{
\draw[-->-, thick] (A) to (B);
\draw[-->-, thick] (A) to (C);
\draw[-->-, thick] (B) to (C);
}{}

\TriGraph{(-0.7,1.1)}{0.55}{
\draw[-->-, thick] (B)--(A);
\draw[-->-, thick] (B)--(C);
\draw[-->-, thick] (A)--(C);
}{}

\TriGraph{(-1.3,0)}{0.55}{
\draw[-->-, thick] (B)--(A);
\draw[-->-, thick] (C)--(A);
\draw[-->-, thick] (B)--(C);
}{}

\TriGraph{(-0.7,-1.1)}{0.55}{
\draw[-->-, thick] (B)--(A);
\draw[-->-, thick] (C)--(B);
\draw[-->-, thick] (C)--(A);
}{}

\TriGraph{(0.7,-1.1)}{0.55}{
\draw[-->-, thick] (A)--(B);
\draw[-->-, thick] (C)--(A);
\draw[-->-, thick] (C)--(B);
}{}

\TriGraph{(0.15,0.1)}{0.4}{
\draw[->-=0.6] (B)--(A);
\draw[->-=0.6] (A)--(C);
\draw[->-=0.6] (C)--(B);
}{}



\TriGraph{(1.05,0.5)}{0.45}{
\draw[thick] (A)--(B);
\draw[thick] (A)--(C);
\draw[thick] (B)--(C);
\Tube{(A)};
}{}

\TriGraph{(0,1.05)}{0.45}{
\draw[thick] (A)--(C);
\draw[thick] (B)--(C);
\draw[thick] (A)--(B);
\Tube[rounded corners=8pt, inner sep=4pt]{(A) (B)};
}{}

\TriGraph{(-1,0.5)}{0.45}{
\draw[thick] (B)--(A);
\draw[thick] (B)--(C);
\draw[thick] (A)--(C);
\Tube{(B)};
}{}

\TriGraph{(-1.05,-0.5)}{0.45}{
\draw[thick]  (B)--(A);
\draw[thick]  (C)--(A);
\draw[thick] (C)--(B);
\Tube[rounded corners=8pt, inner sep=4pt]{(B) (C)};
}{}

\TriGraph{(0,-1.1)}{0.45}{
\draw[thick]  (C)--(A);
\draw[thick] (C)--(B);
\draw[thick] (B)--(A);
\Tube{(C)};
}{}

\TriGraph{(1.05,-0.5)}{0.45}{
\draw[thick]  (A)--(B);
\draw[thick]  (C)--(B);
\draw[thick] (C)--(A);
\Tube{(A) (C)};
}{}

\end{tikzpicture}
\end{minipage}
\hfill
\begin{minipage}{0.48\textwidth}
\centering
\begin{tikzpicture}[
    scale=2.2,
    >=Latex,
    every node/.style={font=\small},
    edge/.style={thick},
]


\coordinate (v1) at (1,0);
\coordinate (v2) at (0.5,0.866);
\coordinate (v3) at (-0.5,0.866);
\coordinate (v4) at (-1,0);
\coordinate (v5) at (-0.5,-0.866);
\coordinate (v6) at (0.5,-0.866);
\coordinate (p) at (0,0);

\fill[blue!10] (v1)--(v2)--(v3)--(v4)--(v5)--(v6)--cycle;
\draw[thick] (v1)--(v2)--(v3)--(v4)--(v5)--(v6)--cycle;


\draw[thick] (v2)--(v3)--(v4)--(p)--cycle;
\draw[thick] (v4)--(v5)--(v6)--(p)--cycle;
\draw[thick] (v6)--(v1)--(v2)--(p)--cycle;


\begin{scope}[shift={($0.45*(p)+0.55*(v3)$)}, scale=0.15]
\TriGraph[
  \node at ($(A)!0.5!(B)+(-0.1,0.06)$) {$a$};
  \node at ($(A)!0.5!(C)+(0,-0.14)$) {$b$};
  \node at ($(B)!0.5!(C)+(0.1,0.06)$) {$c$};
]{(0,0)}{5}{
\draw[double arrow,red,thick] (A)--(B);
\draw[double arrow,red,thick] (A)--(C);
\draw[->-,thick] (B)--(C);
}{}
\end{scope}

\begin{scope}[shift={($0.5*(v4)+0.5*(v6)$)}, scale=0.15]
\TriGraph[
  \node at ($(A)!0.5!(B)+(-0.1,0.06)$) {$a$};
  \node at ($(A)!0.5!(C)+(0,-0.14)$) {$b$};
  \node at ($(B)!0.5!(C)+(0.1,0.06)$) {$c$};
]{(0,0)}{5}{
\draw[double arrow,red,thick] (A)--(B);
\draw[double arrow,red,thick] (B)--(C);
\draw[->-,thick] (C)--(A);
}{}
\end{scope}

\begin{scope}[shift={($0.5*(v6)+0.5*(v2)$)}, scale=0.15]
\TriGraph[
  \node at ($(A)!0.5!(B)+(-0.1,0.06)$) {$a$};
  \node at ($(A)!0.5!(C)+(0,-0.14)$) {$b$};
  \node at ($(B)!0.5!(C)+(0.1,0.06)$) {$c$};
]{(0,0)}{5}{
\draw[double arrow,red,thick] (C)--(B);
\draw[double arrow,red,thick] (A)--(C);
\draw[->-,thick] (A)--(B);
}{}
\end{scope}


\coordinate (base) at (-0.5,-0.866);

\draw[->,violet] (base) -- ++(1,0)
node[font=\scriptsize,pos=0.55,above] {$\mathbf{w}_a$};

\draw[->,violet] (base) -- ++(0.5,0.866)
node[font=\scriptsize,left,pos=0.9,align=center] {$\mathbf{w}_b$};

\draw[->,violet] (base) -- ++(-0.5,0.866)
node[font=\scriptsize,right,pos=0.5,align=center] {$\mathbf{w}_c$};


\TriGraph{(1.3,0)}{0.55}{
\draw[-->-, thick] (A)--(B);
\draw[-->-, thick] (A)--(C);
\draw[-->-, thick] (C)--(B);
}{}

\TriGraph{(0.7,1.1)}{0.55}{
\draw[-->-, thick] (A) to (B);
\draw[-->-, thick] (A) to (C);
\draw[-->-, thick] (B) to (C);
}{}

\TriGraph{(-0.7,1.1)}{0.55}{
\draw[-->-, thick] (B)--(A);
\draw[-->-, thick] (B)--(C);
\draw[-->-, thick] (A)--(C);
}{}

\TriGraph{(-1.3,0)}{0.55}{
\draw[-->-, thick] (B)--(A);
\draw[-->-, thick] (C)--(A);
\draw[-->-, thick] (B)--(C);
}{}

\TriGraph{(-0.7,-1.1)}{0.55}{
\draw[-->-, thick] (B)--(A);
\draw[-->-, thick] (C)--(B);
\draw[-->-, thick] (C)--(A);
}{}

\TriGraph{(0.7,-1.1)}{0.55}{
\draw[-->-, thick] (A)--(B);
\draw[-->-, thick] (C)--(A);
\draw[-->-, thick] (C)--(B);
}{}

\TriGraph{(-0.15,0.1)}{0.4}{
\draw[->-=0.6] (A)--(B);
\draw[->-=0.6] (C)--(A);
\draw[->-=0.6] (B)--(C);
}{}



\TriGraph{(1.05,0.5)}{0.45}{
\draw[thick] (A)--(B);
\draw[thick] (A)--(C);
\draw[thick] (B)--(C);
\Tube{(A)};
}{}

\TriGraph{(0,1.05)}{0.45}{
\draw[thick] (A)--(C);
\draw[thick] (B)--(C);
\draw[thick] (A)--(B);
\Tube[rounded corners=8pt, inner sep=4pt]{(A) (B)};
}{}

\TriGraph{(-1,0.5)}{0.45}{
\draw[thick] (B)--(A);
\draw[thick] (B)--(C);
\draw[thick] (A)--(C);
\Tube{(B)};
}{}

\TriGraph{(-1.05,-0.5)}{0.45}{
\draw[thick]  (B)--(A);
\draw[thick]  (C)--(A);
\draw[thick] (C)--(B);
\Tube[rounded corners=8pt, inner sep=4pt]{(B) (C)};
}{}

\TriGraph{(0,-1.1)}{0.45}{
\draw[thick]  (C)--(A);
\draw[thick] (C)--(B);
\draw[thick] (B)--(A);
\Tube{(C)};
}{}

\TriGraph{(1.05,-0.5)}{0.45}{
\draw[thick]  (A)--(B);
\draw[thick]  (C)--(B);
\draw[thick] (C)--(A);
\Tube{(A) (C)};
}{}

\end{tikzpicture}

\end{minipage}

\caption[Zonotopal tilings]{The two fine zonotopal tilings of $\mathcal{Z}_{\begin{tikzpicture}
    \TriGraph{(0,0)}{0.3}{
    \draw (A)--(B);
    \draw (A)--(C);
    \draw (C)--(B);
    }{} 
\end{tikzpicture}}$, with each fine tile labeled by a sign graph, and the corresponding spanning tree in red.}
\label{fig:ZonotopalTilingTriangle}
\end{figure}

The two fine zonotopal tilings of $\mathcal{Z}_{\begin{tikzpicture}
    \TriGraph{(0,0)}{0.3}{
    \draw (A)--(B);
    \draw (A)--(C);
    \draw (C)--(B);
    }{} 
\end{tikzpicture}}$ are shown in Fig.~\ref{fig:ZonotopalTilingTriangle}, and the spanning tree corresponding to each tile is drawn in red. Each of the tiles corresponds to a simplex in $\mathcal{S}_{\begin{tikzpicture}
    \TriGraph{(0,0)}{0.3}{
    \draw (A)--(B);
    \draw (A)--(C);
    \draw (C)--(B);
    }{} 
\end{tikzpicture}}$. Therefore, $\mathcal{S}_{\begin{tikzpicture}
    \TriGraph{(0,0)}{0.3}{
    \draw (A)--(B);
    \draw (A)--(C);
    \draw (C)--(B);
    }{} 
\end{tikzpicture}}$ has two triangulations\footnote{We only consider triangulations which only use the vertices of the polytope.} and three simplices in each of them. 

From these sign graphs one can easily read the facets bounding the corresponding tile. The minimal cuts defining the facets of each tile are given by the cuts of the corresponding spanning tree. Each of these minimal cuts must cut exactly one edge of the spanning tree $\mathcal{T}$. This results in $2\times (N-1)$ minimal cuts, corresponding to the expected number of facets of a fine tile $2\times \mathrm{dim}(\mathcal{Z}_\Gamma)$. Then, to the cut edge $e\in \mathcal{T}$ we associate $y_e$, while to cut edges $e\notin \mathcal{T}$ oriented in the same direction (resp. opposite direction) of the cut, the corresponding denominator appears with the variable $+y_e$ (resp. $-y_e$).

For example, consider the tile with sign graph \tikz[baseline=-0.5ex]{
\TriGraph[
  \node at ($(A)!0.5!(B)+(-0.22,0.1)$) {a};
  \node at ($(A)!0.5!(C)+(0,-0.2)$) {b};
  \node at ($(B)!0.5!(C)+(0.2,0.1)$) {c};
  \node at ($(A)+(-0.1,-0.05)$) {1};
  \node at ($(B)+(0,0.18)$) {2};
  \node at ($(C)+(0.13,-0.05)$) {3};
]{(0,0)}{1}{
  \draw[double arrow2,red,thick] (A)--(B);
  \draw[double arrow2,red,thick] (B)--(C);
  \draw[->-, thick] (A)--(C);
}{}
}. The four minimal cuts are \tikz[baseline=-0.5ex]{
\TriGraph[
  \Tube[rounded corners=4pt, inner sep=4pt]{(A)};
]{(0,0)}{0.8}{
  \draw[double arrow2,red,thick] (A)--(B);
  \draw[double arrow2,red,thick] (B)--(C);
  \draw[->-] (A)--(C);
}{}
}
, \tikz[baseline=-0.5ex]{
\TriGraph[
  \Tube[rounded corners=8.5pt, inner sep=4pt]{(A) (B)};
]{(0,0)}{0.8}{
  \draw[double arrow2,red,thick] (A)--(B);
  \draw[double arrow2,red,thick] (B)--(C);
  \draw[->-] (A)--(C);
}{}
}, 
\tikz[baseline=-0.5ex]{
\TriGraph[
  \Tube[rounded corners=8.5pt, inner sep=4pt]{(B) (C)};
]{(0,0)}{0.8}{
  \draw[double arrow2,red,thick] (A)--(B);
  \draw[double arrow2,red,thick] (B)--(C);
  \draw[->-] (A)--(C);
}{}
}
and \tikz[baseline=-0.5ex]{
\TriGraph[
  \Tube[rounded corners=4pt, inner sep=4pt]{(C)};
]{(0,0)}{0.8}{
  \draw[double arrow2,red,thick] (A)--(B);
  \draw[double arrow2,red,thick] (B)--(C);
  \draw[->-] (A)--(C);
}{}
}, corresponding to hyperplanes $\{  y_a+y_b+x_1=0 \}$, $\{  y_b+y_c+x_1+x_2=0 \}$, $\{  y_a-y_b-x_1=0 \}$ and $\{  y_c-y_b-x_1-x_2=0 \}$,respectively. The first two are real boundaries of $\mathcal{Z}_{\begin{tikzpicture}
    \TriGraph{(0,0)}{0.3}{
    \draw (A)--(B);
    \draw (A)--(C);
    \draw (C)--(B);
    }{} 
\end{tikzpicture}}$, while the other two are internal boundaries that separate tiles, giving rise to spurious poles in this representation of the integral. The canonical form of this tile is 
\begin{equation}
\tilde{\Omega}\big(\tikz[baseline=-0.5ex]{
\TriGraph{(0,0)}{1}{
  \draw[double arrow2,red,thick] (A)--(B);
  \draw[double arrow2,red,thick] (B)--(C);
  \draw[->-] (A)--(C);
}{}
}\big)=\frac{4y_ay_c}{\left(y_a^2-(y_b+x_1)^2\right)\left(y_c^2-(y_b+x_1+x_2)^2\right)}.
\end{equation}
Summing over all tiles and multiplying by the canonical form of the base simplex gives 
\begin{align}
\tilde{\Omega}\big(\mathcal{S}_{\begin{tikzpicture}
    \TriGraph{(0,0)}{0.3}{
    \draw[thick] (A)--(B);
    \draw[thick] (A)--(C);
    \draw[thick] (C)--(B);
    }{} 
\end{tikzpicture}}\big)=&\frac{1}{2y_b}\frac{1}{\left(y_a^2-(y_b+x_1)^2\right)\left(y_c^2-(y_b+x_1+x_2)^2\right)}+\frac{1}{2y_a}\frac{1}{\left(y_b^2-(y_a+x_2+x_3)^2\right)\left(y_c^2-(y_a+x_2)^2\right)} \notag \\
&+\frac{1}{2y_c}\frac{1}{\left(y_a^2-(y_c+x_2)^2\right)\left(y_b^2-(y_c+x_3)^2\right)}.
\end{align}
This corresponds (up to the $(-1)^E$ factor) to the expression obtained from LTD, by iterated integration of loop energy variables. The original representation of LTD~\cite{Catani:2008xa,Bierenbaum:2010cy,Bierenbaum:2012th,Buchta:2014dfa,Buchta:2015wna} is written as a sum over spanning trees of the graph. It is nothing more than a sum over the simplices of a triangulation of the SF. The canonical functions obtained from the different triangulations can be found by choosing a different order of loops in the integration, although there is not a one-to-one correspondence.

\subsection{Graphical method for triangulations \label{eq:ExternalAtlas}}

The combinatorics we need in order to find triangulations of $\mathcal{S}_\Gamma$ mostly appear in the literature in the study of the \textit{tropical Jacobian} $J(\Gamma^Y)$ of a metric graph $\Gamma^Y$~\cite{ding2023framework,BackmanBakerYuen2017,BackmanBakerYuen2019}. Simplifying, a metric graph can be seen as a graph with lengths $l_e $ for each of its edges, so that $l_e\rightarrow 0$ corresponds to an edge contraction. These lengths are related to variables $y_e$, so that each metric graph can be associated to a deformed graphical zonotope $\mathcal{Z}_\Gamma^Y$. A tropical Jacobian is a torus in real space obtained as the field theory limit $\alpha' \rightarrow 0$ in string theory of the Jacobian variety of a Riemann surface, which appears in the calculations of string amplitudes. It is therefore not surprising that this geometry appears in the context of Feynman integrals. The geometry of the tropical Jacobian is in fact closely related to that of the (deformed) graphical zonotope, and its fine zonotopal tilings are used to simplify the study of the geometry of the tropical Jacobian. In that sense, $\mathcal{S}_\Gamma$ encodes, through the deformations of the graphical zonotope, the geometry of all tropical Jacobians of all possible metric graphs for a given graph. However, we will not develop on the connection of the tropical Jacobian to the SF and Feynman integrals. Moreover, because the combinatorics and triangulations do not depend on deformations, provided that there is no degeneration, we just consider (undeformed) graphical zonotopes, corresponding to tropical Jacobians of graphs with unit lengths. 

Most of the work in the literature~\cite{ding2023framework,crepeau2025generalizedbreakdivisorstriangulations} is also related to the Lawrence polytope of the graphic oriented matroid of a directed graph $\vec{\Gamma}$, $\Lambda (\vec{\Gamma})$. It is obtained from Cayley embedding of segments $[0,\mathbf{e}_j-\mathbf{e}_i]$ for each directed edge $i\rightarrow j$ in the directed graph $\vec{\Gamma}$ defining the polytope. $\Lambda (\vec{\Gamma})$ is the convex hull of $\bigcup_{e=ij\in E(\Gamma)} \{\mathbf{y_e}, \mathbf{x_j}-\mathbf{x_i} +\mathbf{y_e} \}$. As $\mathcal{S}_\Gamma$, it also has the base-fiber structure, with the only difference that the fiber is a different realization of the deformed graphical zonotope, $\sum_{i\rightarrow j =e\in \vec{\Gamma}} y_e [0,\mathbf{e}_j-\mathbf{e}_i]$. The Lawrence polytope and the SF are linearly equivalent because they differ by an invertible linear transformation, i.e. a linear change of coordinates. Then, triangulations of $\mathcal{S}_\Gamma$ are also obtained from triangulations of $\Lambda (\vec{\Gamma})$, for any choice of orientation $\vec{\Gamma}$.

Let an \textit{externally oriented basis} $\vec{B}$ be a sign graph where all edges in a spanning tree are bioriented and all edges not in that spanning tree are oriented in only one direction. From Sec.~\ref{sec:zonotopaltilings}, this corresponds to a fine tile of $\mathcal{Z}_\Gamma$, which can equally be represented by a sign vector $\sigma$ such that $\sigma^0=E(\mathcal{T})$, $\mathcal{T}$ being the corresponding spanning tree. We call \textit{external atlas} a collection $\mathcal{A}$ of externally oriented bases where there is exactly one for each spanning tree in $\Gamma$. It is said to be \textit{triangulating} if all $\vec{B} \in \mathcal{A}$ corresponds to a different simplex in the same triangulation of $S_\Gamma$. 

A simple condition for $\mathcal{A}$ to be triangulating is given in Ref.~\cite{ding2023framework}: for any pair $\vec{B}_1,\vec{B}_2\in \mathcal{A}$, $\vec{B}_1 \cap (- \vec{B}_2)$ has no cocircuit, i.e. no cut compatible with the orientation of the edges. Here, an edge with no orientation is considered to be incompatible with any cut. One could find triangulating external atlases by going one externally oriented basis after the other. 

\paragraph{Acyclic cycle signatures} There is a simpler way to proceed in practice~\cite{BackmanBakerYuen2019} involving \textit{acyclic cycle signatures}~\footnote{There may be cycle signatures that are not acyclic but can also give non-regular triangulations, but we restrict here to regular triangulations.}. Call $\mathcal{C}(\Gamma)$ the set of cycles of $\Gamma$; all of them, not just a basis of cycles. A \textit{cycle signature} is a choice of cyclic orientation for each of these cycles. By comparing it to a reference orientation, the cycle signature of $C\in\mathcal{C}(\Gamma)$ is given by a sign vector $\bar{\sigma}(C)$ with $\bar{\sigma}(C)^+ \cup \bar{\sigma}(C)^-=E(C)$. Then, the cycle signature is said to be acyclic if
\begin{equation}
    \sum_{C \in \mathcal{C}(\Gamma)} a_C \bar{\sigma}(C)=0 \text{ with } a_C \geq 0 \implies a_C=0 \text{ for all }  C \in \mathcal{C}(\Gamma),
    \label{eq:AcyclicCycleSignature}
\end{equation}
where we identify $\bar{\sigma}(C)\in \{-1,0,+1\}^E$. This is equivalent to saying that the "sum" of any two directed cycles in $\mathcal{C}(\Gamma)$ must also be a directed cycle (or union of directed cycles) appearing in $\mathcal{C}(\Gamma)$.

A method to easily find an acyclic cycle signature of $\Gamma$ requires to order the edges in $\Gamma$ and choose an orientation $\vec{\Gamma}'$ of $\Gamma$.
For each $C \in \mathcal{C}(\Gamma)$ direct the cycle according to the orientation of the smallest edge belonging to the cycle. Then, the resulting signature $\bar{\sigma}$ is acyclic. There happens to be a bijection between acyclic signatures $\bar{\sigma}$ of $\Gamma$ and regular triangulations of $\mathcal{S}_\Gamma$~\cite{ding2023framework}, that we may then label $T_{\bar{\sigma}}$.

Let us illustrate this procedure with the graph \tikz[baseline=-0.5ex]{
\BoxGraph{1}{
  \node[font=\scriptsize] at ($(A)+(-0.15,0)$) {1};
  \node[font=\scriptsize] at ($(B)+(0,0.15)$) {2};
  \node[font=\scriptsize] at ($(C)+(0.15,0)$) {3};
  \node[font=\scriptsize] at ($(D)+(0,-0.15)$) {4};
  \draw[thick] (A)--(D);
  \draw[thick] (A)--(B);
  \draw[thick] (A)--(C);
  \draw[thick] (B)--(C);
  \draw[thick] (C)--(D);
}
} . We choose the orientation with ordering of edges \tikz[baseline=-0.5ex]{
\BoxGraph{1}{
  \node[font=\scriptsize,red] at ($(A)!0.5!(B)+(-0.12,0.1)$) {1};
  \node[font=\scriptsize,red] at ($(B)!0.5!(C)+(0.1,0.1)$) {2};
  \node[font=\scriptsize,red] at ($(A)!0.5!(C)+(0,0.14)$) {3};
  \node[font=\scriptsize,red] at ($(D)!0.5!(C)+(0.1,-0.1)$) {4};
  \node[font=\scriptsize,red] at ($(A)!0.5!(D)+(-0.1,-0.1)$) {5};
  \draw [->-=0.65, thick] (A)--(D);
  \draw [->-=0.65, thick] (A)--(B);
  \draw [->-=0.65, thick] (A)--(C);
  \draw [->-=0.65, thick] (B)--(C);
  \draw [->-=0.65, thick] (C)--(D);
}
}. Then, the orientations of the three cycles according to the smallest edge are
\begin{equation}
    \begin{tikzpicture}[scale=2]
        \coordinate (A) at (-0.3,0);
        \coordinate (D) at (0,-0.3);
        \coordinate (C) at (0.3,0);
        \fill[black] (A) circle[radius=0.025];
        \fill[black] (D) circle[radius=0.025];
        \fill[black] (C) circle[radius=0.025];
        \draw [->-=1.2, thick] (A)--(C);
        \draw [->-=1.2, thick] (C)--(D);
        \draw [->-=1.2, thick] (D)--(A); 
    \end{tikzpicture}\quad , \qquad 
    \begin{tikzpicture}[scale=2]
        \coordinate (A) at (-0.3,0);
        \coordinate (B) at (0,0.3);
        \coordinate (C) at (0.3,0);
        \fill[black] (A) circle[radius=0.025];
        \fill[black] (B) circle[radius=0.025];
        \fill[black] (C) circle[radius=0.025];
        \draw [->-=1.2, thick] (A)--(B);
        \draw [->-=1.2, thick] (B)--(C);
        \draw [->-=1.2, thick] (C)--(A);
    \end{tikzpicture}\quad , \qquad
     \begin{tikzpicture}
        \BoxGraph{2}{
        \draw [->-=1.2, thick] (D)--(A);
        \draw [->-=1.2, thick] (A)--(B);
        \draw [->-=1.2, thick] (B)--(C);
        \draw [->-=1.2, thick] (C)--(D);
        } 
    \end{tikzpicture}\quad,
\end{equation}
which constitute the acyclic cycle signature $\bar{\sigma}$. In fact, we can see that the sum of the first two directed cycles equals the third one, satisfying thus condition (\ref{eq:AcyclicCycleSignature}).

\paragraph{Triangulating external atlas} Note that a spanning tree $\mathcal{T}$ defines a basis of cycles of a connected graph $\Gamma$, called the fundamental cycle basis. In fact, for each $e\notin \mathcal{T}$ there is exactly one cycle contained in $\{e\}\cup \mathcal{T}$, and the cycles corresponding to different edges constitute a linearly independent subset. We call $C(\mathcal{T},e)$ the cycle defined by $\mathcal{T}$ and $e \notin \mathcal{T}$. Now we can get the triangulating external atlas $\mathcal{A}_{\bar{\sigma}}$. The externally oriented basis $\vec{B}_{\mathcal{T}} \in \mathcal{A}_{\bar{\sigma}}$ corresponding to spanning tree $\mathcal{T}$ is obtained by orienting each edge $e \notin \mathcal{T}$ according to the orientation of $\mathcal{C}(\mathcal{T},e)$ in $\bar{\sigma}$. 

Following the previous example, the externally oriented bases in $\mathcal{A}_{\bar{\sigma}}$ are
\begin{align}
    &\begin{tikzpicture}
        \BoxGraph{2}{
        \draw [->-,thick] (C)--(A);
        \draw [->-,thick] (D)--(A);
        \draw[double arrow,red,thick] (A)--(B);
        \draw[double arrow,red,thick] (B)--(C);
        \draw[double arrow,red,thick] (C)--(D);
        } 
    \end{tikzpicture}\quad, \qquad
    \begin{tikzpicture}
        \BoxGraph{2}{
        \draw [->-,thick] (C)--(A);
        \draw [->-,thick] (C)--(D);
        \draw[double arrow,red,thick] (A)--(B);
        \draw[double arrow,red,thick] (B)--(C);
        \draw[double arrow,red,thick] (A)--(D);
        } 
    \end{tikzpicture}\quad, \qquad
    \begin{tikzpicture}
        \BoxGraph{2}{
        \draw [->-,thick] (A)--(C);
        \draw [->-,thick] (B)--(C);
        \draw[double arrow,red,thick] (A)--(B);
        \draw[double arrow,red,thick] (A)--(D);
        \draw[double arrow,red,thick] (C)--(D);
        } 
    \end{tikzpicture}\quad, \qquad
    \begin{tikzpicture}
        \BoxGraph{2}{
        \draw [->-,thick] (A)--(C);
        \draw [->-,thick] (A)--(B);
        \draw[double arrow,red,thick] (A)--(D);
        \draw[double arrow,red,thick] (B)--(C);
        \draw[double arrow,red,thick] (C)--(D);
        } 
    \end{tikzpicture}\quad, \notag \\
    &\begin{tikzpicture}
        \BoxGraph{2}{
        \draw [->-,thick] (B)--(C);
        \draw [->-,thick] (C)--(D);
        \draw[double arrow,red,thick] (A)--(B);
        \draw[double arrow,red,thick] (A)--(C);
        \draw[double arrow,red,thick] (A)--(D);
        } 
    \end{tikzpicture}\quad, \qquad
    \begin{tikzpicture}
        \BoxGraph{2}{
        \draw [->-,thick] (A)--(B);
        \draw [->-,thick] (D)--(A);
        \draw[double arrow,red,thick] (B)--(C);
        \draw[double arrow,red,thick] (A)--(C);
        \draw[double arrow,red,thick] (D)--(C);
        } 
    \end{tikzpicture}\quad, \qquad
    \begin{tikzpicture}
        \BoxGraph{2}{
        \draw [->-,thick] (B)--(C);
        \draw [->-,thick] (D)--(A);
        \draw[double arrow,red,thick] (A)--(B);
        \draw[double arrow,red,thick] (A)--(C);
        \draw[double arrow,red,thick] (D)--(C);
        } 
    \end{tikzpicture}\quad, \qquad
    \begin{tikzpicture}
        \BoxGraph{2}{
        \draw [->-,thick] (A)--(B);
        \draw [->-,thick] (C)--(D);
        \draw[double arrow,red,thick] (B)--(C);
        \draw[double arrow,red,thick] (A)--(C);
        \draw[double arrow,red,thick] (A)--(D);
        } 
    \end{tikzpicture}\quad.
    \label{eq:TriangulatingExternalAtlas}
\end{align}
This gives us all information we need about the triangulation $T_{\bar{\sigma}}$. For example, the first externally oriented basis (top-left) is the convex hull of $\{\mathbf{y_{12}}+\mathbf{x_1}-\mathbf{x_2},\mathbf{y_{12}}+\mathbf{x_2}-\mathbf{x_1},\mathbf{y_{23}}+\mathbf{x_2}-\mathbf{x_3},\mathbf{y_{23}}+\mathbf{x_3}-\mathbf{x_2},\mathbf{y_{34}}+\mathbf{x_3}-\mathbf{x_4},\mathbf{y_{34}}+\mathbf{x_4}-\mathbf{x_3} ,\mathbf{y_{13}}+\mathbf{x_1}-\mathbf{x_3},\mathbf{y_{14}}+\mathbf{x_1}-\mathbf{x_4} \}$. In the same way we did in Sec.~\ref{sec:zonotopaltilings} for the triangle graph, we find that the term of the canonical function corresponding to this tile in the spanning tree representation is 
\begin{equation}
    \frac{1}{4y_{13}y_{14}}\frac{1}{(y_{12}^2-(y_{13}+y_{14}-x_1)^2)(y_{23}^2-(y_{13}+y_{14}-x_1-x_2)^2)(y_{34}^2-(y_{14}+x_4)^2)}.
    \label{eq:ExampleTesmExternalAtlas}
\end{equation}

\paragraph{Adjacency graph} We also want to look at the pairs of simplices in a triangulation $T_{\bar{\sigma}}$, or equivalently tiles in a fine zonotopal tiling, that share a common internal facet. This information will be useful for future work on non-scalar Feynman integrals. It is now more convenient to reason in terms of zonotopal tilings.  This adjacency of facets in a fine tiling can be encoded in an \textit{adjacency graph} $G_\sigma$. Its vertices are fine tiles of $\mathcal{Z}_\Gamma$, given by externally oriented bases of $\mathcal{A}_{\bar{\sigma}}$, and an edge connecting two vertices means that the two tiles have a facet in common. We note $\tau_\mathcal{T}^\sigma$ the fine tile corresponding to the spanning tree $\mathcal{T}$ in the tiling given by the triangulating external atlas $\mathcal{A}_\sigma$.

A facet of a tile $\tau_\mathcal{T}^\sigma$ is also a parallelotope, i.e. a zonotope generated by linearly independent vectors, but with one dimension less. It is generated by all the generators $\tau_\mathcal{T}^\sigma$ but one. Then, it corresponds to a spanning 2-forest $\mathcal{T}\backslash \{e\}$, with $\{e\} \in \mathcal{T}$. The orientation of $e$ specifies which of the two opposite facets of $\tau_\mathcal{T}^\sigma$ we are referring to. Then, from an externally oriented basis $\vec{B}_{\mathcal{T}}\in \mathcal{A}_{\bar{\sigma}}$ corresponding to tile $\tau_\mathcal{T}^\sigma$, one gets the sign graph of one of its facets by changing a biorientation of one edge $e$ in $\mathcal{T}$ by a specific orientation. The $2(N-1)$ possible graphs obtained from doing so to $\vec{B}_{\mathcal{T}}$ are in bijection with the facets of $\tau_\mathcal{T}^\sigma$. 

Then, two tiles $\tau_{\mathcal{T}_1}^\sigma$ and $\tau_{\mathcal{T}_2}^\sigma$ are adjacent if and only if one can obtain the same sign graph by doing the above described procedure in some way to $\vec{B}_{\mathcal{T}_1}$ and $\vec{B}_{\mathcal{T}_2}$, both in $\mathcal{A}_{\bar{\sigma}}$. For this to be possible, a necessary condition is $\mathcal{T}_1\backslash \{e_1\}=\mathcal{T}_2\backslash \{e_2\}$ for some $e_1 \in \mathcal{T}_1$, $e_2 \in \mathcal{T}_2$, i.e. $\mathcal{T}_1 \cap \mathcal{T}_2$ is a spanning 2-forest. In that case, one can orient $e_1$ in $\vec{B}_{\mathcal{T}_1}$ according to its orientation in $\vec{B}_{\mathcal{T}_2}$ and  $e_2$ in $\vec{B}_{\mathcal{T}_2}$ according to its orientation in $\vec{B}_{\mathcal{T}_1}$. We now have the same sign graph in $E(\mathcal{T}_1 \cup\mathcal{T}_2)$. Therefore, we require the edges in $\Gamma \backslash(\mathcal{T}_1 \cup\mathcal{T}_2)$ to be oriented in the same way in $\vec{B}_{\mathcal{T}_1}$ and $\vec{B}_{\mathcal{T}_2}$ in order for the resulting sign graphs to be the same.

Summing up, $\tau_{\mathcal{T}_1}^\sigma$ and $\tau_{\mathcal{T}_2}^\sigma$ are adjacent if and only if $\mathcal{T}_1\backslash \{e_1\}=\mathcal{T}_2\backslash \{e_2\}$ for some $e_1 \in \mathcal{T}_1$, $e_2 \in \mathcal{T}_1$ and $\vec{B}_{\mathcal{T}_1}|_{\Gamma \backslash(\mathcal{T}_1 \cup\mathcal{T}_2)}=\vec{B}_{\mathcal{T}_2}|_{\Gamma \backslash(\mathcal{T}_1 \cup\mathcal{T}_2)}$.

\begin{figure}
    \centering
    \begin{tikzpicture}[
            metaedge/.style={thick},
        ]       
        \node[circle,minimum size=2cm,inner sep=0pt] (G1) at (-4,4) {};
        \node[circle,minimum size=2cm,inner sep=0pt] (G2) at (4,4) {};
        \node[circle,minimum size=2cm,inner sep=0pt] (G3) at (4,-4) {};
        \node[circle,minimum size=2cm,inner sep=0pt] (G4) at (-4,-4) {};
        \node[circle,minimum size=2cm,inner sep=0pt] (G5) at (-2,0) {};
        \node[circle,minimum size=2cm,inner sep=0pt] (G6) at (0,2) {};
        \node[circle,minimum size=2cm,inner sep=0pt] (G7) at (2,0) {};
        \node[circle,minimum size=2cm,inner sep=0pt] (G8) at (0,-2) {};
        \begin{scope}[shift={(G1.center)}]
            \BoxGraph{2.5}{
            \draw [->-,very thick] (B)--(C);
            \draw [->-,very thick] (D)--(A);
            \draw[double arrow,red,very thick] (A)--(B);
            \draw[double arrow,red,very thick] (A)--(C);
            \draw[double arrow,red,very thick] (D)--(C);
            } 
        \end{scope}
        \begin{scope}[shift={(G2.center)}]
           \BoxGraph{2.5}{
            \draw [->-,very thick] (B)--(C);
            \draw [->-,very thick] (C)--(D);
            \draw[double arrow,red,very thick] (A)--(B);
            \draw[double arrow,red,very thick] (A)--(C);
            \draw[double arrow,red,very thick] (A)--(D);
            }
        \end{scope}
        \begin{scope}[shift={(G3.center)}]
            \BoxGraph{2.5}{
            \draw [->-,very thick] (A)--(B);
            \draw [->-,very thick] (C)--(D);
            \draw[double arrow,red,very thick] (B)--(C);
            \draw[double arrow,red,very thick] (A)--(C);
            \draw[double arrow,red,very thick] (A)--(D);
            } 
        \end{scope}
        \begin{scope}[shift={(G4.center)}]
            \BoxGraph{2.5}{
            \draw [->-,very thick] (A)--(B);
            \draw [->-,very thick] (D)--(A);
            \draw[double arrow,red,very thick] (B)--(C);
            \draw[double arrow,red,very thick] (A)--(C);
            \draw[double arrow,red,very thick] (D)--(C);
            }
        \end{scope}
        \begin{scope}[shift={(G5.center)}]
            \BoxGraph{2.5}{
            \draw [->-,very thick] (C)--(A);
            \draw [->-,very thick] (D)--(A);
            \draw[double arrow,red,very thick] (A)--(B);
            \draw[double arrow,red,very thick] (B)--(C);
            \draw[double arrow,red,very thick] (C)--(D);
            } 
        \end{scope}
        \begin{scope}[shift={(G6.center)}]
           \BoxGraph{2.5}{
            \draw [->-,very thick] (A)--(C);
            \draw [->-,very thick] (B)--(C);
            \draw[double arrow,red,very thick] (A)--(B);
            \draw[double arrow,red,very thick] (A)--(D);
            \draw[double arrow,red,very thick] (C)--(D);
            }
        \end{scope}
        \begin{scope}[shift={(G7.center)}]
            \BoxGraph{2.5}{
            \draw [->-,very thick] (C)--(A);
            \draw [->-,very thick] (C)--(D);
            \draw[double arrow,red,very thick] (A)--(B);
            \draw[double arrow,red,very thick] (B)--(C);
            \draw[double arrow,red,very thick] (A)--(D);
            } 
        \end{scope}
        \begin{scope}[shift={(G8.center)}]
            \BoxGraph{2.5}{
            \draw [->-,very thick] (A)--(C);
            \draw [->-,very thick] (A)--(B);
            \draw[double arrow,red,very thick] (A)--(D);
            \draw[double arrow,red,very thick] (B)--(C);
            \draw[double arrow,red,very thick] (C)--(D);
            } 
        \end{scope}
        \draw[metaedge] (G1) -- (G2);
        \draw[metaedge] (G2) -- (G3);
        \draw[metaedge] (G3) -- (G4);
        \draw[metaedge] (G1) -- (G4);
        \draw[metaedge] (G1) -- (G5);
        \draw[metaedge] (G1) -- (G6);
        \draw[metaedge] (G2) -- (G6);
        \draw[metaedge] (G2) -- (G7);
        \draw[metaedge] (G3) -- (G7);
        \draw[metaedge] (G3) -- (G8);
        \draw[metaedge] (G4) -- (G5);
        \draw[metaedge] (G4) -- (G8);
        \draw[metaedge] (G5) -- (G7);
        \draw[metaedge] (G6) -- (G8);
    \end{tikzpicture}  
    \caption[Adjacency graph]{Adjacency graph of the tiling of $\mathcal{Z}^Y_{\begin{tikzpicture}
    \BoxGraph[black][1.2]{0.3}{}{
    \draw (A)--(B);
    \draw (A)--(D);
    \draw (B)--(C);
    \draw (D)--(C);
    }{}
    \end{tikzpicture}}$ given by the triangulating external atlas $\mathcal{A}_{\bar{\sigma}}$ in (\ref{eq:TriangulatingExternalAtlas}).}
    \label{fig:AdjacencyGraph}
\end{figure}

\subsection{Jeffrey-Kirwan residues \label{sec:JeffreyKirwan}}

The spanning tree representation can also be obtained simply by using an equivalent method with linear algebra and cones, without the need of graph combinatorics. It is called the Jeffrey-Kirwan~(JK) residue method, originally proposed in Ref.~\cite{Jeffrey:1993cun}. 

Let us first motivate how a fan naturally appears from Feynman energy integration. We already discussed it briefly at the beginning of Sec.~\ref{sec:CausalRepresentation}. According to Eqs. (\ref{eq:MomentaAssig}) and (\ref{eq:DenomGf+-}), denominators are of the form
\begin{equation}
    G_F^{\kappa}(q_e)=\frac{1}{\tilde{\beta}_e^\kappa . \ell_{0}+\alpha_e^\kappa - i\epsilon_e^\kappa},
\end{equation}
with $\kappa\in\{+,-\}$, $\tilde{\beta}_e^\kappa=-\kappa\beta_e \in \{ -1,0,1 \}^L$, $\ell_0=(\ell_{1,0}\, \, \dots \, \, \ell_{L,0})$ and some $\alpha_e^\kappa \in \mathbb{R}$.

The $2E$ vectors $\tilde{\beta}_e^\kappa$ are called \textit{charges}, and we call $C$ the vector configuration of all these charges. Note that, for the denominator $G_F^{\kappa}(q_e)$ to vanish, we require $\mathrm{Im}(\tilde{\beta}_e^\kappa \ell_{0})=\epsilon_e^\kappa >0$. In fact, here we should view $\mathbb{R}^L$ as "imaginary loop space" $\mathrm{Im}(\mathbb{C}^L)$ along which contour integration deformations occur in the complexified loop momenta space $\mathbb{C}^L$. The charge vectors indicate the directions along which the poles of the integrand lie.

We now choose a vector $\eta\in \mathbb{R}^L$, which should be seen as the deformation direction of the integration contour $\gamma$. In fact, $\gamma$ lies, before deformation, in real space, i.e. at the origin $O\in\mathbb{R}^L$. The result of the integral can be found by summing all the iterated residues corresponding to poles that $\gamma$ crosses until it reaches infinity, where it finally vanishes, unless there is a pole at infinity. We explained at the beginning of Sec.~\ref{sec:CausalRepresentation} that terms corresponding to cyclic configurations vanish because in those cases the integration contour can be deformed to infinity without crossing any pole. We need, for the whole integrand, a condition that tells us which poles $\gamma$ encounters depending on the direction given by $\eta$.

Note that the charges form a Gale dual realization of the vector configuration given by the row space of $B_{\vec{\Gamma}}$, passing thus from the "vertex space" $\mathbb{R}^{N-1}$ in which the graphical zonotope lives to "loop space" $\mathbb{R}^L$. We explained this at the end of Sec.~\ref{sec:ContourIntegralRep}. The Gale dual of the vector configuration defined by the incidence matrix $B_{\vec{\gamma}}$ of some directed graph $\vec{\Gamma}$, with associated sign vector $\sigma \in \{+,-\}^E$ determining the orientation of its edges, is the vector configuration $\{ \tilde{\beta}_1^{\sigma_1},\dots,\tilde{\beta}_E^{\sigma_E} \}$. It corresponds to a given term of the causal representation in \Eq{eq:ContourIntegralCausal2}.

This basis $\{ \tilde{\beta}_1^{\sigma_1},\dots,\tilde{\beta}_E^{\sigma_E} \}$ is a realization of the cographic oriented matroid $\mathcal{M}(\Gamma)$. Let $\mathcal{B}(C)$ be the set of bases of $C$, i.e. the collections of linearly independent vectors in $C$ that span $\mathbb{R}^L$. These correspond to bases of the cographic matroid $\mathcal{M}(\Gamma)$. Each basis $A=\{\tilde{\beta}_1^A,\dots,\tilde{\beta}_L^A\}\in \mathcal{B}(C)$, defines the top-dimensional simplicial cone $\sum_{i=1}^L a_i \tilde{\beta}_i^A \in \mathbb{R}^L$, with $a_i\geq0$, and such that all coefficients do not vanish simultaneously. 

On the other hand, each of these bases corresponds to a subset of $L$ linearly independent denominators $G_F^{\kappa}(q_e)$. Therefore, they intersect transversely in an isolated point in $\mathbb{C}^L$, which can trap $\gamma$ at some given values of the external parameters. They contribute to the amplitude via $L$-fold iterated residues. Then, these multidimensional residues are controlled by bases of the cographic oriented matroid, i.e. simplicial cones in the fan. But an integration contour $\gamma$ deformed in the direction of $\eta$ only encounters poles corresponding to the simplicial cones that contain $\eta$. Consequently, $\gamma$ only crosses these residues when deformed to infinity in the direction of $\eta$.

This is exactly what JK residues describe~\cite{Ferro:2018vpf}. We start from the rational differential form
\begin{equation}
    \frac{d\ell_{1,0}\wedge \dots \wedge d\ell_{L,0}}{\prod_{e=1}^E\left((\tilde{\beta}_e^+ . \ell_{0}+\alpha_e - i\epsilon_e^+)(\tilde{\beta}_e^- . \ell_{0}+\alpha_e - i\epsilon_e^-)\right)}.
\end{equation}
For generic values of the external parameters, all denominators are generic, meaning that any $L$ denominators whose charge vectors form a basis of $\mathbb{R}^L$ intersect transversely at a unique point, whereas $L+1$ generic denominators have no common intersection. We also choose $\eta$ not to lie in the boundary of any cone in $\mathcal{B}(C)$.

The integration of the differential form over $\gamma$ gives~\cite{Ferro:2018vpf}
\begin{align}
    \mathrm{JKRes}^{C, \eta} \omega=\sum_{A \in \mathcal{B}(C)} \mathrm{JKRes}_{\ell_0=\ell_{0_A}}^{C, \eta} \omega,
    \label{eq:JKRes1}
\end{align}
with
\begin{align}
    &\mathrm{JKRes}_{\ell_0=\ell_{0_A}}^{C, \eta} \omega=\left\{\begin{array}{ll}
     \frac{1}{\prod_{\tilde{\beta}_e^\kappa \notin A} \left(\tilde{\beta}_e^\kappa .\ell_{0_A}+\alpha_e^\kappa- \imath \epsilon_e^\kappa\right)}, & \eta \in \mathrm{Cone}_A \\
    0, & \text { otherwise }
    \end{array} ,\right.
    \label{eq:JKRes2}
\end{align}
with $\kappa \in \{-,+\}$. In the expression for $\mathrm{JKRes}_{\ell_0=\ell_{0_A}}^{C, \eta} \omega$, a prefactor $\frac{1}{\left|\operatorname{det}\left(\tilde{\beta}_1^A \ldots \tilde{\beta}_L^A\right)\right|}$ appears in general. It does not in our case, because this fundamental cycle matrix is totally unimodular, so every nonsingular maximal minor is $\pm1$. 

A JK basis $A$ has $L$ linearly independent charges. The $L$ corresponding edges in the graph form a basis of the cographic matroid. Therefore, the complement is a basis of the graphic matroid, i.e. a spanning tree. Actually, a basis $A \in \mathcal{B}(C)$ defines an orientation for each edge $e \notin \mathcal{T}$, which coincides with the orientations of external edges in the externally oriented basis $\vec{B}_\mathcal{T}$. Moreover, one can see from the formula for JK residues that two vectors $\eta_1$ and  $\eta_2$ give the same form for the result if the subset of simplicial cones they belong to is the same. This defines an equivalence class of $\eta$ vectors. The subspaces of equivalent $\eta$ vectors create a subdivision of $\mathbb{R}^L$ into cones, that we call chambers. 

Two $\eta$ vectors from different chambers select different spanning-tree representations of the same original contour integral. Assuming that the contour deformation has no contribution at infinity, the total result in the sum in \Eq{eq:JKRes1} does not depend on $\eta$~\cite{Jeffrey:1993cun}. It turns out that each chamber corresponds to a different triangulation of $\mathcal{S}_\Gamma$, and they are therefore in bijection with acyclic cycle signatures of $\Gamma$. \Eq{eq:JKRes2} is the result of taking the iterated residues of the hyperplanes corresponding to each cone $A$ such that $\eta \in \mathrm{Cone}_A$. As we explained previously, these are the singularities that $\gamma$ may encounter when deformed in the direction of $\eta$.

The JK residues method is a fast way to find the spanning tree representation of the canonical form of $\mathcal{S}_\Gamma$ only from linear algebra. Avoiding combinatorial methods on graphs can be interesting for a systematic and efficient computation. Now let us discuss how to translate all of this into calculations with matrices. We need to encode the denominators of the integrand into a couple matrices $C$ and $m$. $C$ is simply the $L \times (2E)$ matrix of the vector configuration $C$, i.e. the matrix obtained from the $2E$ charges $\tilde{\beta}_e^\kappa \in \mathbb{R}^L$. Then, the entries of the vector $C^T \ell_0$, with $\ell_0=(\ell_{1,0}\, \, \dots \, \, \ell_{L,0})$, are the dependencies of each denominator on the loops' energies.

Columns from $1$ to $E$ are the vectors $\tilde{\beta}_e^+$, $e=1,\dots,E$, and columns from $E+1$ to $2E$ are vectors $\tilde{\beta}_e^-$. $m$ is a row vector with $2E$ entries, where $(m)_i=y_i-(q_{i,0}-(C^T \ell_0)_i)$ and $(m)_{E+i}=y_i+q_{i,0}-(C^T \ell_0)_i$ for $i=1,\dots,E$. Here $q_{i,0}-(C^T \ell_0)_i$ are simply linear combinations of $x_i$ variables which appear in the expression for the energies of internal particles and depend on the assignment of loop momenta to the Feynman diagram.
Then, we have 
\begin{equation}
    G_F^{+}(q_i)=\frac{1}{(C^T \ell + m)_i}, \quad G_F^{-}(q_i)=\frac{1}{(C^T \ell + m)_{E+i}}.
\end{equation}
Once the vector $\eta$ is chosen, the first step is to determine in which simplicial cones of the fan generated by $C$ it is contained. Simplicial top-dimensional cones are generated by $L\times L$ minors of $C$, $C_A=\left[\begin{array}{ccc}\mid & & \mid \\ C_{i_1} & \cdots & C_{i_L} \\ \mid & & \mid\end{array}\right]$, with the subset $A=\{C_{i_1},\dots,C_{i_L}\}$. We also use $A$ to denote the cone generated by this subset. Then, $\eta$ belongs to $A$ if and only if there exists a vector $d\neq (0 \, \dots \, 0 )$ with all components nonnegative such that $\eta=C_A d$. For this to happen, $C_A$ needs to be invertible and then one checks whether all the components in $(C_A)^{-1}\eta$ are positive. This can be applied to all $2E \choose L $ minors of $C$ to get the desired cones.

This simplicial cone is related to a term in the spanning tree representation coming from a pole in the intersection of hyperplanes $\{ (  C^T \ell + m)_{i_1}=0 \}\cap \dots \cap \{ (  C^T \ell+ m)_{i_L}=0 \}$. This gives a linear system of equations with variables $\ell_{1,0},\dots,\ell_{L,0}$, that we can solve for: $\ell_{0_A}\equiv-(C_A^T)^{-1}m_A$, with $(m_A)_j=(m)_{i_j}$. We finally get the corresponding term in the spanning tree representation by evaluating the denominators $(C^T\ell+m)_k$ with $k\neq i_1,\dots,i_L$ at $\ell_0=\ell_{0_A}$, as in Eq. (\ref{eq:JKRes2}).

\textit{Example} : We study the same 2-loop graph as before with the following momenta assignment.

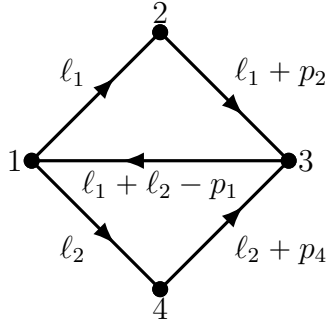
\begin{equation}
    \begin{tikzpicture}
        \coordinate (A) at (-1.7,0);
        \coordinate (B) at (0,1.7);
        \coordinate (C) at (1.7,0);
        \coordinate (D) at (0,-1.7);
        
        \fill (A) circle[radius=3pt];
        \fill (B) circle[radius=3pt];
        \fill (C) circle[radius=3pt];
        \fill (D) circle[radius=3pt];
        \node[font=\large, anchor=east]  at (A) {1};
        \node[font=\large, anchor=south] at (B) {2};
        \node[font=\large, anchor=west]  at (C) {3};
        \node[font=\large, anchor=north] at (D) {4};
    
        \fill (A) circle[radius=3pt];
        \fill (B) circle[radius=3pt];
        \fill (C) circle[radius=3pt];
        \fill (D) circle[radius=3pt];
    
    
        \draw[font=\large,->-,very thick] (B)--(C)
            node[midway, above right] {$\ell_1+p_2$};
    
        \draw[font=\large,->-,very thick] (A)--(B)
            node[midway, above left] {$\ell_1$};
    
        \draw[font=\large,->-,very thick] (A)--(D)
            node[midway, below left] {$\ell_2$};
    
        \draw[font=\large,->-,very thick] (D)--(C)
            node[midway, below right] {$\ell_2+p_4$};
    
        \draw[font=\large,->-,very thick] (C)--(A)
            node[midway, below] {$\ell_1+\ell_2-p_1$};
            
    \end{tikzpicture} \notag
\end{equation}

With the ordering of edges $\{(12),(23),(14),(34),(13)\}$, we have the matrices
\begin{align}
    C&=\begin{pmatrix}
        -1 & -1 & 0 & 0 & -1 & 1 & 1 & 0 & 0 & 1 \\
        0 & 0 & -1 & -1 & -1 & 0 & 0 & 1 & 1 & 1
    \end{pmatrix} \\
    m&=(y_{12} \, \, \, y_{23}-x_2\, \, \, y_{14}\, \, \, y_{34}-x_4\, \, \, y_{13}+x_1 \, \, \, y_{12} \, \, \, y_{23}+x_2\, \, \, y_{14}\, \, \, y_{34}+x_4\, \, \, y_{13}-x_1).
\end{align}
We make the choice $\eta=(-1\,\,\,1)$. $\eta$ belongs to eight cones, $\{C_1,C_8\}$, $\{C_1,C_9\}$, $\{C_2,C_8\}$, $\{C_2,C_9\}$, $\{C_1,C_{10}\}$, $\{C_2,C_{10}\}$, $\{C_5,C_8\}$ and $\{C_5,C_9\}$.

Let us choose the subset $A=\{C_5,C_8\}$, with minor $C_A=\begin{pmatrix}
         -1  & 0  \\
         -1 &  1 
    \end{pmatrix}$, and $m_A=(y_{13}+x_1 \, \, \, y_{14})$. We get 
\begin{equation}
(C_A)^{-1}\eta=\begin{pmatrix}
         1  \\
         2 
    \end{pmatrix},
\end{equation}
so that $\eta \in A$. Then, 
\begin{equation}
    \ell_{0_A}=-(C_A^T)^{-1}m_A=\begin{pmatrix}
         y_{13}+y_{14}+x_1  \\
         -y_{14}
    \end{pmatrix}.
\end{equation}
Applying \Eq{eq:JKRes2}, the resulting term of the spanning tree representation is 
\begin{equation}
    \frac{1}{4y_{13}y_{14}}\frac{1}{(y_{12}^2-(y_{13}+y_{14}+x_1)^2)(y_{23}^2-(y_{13}+y_{14}+x_1+x_2)^2)(y_{34}^2-(y_{14}-x_4)^2)},
\end{equation}
which is equal to the term obtained in \Eq{eq:ExampleTesmExternalAtlas} up to the transformation $x_i \leftrightarrow -x_i$, corresponding to a time reversal of the scattering process. One can in fact see that this transformation on a fine zonotopal tiling also gives a fine zonotopal tiling. This means that this transformation relates two different simplices in triangulations (or the same triangulation) of $\mathcal{S}_\Gamma$. In fact the choice of $\eta$ vector here gives the triangulating external atlas obtained from the one in (\ref{eq:TriangulatingExternalAtlas}) by reversing the orientation of the non-tree edges.

\section{Conclusions and outlook}

The geometrical description of the SF described in this article introduces novel, efficient tools to understand its combinatorics in a simpler way. The correspondence between the canonical form of the SF and the scalar Feynman energy integral provides useful information on multiloop Feynman integrals, which is closely related to the LTD causal representation of Feynman integrals and scattering amplitudes. The method proposed in this article bootstraps both the causal representation and the spanning tree representations of LTD simply by applying a combinatorial approach to graphs. 

The Cayley polytope nature of the SF underlies its structure as a base space on the on-shell internal energies $y_e$, which parametrizes a family of deformations of the graphical zonotope of the Feynman diagram. The face structure of the graphical zonotope reflects the factorization channels and the iterated residues of the Feynman energy integral.

Applying known mathematical results on Cayley polytopes allowed us first to study the triangulations of the SF from the fine graphical zonotopal tilings, and then to characterize these tilings and find the spanning-tree representation of the canonical form. The drawback of this expression is the presence of denominators corresponding to internal facets in the zonotopal tiling, which appear as spurious poles of the integrand. The causal representation avoids this, as it is obtained from the normal fan of the graphical zonotope. Two methods to get the expression were proposed, and they result into efficient algorithms.

A possible future direction beyond the work carried out in this article is the study of non-scalar Feynman energy integrals. It would be useful to find a method based on the geometry of the SF to obtain the non-scalar LTD expression without spurious poles.

Moreover, positive geometries offer the possibility of treating jointly in one geometry several Feynman diagrams with a certain number of loops. The loop cosmohedron, recently introduced in Ref.~\cite{Arkani-Hamed:2024jbp}, encodes the full cosmological wavefunction in $\mathrm{Tr}(\phi^3)$ theory in one polytope. Surfaceology~\cite{Arkani-Hamed:2023lbd} also generalizes associahedra to loop level scattering amplitudes. It would be interesting to study the total energy residue of the cosmohedron as well as its possible representations. The possibility of deriving compact LTD expressions for scattering amplitudes of physical processes could be very impactful for precision calculations in particle physics. Geometrical structures similars to those described in this article may actually appear in more complicated geometries and facilitate obtaining their canonical forms. For example, deformed zonotopal generalized permutohedra appear in the study of the ABHY associahedron~\cite{Early:2018zuw}.

Therefore, the geometric study of the SF provides an efficent method to describe scalar Feynman energy integral based on combinatorics. In addition, it is also a first step for further work that could be of interest to the calculation of multiloop Feynman integrals and scattering amplitudes.

\section{Acknowledgements}

This work is supported by the Spanish Government and ERDF/EU - Agencia Estatal de Investigaci\'on MCIN/AEI/10.13039/501100011033,  Grants No. PID2023-146220NB-I00, No. EUR2025-164820, and No. CEX2023-001292-S. 
The work of JRS is funded by AEI, Grant No. PREP2023‐001474.

\begin{appendices}

\section{Graphic and cographic oriented matroids \label{app:OrientedMatroids}}

See, for example, Ref.~\cite{Reiner2005Matroids} for an introduction to matroids and oriented matroids.

\subsection{Matroids}

Matroids generalize the concept of linear dependencies between vectors in a vector space. A \textit{matroid} $\mathcal{M}$ is a pair $(\mathsf{E},\mathcal{I})$, where $\mathsf{E}$ is the \textit{ground} set and $\mathcal{I}$ is a collection of subsets of $\mathsf{E}$, called \textit{independent sets}, that satisfy
\begin{enumerate}
    \item 
 $\emptyset\in \mathcal{I}$.

\item
 If $I\in \mathcal{I}$ and $J \subseteq I$, then $J\in \mathcal{I}$.

\item
 If $A,B\in \mathcal{I}$ and $|A|<|B|$, then there exists $x\in B\backslash A$ such that $A \cup \{x\} \in \mathcal{I}$.
\end{enumerate}
The subsets of $\mathsf{E}$ that are not independent are called dependent sets.
\textit{Bases} of a matroid $\mathcal{M}=(\mathsf{E},\mathcal{I})$, also called \textit{maximal independent sets}, are independent sets $B$ such that $B\cup\{x\}$ is dependent for any $x\in \mathsf{E}\backslash B$.
A \textit{circuit} $C$ of $\mathcal{M}$ is a minimal dependent subset of $\mathsf{E}$, i.e. $I$ is independent for any $I\subsetneq C$. 

One can define a matroid by its collection of bases $\mathcal{B}$, $\mathcal{M}=(\mathsf{E},\mathcal{B})$, or by its collection of circuits $\mathcal{C}$, $\mathcal{M}=(\mathsf{E},\mathcal{C})$, instead of by the independent sets. $\mathcal{B}$ is a collection of subsets of $\mathsf{E}$ such that
\begin{enumerate}
\item  $\mathcal{B}$ is nonempty.
\item
If $B_1,B_2 \in \mathcal{B}$ and $x_1\in B_1\backslash B_2$, then there exists $x_2 \in B_2\backslash B_1$ such that $(B_1 \backslash\{x_1\})\cup \{x_2\}\in \mathcal{B}$.
\end{enumerate}
From a matroid $\mathcal{M}=(\mathsf{E},\mathcal{B})$ defined from its set of  bases $\mathcal{B}$, there exists a \textit{dual matroid} $\mathcal{M}^\ast=(\mathsf{E},\mathcal{B}^\ast)$, such that $\mathcal B^\ast=\{\mathsf{E}\setminus B \mid B\in\mathcal B\}$ with $(\mathcal{M}^\ast)^\ast=\mathcal{M}$.
Then, $I$ is an independent set in $\mathcal{M}^\ast$ if and only if $\mathsf{E}\backslash I$ spans $\mathcal{M}$ (it contains a basis of $\mathcal{M}$). The circuits of the dual $\mathcal{M}^\ast$ are called the \textit{cocircuits} of $\mathcal{M}$. Conversely, the cocircuits of $\mathcal{M}^\ast$ are the circuits of $\mathcal{M}$.

A matroid $\mathcal{M}$ is called \textit{representable} if it is isomorphic to a \textit{vector matroid}, i.e. a matroid whose ground set is a finite family of vectors $\mathsf{E}$ in a vector space $V$ and whose independent sets are the linearly independent subsets of $\mathsf{E}$ in $V$.
We are especially interested in the matroids defined from an undirected connected graph $\Gamma$, the \textit{graphic matroid} $\mathcal{M}_\Gamma$ and its dual $\mathcal{M}_\Gamma^\ast$, the \textit{cographic matroid}. Their ground set is the set of edges of $\Gamma$: $\mathsf{E}=E(\Gamma)$. 
The independent sets of $\mathcal{M}_\Gamma$ are the forests of $\Gamma$. One can see that this satisfies the previously mentioned conditions. Then, the bases of $\mathcal{M}_\Gamma$ are its spanning trees and its circuits correspond to the simple cycles of $\Gamma$.

The incidence matrix of a directed graph $\vec{\Gamma}$ with $E$ edges and $N$ vertices is an $N\times E$  matrix $B$ with entries
\begin{equation}
B_{i j}=\left\{\begin{aligned}
1 & \text { if edge } e_j \text { is incoming to vertex } v_i \\
-1 & \text { if edge } e_j \text { is outgoing from vertex } v_i \\
0 & \text { otherwise. }
\end{aligned}\right.
\end{equation}
It has rank $N-c$, where $c$ is the number of connected components of $\Gamma$.
The graphic matroid $\mathcal{M}_\Gamma$ is represented by the set of vectors forming the columns of the incidence matrix of any orientation $\vec{\Gamma}$ of $\Gamma$. A subset of edges inducing a subgraph that contains a cycle corresponds to a subset of columns of $B$ that are linearly dependent.

The cycle space of $\Gamma$ is $\ker B$, of dimension $L$. Consider a $L \times E$ matrix $\beta$, called a cycle matrix, satisfying 
\begin{equation}
    B\beta^T=0.
\end{equation}
Its row space is $\ker B$. Then, the $E$ columns of $\beta$ give a representation of $\mathcal{M}_\Gamma^\ast$.
The circuits of $\mathcal{M}_\Gamma^\ast$, i.e. the cocircuits of $\mathcal{M}_\Gamma$, are the subsets of edges cut by minimal cuts of the graph. Here, by minimal cuts we mean the cuts of edges inducing a partition of the vertex set of the graph into two subsets, both inducing connected subgraphs. 
$\mathcal{M}_\Gamma^\ast$ is represented by the columns of any matrix whose row space is $\ker B$. Additionally, there is a theorem by Tutte stating that the dual matroid
$\mathcal M_\Gamma^\ast$ is graphic if and only if $\Gamma$
is planar. In that case, $\mathcal M_\Gamma^\ast \cong \mathcal M_{\Gamma^\ast}$, where $\Gamma^\ast$ is a planar dual graph of $\Gamma$.

\subsection{Oriented matroids}

A \textit{signed set} $X$ is a subset $\underline{X}$ of the ground set $\mathsf{E}$, called the \textit{support} of $X$, along with a partition of it into two subsets $\underline{X}=X^+\sqcup X^-$, called the positive and negative elements respectively. By assigning "$0$" to the elements in $X^0 \equiv \mathsf{E}\backslash \underline{X}$ and "$+$" and "$-$" to the elements in $X^+$ and $X^-$ respectively, $X$ can be written as an element of $\{+,-,0\}^\mathsf{E}$. We call the \textit{opposite} of $X$, $-X$, the signed set such that $(-X)^+=X^-$ and $(-X)^-=X^+$.

A way to define an oriented matroid $\mathcal{M}$ is via a collection $\mathcal{C}$ of signed sets on $\mathsf{E}$ called the \textit{signed circuits} of $\mathcal{M}$ that satisfy
\begin{enumerate}
\item $0^E \notin \mathcal{C}$.
\item
$-\mathcal{C}=\mathcal{C}$ (the opposite of a signed circuit is also a signed circuit).
\item
 For all $C_1,C_2 \in \mathcal{C}$ with $\underline{C_1} \subseteq \underline{C_2}$, then $C_1=\pm C_2$.
\item
 For all $C_1,C_2 \in \mathcal{C}$, with $C_1\neq\pm C_2$ and $e \in C_1^+ \cap C_2^-$, then there exists $C_3 \in \mathcal{C}$ such that $C_3^+ \subseteq (C_1^+ \cup C_2^+)\backslash \{e\}$ and $C_3^- \subseteq (C_1^- \cup C_2^-)\backslash \{e\}$.
\end{enumerate}

We can extend vector matroids to \textit{vector oriented matroids} that encode the linear dependence of a finite family of vectors $\mathsf{E}$ in a vector space $V$ taking into account the signs of the coefficients. For a set of vectors $\{v_1,\dots,v_n\}$ consider 
\begin{equation}
    \sum_{i=1}^n \lambda_i v_i=0~.
\end{equation}
Then any minimal linear dependence solving this equation defines a signed circuit such that $X^+=\{i \mid \lambda_i>0\}$ and $X^-=\{i \mid \lambda_i<0\}$. This collection of signed circuits defines the vector oriented matroid. An oriented matroid that can be realized in this way is called \textit{representable}. 

 Two signed sets are orthogonal, denoted $X \perp Y$, if $\underline{X} \cap \underline{Y}=\emptyset$ or if $X\mid_{\underline{X}\cap \underline{Y}}$ and $Y\mid_{\underline{X}\cap \underline{Y}}$ are neither opposite nor equal. It is known that there is a unique oriented matroid $\mathcal{M}^\ast=(\mathsf{E},\mathcal{C}^\ast)$, called the \textit{dual oriented matroid} of $\mathcal{M}$, whose signed circuits are the support-minimal nonzero signed sets $D$ satisfying $D \perp C$ for all $C\in \mathcal{C}$. We have $(\mathcal{M^\ast})^\ast=\mathcal{M}$. The elements of $\mathcal{C}^\ast$, circuits of $\mathcal{M}^\ast$, are called the \textit{cocircuits} of $\mathcal{M}$.

  Let $X_1$ and $X_2$ be two signed sets. Their \textit{composition} is defined to be $X_1 \circ X_2$ such that $(X_1 \circ X_2)^+=X_1^+ \cup (X_2^+ \backslash \underline{X_1})$, $(X_1 \circ X_2)^-=X_1^- \cup (X_2^- \backslash \underline{X_1})$. Then any composition of signed circuits (resp. signed cocircuits) of $\mathcal{M}_\Gamma$ is called a \textit{signed vector} (resp. \textit{signed covector}). For loopless oriented matroids, maximal covectors, i.e. covectors such that $\underline{X}=\mathsf{E}$, are called \textit{topes}. 

 To define the graphic oriented matroid $\mathcal{M}_\Gamma$, one first picks a reference orientation $\vec{\Gamma}$ of $\Gamma$. Then to every directed cycle $\vec{C}$ (a cycle with a cyclic orientation) that one can obtain from $\Gamma$, we associate a signed circuit. It is the sign vector $X$ such that $\underline{X}$ is the set of edges in $\vec{C}$ and an element of $X^+$ (resp $X^-$) is an edge in $\vec{C}$ whose orientation agrees (resp. disagrees) with the orientation of that edge in $\vec{\Gamma}$. The resulting collection of signed circuits $\mathcal{C}$ satisfies the requirements mentioned above. 

 In the cographic case, the signed circuits of $\mathcal{M}^\ast_\Gamma$ (signed cocircuits of $\mathcal{M}_\Gamma$) correspond to the bonds of $\Gamma$, i.e., minimal cuts such that all cut edges are oriented from a partition subset to the other. For some $S \subset V(\Gamma)$, a bond corresponds to orienting all the edges in the graph which have one endpoint in $S$ and one in $ V\backslash S$ from $S$ to $ V\backslash S$. Comparing this partial orientation of the graph with the fixed reference orientation gives the corresponding signed cocircuit.
 The topes of $\mathcal{M}_\Gamma$ are in bijection with the acyclic orientations of $\Gamma$, the maximal signed covectors containing no directed cycles. 

We summarize in the following table how the concepts in the theory of matroids and oriented matroids in the graphic case translate into the language of graphs for a connected graph.

\begin{table}[h]
\centering
\begin{tabular}{c|c}
\textbf{Graphic (oriented) matroid $\mathcal{M}_\Gamma$} & \textbf{Graph-theoretic correspondence} \\
\hline
Independent set & Forest \\[0.2em]
Basis & Spanning tree \\[0.2em]
Circuit & Simple cycle (loop) \\[0.2em]
Cocircuit & Minimal cut \\[0.2em]
Signed circuit & Directed simple cycle (loop with cyclic orientation) \\[0.2em]
Signed cocircuit & Oriented bond \\[0.2em]
Signed covector & Ordered vertex partition (not always binary) of the graph  \\[0.2em]
Tope & Acyclic orientation
\end{tabular}
\label{tab:graphic_matroid_dictionary}
\end{table}

\section{Benchmark multiloop graphs \label{app:Examples}}

 \subsection{Multiloop Banana graph}

 Let us study first the geometry related to the multiloop graph with $2$ vertices connected by $E=L+1$ edges, the so-called banana graph or Maximal Loop Topology~(MLT)~\cite{Aguilera-Verdugo:2020set}. We denote by $\mathcal{B}_L$ the banana graph with $L$ loops.

 
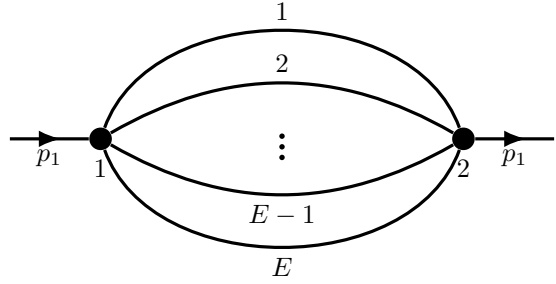
\begin{equation}
    \begin{tikzpicture}[scale=0.8]
    
    \node[circle,fill=black,inner sep=3pt,label=below:$1$] (A) at (0,0) {};
    \node[circle,fill=black,inner sep=3pt,label=below:$2$] (B) at (6,0) {};
    
    \draw[bend right=70, very thick] (A) to node[midway,below] {$E$} (B);
    \draw[bend right=30, very thick] (A) to node[midway,below] {$E-1$} (B);
    
    \node at (3,0) {\Huge $\vdots$};    
    \draw[bend left=30, very thick] (A) to node[midway,above] {$2$} (B);
    \draw[bend left=70, very thick] (A) to node[midway,above] {$1$} (B);
    
    \draw[->-, very thick] (-1.5,0) -- (A)
        node[midway,below] {$p_1$};
    
    \draw[->-, very thick] (B) -- (7.5,0)
        node[midway,below] {$p_1$};
    \end{tikzpicture} \notag
 \end{equation}

We will consider all $y_e>0$ without loss of generality. The deformed graphical zonotope satisfying $\sum y_e=1$ is 
 \begin{equation}
     \mathcal{Z}_{\mathcal{B}_L}^Y=\sum_{e=1}^{E} y_e [\mathbf{x_1}-\mathbf{x_2},\mathbf{x_2}-\mathbf{x_1}]= [\mathbf{x_1}-\mathbf{x_2},\mathbf{x_2}-\mathbf{x_1}].
 \end{equation}
 Therefore the fiber $\mathcal{Z}_{\mathcal{B}_L}^Y=\mathcal{Z}_{\mathcal{B}_L}$ is independent of the point in the base simplex $\Delta^{E-1}$. It is a line segment. Then, $\mathcal{S}_{\mathcal{B}_L}$ factorizes into $\Delta^{E-1}\times \mathcal{Z}_{\mathcal{B}_L}$. So $\mathcal{S}_{\mathcal{B}_L}$ is a prism whose base is a simplex of dimension $L$.

 One can read the causal representation of its canonical function easily:
 \begin{equation}
     \tilde{\Omega}(\mathcal{S}_{\mathcal{B}_L})=\frac{1}{\prod_e (2y_e)} \tilde{\Omega}(\mathcal{Z}_{\mathcal{B}_L})=\frac{1}{\prod_e (2y_e)}\left(\frac{1}{1+x_1}+\frac{1}{1-x_1}\right)=\frac{1}{\prod_e (2y_e)}\left(\frac{1}{\sum_e y_e +x_1}+\frac{1}{\sum_e y_e -x_1}\right),
 \end{equation}
where in the last equality we recovered the homogeneous form of the canonical function to get the dependence on $\sum_e y_e$. We also used conservation of the total energy $x_1+x_2=0$ to work only with the $x_1$ coordinate on $\mathcal{Z}_{\mathcal{B}_L}$.
Each term corresponds to one of the two acyclic orientations of the banana graph. All propagators flowing from vertex $1$ to vertex $2$ and vice versa.

Now let us study the triangulations of the prism $\mathcal{S}_{\mathcal{B}_L}$. Resorting to the fine zonotopal tilings of $\mathcal{Z}_{\mathcal{B}_L}$ is the easiest way to do so. All fine zonotopal tilings of $\mathcal{Z}_{\mathcal{B}_L}$ correspond to concatenating the $E$ segments $y_e [\mathbf{x_1}-\mathbf{x_2},\mathbf{x_2}-\mathbf{x_1}]$ in any of the $E!$ possible orderings. 
Let us see this using the triangulating external atlases described in Sec.~\ref{eq:ExternalAtlas}. The first step is to get an acyclic cycle signature, in bijection with regular triangulations, that we know can be done with an ordering of edges of the graph, which corresponds to a permutation $P\in S_E$ on the original reference labelling of edges, $e\rightarrow P^{-1}(e)$. Each such permutation gives a regular triangulation of $\mathcal{S}_{\mathcal{B}_L}$. 

To visualize it better, imagine reordering the edges so that $P(1)$ is the edge above, then $P(2)$, and so on, up to $P(E)$ which is the edge on the bottom of the graph. We choose as reference orientation the acyclic orientation where edges are oriented from vertex $1$ to vertex $2$. Then the acyclic cycle signature is obtained for each simple cycle, by orienting from vertex $1$ to vertex $2$ the edge above and in the opposite direction the edge below. 

This gives us the rule to find the triangulating external atlas given by permutation $P$, which we denote by $\mathcal{A}_P$. Simplices of a triangulation are in bijection with edges of the graph, since those are the spanning trees of the banana graph. Then, the externally oriented basis $\vec{B}_t \in \mathcal{A}_P$ associated to edge $t$ is given by orienting all edges above it (edges $e'$ such that $P^{-1}(e')<P^{-1}(t)$ from vertex $1$ to vertex $2$, and all edges below it in the opposite direction, as shown below.

 
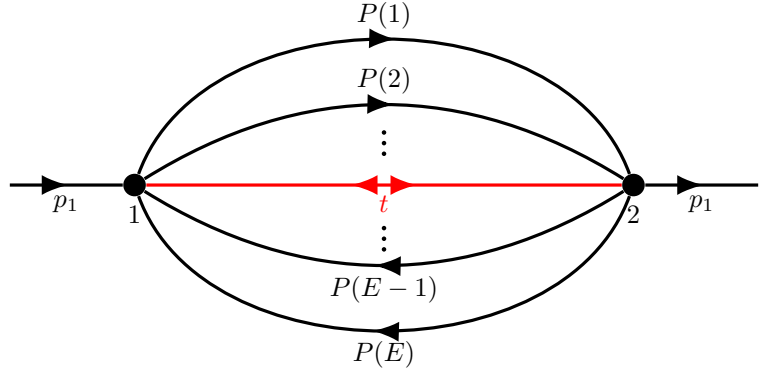
\begin{equation} 
    \begin{tikzpicture}[scale=1.1]
    
    \node[circle,fill=black,inner sep=3pt,label=below:$1$] (A) at (0,0) {};
    \node[circle,fill=black,inner sep=3pt,label=below:$2$] (B) at (6,0) {};
    
    \draw[->--=1.2,bend right=-70, very thick] (B) to node[midway,below] {$P(E)$} (A);
    \draw[->--=1.2,bend right=-32, very thick] (B) to node[midway,below] {$P(E-1)$} (A);
    
    \draw[double arrow-=1.2,red, very thick] (A) to node[midway,below] {$t$} (B);
    
    \node at (3,0.6) {\huge $\vdots$};
    \node at (3,-0.55) {\huge $\vdots$};
    
    \draw[->--=1.2,bend left=32, very thick] (A) to node[midway,above] {$P(2)$} (B);
    \draw[->--=1.2,bend left=70, very thick] (A) to node[midway,above] {$P(1)$} (B);
    
    \draw[->--=1.2, very thick] (-1.5,0) -- (A)
        node[midway,below] {$p_1$};
    
    \draw[->--=1.2, very thick] (B) -- (7.5,0)
        node[midway,below] {$p_1$};
    
    \end{tikzpicture} \notag
 \end{equation}

The corresponding fine tile $\tau_P^t$ is then 
\begin{equation}
    \tau_P^t=y_{t}[\mathbf{x_1}-\mathbf{x_2},\mathbf{x_2}-\mathbf{x_1}]+\left( \sum_{i=1}^{P^{-1}(t)-1} y_{P(i)} - \sum_{i=P^{-1}(t)+1}^{E} y_{P(i)} \right) (\mathbf{x_2}-\mathbf{x_1}).
\end{equation}
Such segments $\tau_P^t$ for all $t=1,\dots,E$ give a zonotopal tiling of $\mathcal{Z}_{\mathcal{B}_L}$, and, equivalently, a regular triangulation of $\mathcal{S}_{\mathcal{B}_L}$. One can easily read the simplex $\Delta_P^t$ of the prism that corresponds to $\tau_P^t$ from the corresponding externally oriented basis. We have
\begin{equation}
\Delta_P^t=
\operatorname{Conv}\!\left(
\{v_t^+,v_t^-\}
\cup
\{v_{P(i)}^+ : 1\leq i<P^{-1}(t)\}
\cup
\{v_{P(i)}^- : P^{-1}(t)<i\leq E\}
\right).
\end{equation}
where we define $\mathbf{v_e^+}= \mathbf{y_e}+\mathbf{x_2}-\mathbf{x_1} $ and $\mathbf{v_e^-}= \mathbf{y_e}+\mathbf{x_1}-\mathbf{x_2} $, the vertices of $\mathcal{S}_{\mathcal{B}_L}$. In fact $\mathrm{dim}(\mathcal{S}_{\mathcal{B}_L})=E$, so that the simplices of its triangulations are the convex hull of $E+1$ points.

\begin{figure}[h]
\centering

\begin{minipage}{0.48\textwidth}
\centering
    \begin{tikzpicture}[scale=5,line join=round,line cap=round]
        \coordinate (v1-) at (-0.57735,-0.1005);
        \coordinate (v2-) at (0.78868,-0.03679);
        \coordinate (v3-) at (-0.21132,0.13729);
        \coordinate (v1+) at (-0.57735,0.88423);
        \coordinate (v2+) at (0.78868,0.94794);
        \coordinate (v3+) at (-0.21132,1.12202);
        \coordinate (z+) at (-0.071325, 1.01321);
        \coordinate (z-) at (-0.071325, 0.02848);
        \coordinate (z1) at (-0.071325, 0.6192);
        \coordinate (z2) at (-0.071325, 0.3731);
        \coordinate (z++) at (-0.071325, 1.41321);
        \coordinate (z--) at (-0.071325, -0.25848);
        \fill[red!60,opacity=.7] (v3+)--(v1+)--(v3-)--(v2+)--cycle;
        \fill[violet!60,opacity=.85] (v2+)--(v2-)--(v3-)--(v1+)--cycle;
        \fill[blue!60,opacity=.7] (v1+)--(v2-)--(v1-)--cycle;
        \draw[dash pattern=on 3pt off 2pt,black!60,line width=.7pt] (v1-)--(v3-);
        \draw[dash pattern=on 3pt off 2pt,black!60,line width=.7pt] (v3-)--(v2-);
        \draw[dash pattern=on 3pt off 2pt,black!60,line width=.7pt] (v3+)--(v3-);
        \draw[dash pattern=on 3pt off 2pt,black!60,line width=.45pt] (v1+)--(v3-);
        \draw[dash pattern=on 3pt off 2pt,black!60,line width=.45pt] (v2+)--(v3-);
        \draw[red!70,line width=3pt] (z+)--(z1);
        \draw[violet!85,line width=3pt] (z1)--(z2);
        \draw[blue!80,line width=3pt] (z2)--(z-);
        \draw[dash pattern=on 3pt off 2pt,black!60,line width=.7pt] (z+)--(z++);
        \draw[dash pattern=on 3pt off 2pt,black!60,line width=.7pt] (z-)--(z--);
        \draw[black,line width=.9pt] (v2-)--(v1-);
        \draw[black,line width=.9pt] (v2+)--(v3+);
        \draw[black,line width=.9pt] (v3+)--(v1+);
        \draw[black,line width=.9pt] (v1+)--(v2+);
        \draw[black,line width=.9pt] (v1+)--(v1-);
        \draw[black,line width=.9pt] (v2-)--(v2+);
        \draw[dash pattern=on 3pt off 2pt,black!60,line width=.45pt] (v1+)--(v2-);
        \node[below right]  at (z++) {$(y_1,y_2,y_3)=(0.35,0.25,0.4)$};
        \node[above left]  at (v1+) {$\mathbf y_{1}+\mathbf x_1-\mathbf x_2$};
        \node[below]  at (v1-) {$\mathbf y_{1}+\mathbf x_2-\mathbf x_1$};
        \node[below right]  at (v2+) {$\mathbf y_{2}+\mathbf x_1-\mathbf x_2$};
        \node[below]  at (v2-) {$\mathbf y_{2}+\mathbf x_2-\mathbf x_1$};
        \node[above left]  at (v3+) {$\mathbf y_{3}+\mathbf x_1-\mathbf x_2$};
        \node[above left, align=center] at (v3-) 
        {$\mathbf y_3+$\\$\mathbf x_2-\mathbf x_1$};
        \fill (v1+) circle (.4pt);
        \fill (v2+) circle (.4pt);
        \fill (v3+) circle (.4pt);
        \fill (v1-) circle (.4pt);
        \fill (v2-) circle (.4pt);
        \fill (v3-) circle (.4pt);
        \node[font=\LARGE,left]  at (z++) {$\mathcal{S}_{\mathcal{B}_2}$};
        \fill (z+) circle (.4pt);
        \fill (z-) circle (.4pt);
        \fill (z1) circle (.4pt);
        \fill (z2) circle (.4pt);
    \end{tikzpicture}
    
\end{minipage}
\hfill
\begin{minipage}{0.48\textwidth}
\centering
    \begin{tikzpicture}[scale=5,line join=round,line cap=round]
        \coordinate (z+) at (-0.071325, 1.01321);
        \coordinate (z++) at (-0.071325, 1.11321);
        \coordinate (z-) at (-0.071325, 0.02848);
        \coordinate (z1) at (-0.071325, 0.6192);
        \coordinate (z2) at (-0.071325, 0.3731);
        \draw[red!70,line width=3pt] (z+) to node[midway,right] {$y_{3}\mathcal{Z}_{\mathcal{B}_2}$} (z1);
        \draw[violet!85,line width=3pt] (z1) to node[midway,right] {$y_{2}\mathcal{Z}_{\mathcal{B}_2}$} (z2);
        \draw[blue!80,line width=3pt] (z2) to node[midway,right] {$y_{1}\mathcal{Z}_{\mathcal{B}_2}$} (z-);
        \node[below]  at (z-) {$\mathbf x_2-\mathbf x_1$};
        \node[right]  at (z+) {$\mathbf x_1-\mathbf x_2$};
        \node[font=\LARGE,above]  at (z++) {$\mathcal{Z}_{\mathcal{B}_2}$};

        \fill (z+) circle (.4pt);
        \fill (z-) circle (.4pt);
        \fill (z1) circle (.4pt);
        \fill (z2) circle (.4pt);
        
    \end{tikzpicture}
\end{minipage}

\caption{Triangulation of $\mathcal{S}_{\mathcal{B}_2}$ given by the identity permutation on the edges of $\mathcal{B}_2$ (left). The fiber over each point of the simplex $\Delta^2$ is $\mathcal{Z}_{\mathcal{B}_2}$, illustrated with a subdivision corresponding to the triangulation of $\mathcal{S}_{\mathcal{B}_2}$ (right). }
\label{fig:Prisma}
\end{figure}
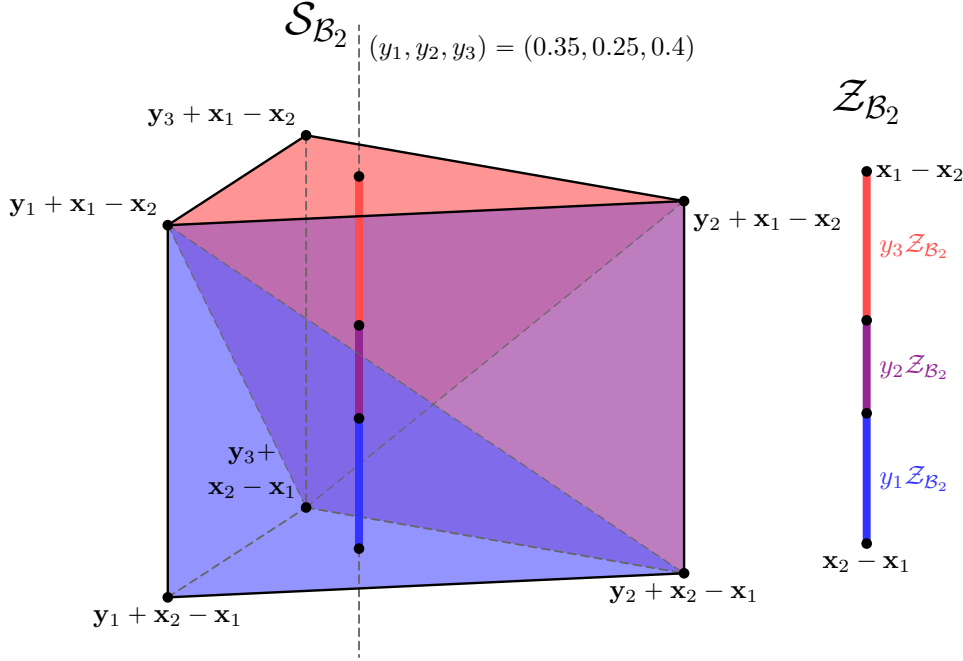

In Fig.~\ref{fig:Prisma}, we represented the triangulation of $\mathcal{S}_{\mathcal{B}_2}$ given by the identity permutation $Id$ on the edges of the graph. The fiber over an arbitrary point $(y_1,y_2,y_3)=(0.35,0.25,0.4)$ of the simplex $\Delta^2$ in edge space is the graphical zonotope $\mathcal{Z}_{\mathcal{B}_2}$, which is just the line segment $[\mathbf{x_1}-\mathbf{x_2},\mathbf{x_2}-\mathbf{x_1}]$. The intersection with the tetrahedra in the triangulation of $\mathcal{S}_{\mathcal{B}_2}$ gives a tiling of $\mathcal{Z}_{\mathcal{B}_2}$, each tile corresponding to an edge of $\mathcal{B}_2$, i.e. a spanning tree. In fact, $\tau_{Id}^i=\Delta_{Id}^i \mid_{(y_1,y_2,y_3)=(0.35,0.25,0.4)}$.
One can easily see that such a tiling of $\mathcal{Z}_{\mathcal{B}_2}$ is given by an ordering of edges of $\mathcal{B}_2$. Because regular triangulations of $\mathcal{S}_{\mathcal{B}_2}$ are in bijection with fine tilings of $\mathcal{Z}_{\mathcal{B}_2}$, each of them corresponds to a permutation in $S_3$. 

\subsection{Three-loop Mercedes with one extra propagator  }

Let us study the three-loop Mercedes graph with one extra propagator, which we denote by $\mathcal{G}$. This graph belongs to the category Next-to-Next to Maximal Loop Topology (N$^2$MLT) in Ref.~\cite{Aguilera_Verdugo_2021}.
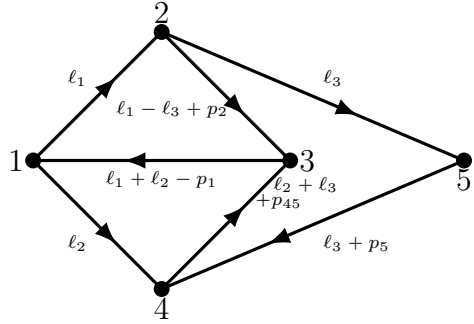
\begin{equation}
     \begin{tikzpicture}
    \coordinate (A) at (-1.7,0);
    \coordinate (B) at (0,1.7);
    \coordinate (C) at (1.7,0);
    \coordinate (D) at (0,-1.7);
    \coordinate (E) at (4,0);
    
    \fill (A) circle[radius=3pt];
    \fill (B) circle[radius=3pt];
    \fill (C) circle[radius=3pt];
    \fill (D) circle[radius=3pt];
    \fill (E) circle[radius=3pt];
    \node[font=\large, anchor=east]  at (A) {1};
    \node[font=\large, anchor=south] at (B) {2};
    \node[font=\large, anchor=west]  at (C) {3};
    \node[font=\large, anchor=north] at (D) {4};
    \node[font=\large, anchor=north] at (E) {5};


    \draw[font=\large,->-,very thick] (B)--(C)
        node[font=\scriptsize,pos=0.6, left] {$\ell_1-\ell_3+p_2$};

    \draw[font=\large,->-,very thick] (A)--(B)
        node[font=\scriptsize,midway, above left] {$\ell_1$};

    \draw[font=\large,->-,very thick] (A)--(D)
        node[font=\scriptsize,midway, below left] {$\ell_2$};

    \draw[font=\large,->-,very thick] (D)--(C)
        node[font=\scriptsize,pos=0.75,right,xshift=-5pt,align=left]{$\,\,\,\,\,\ell_2+\ell_3$\\[-3pt]$+p_{45}$};

    \draw[font=\large,->-,very thick] (C)--(A)
        node[font=\scriptsize,midway, below] {$\ell_1+\ell_2-p_{1}$};

    \draw[font=\large,->-,very thick] (B)--(E)
        node[font=\scriptsize,midway, above right] {$\ell_3$};

    \draw[font=\large,->-,very thick] (E)--(D)
        node[font=\scriptsize,midway, below right] {$\ell_3+p_5$};
        
    \end{tikzpicture} \notag
\end{equation}

The graphical zonotope $\mathcal{Z}_\mathcal{G}$ is $4$-dimensional. From the factorization property of graphical zonotopes one can illustrate its facets and faces. For example, $\mathcal{Z}_\mathcal{G}\cap L_{(1,2,3,4)}\cong \mathcal{Z}_\mathcal{G}\cap L_{(5)}\cong \mathcal{Z}_{\begin{tikzpicture}
    \BoxGraph[black][1.2]{0.3}{}{
    \draw (A)--(B);
    \draw (A)--(D);
    \draw (A)--(C);
    \draw (B)--(C);
    \draw (D)--(C);
    }{}
    \end{tikzpicture}}\times \mathcal{Z}_.=\mathcal{Z}_{\begin{tikzpicture}
    \BoxGraph[black][1.2]{0.3}{}{
    \draw (A)--(B);
    \draw (A)--(D);
    \draw (A)--(C);
    \draw (B)--(C);
    \draw (D)--(C);
    }{}
    \end{tikzpicture}}$
(remember that $L_{(\dots)}$ is the facet of the SF defined by $\{\lambda_{(\dots)}=0\}\cap \mathcal{S}_\Gamma$). A deformation of this graphical zonotope is illustrated in Fig.~\ref{fig:DeformedGZSquareChord}.  
Another example is given by $\mathcal{Z}_\mathcal{G}\cap L_{(1,2,3)}\cong\mathcal{Z}_\mathcal{G}\cap L_{(4,5)}\cong \mathcal{Z}_{\begin{tikzpicture}
    \TriGraph{(1.1,0.55)}{0.3}{
    \draw (A)--(B);
    \draw (A)--(C);
    \draw (B)--(C);
    }{}
\end{tikzpicture}} \times \mathcal{Z}_{\begin{tikzpicture}
    \coordinate (B) at (-0.1,0);
    \coordinate (A) at (0.1,0);
    \fill (A) circle (0.7pt);
    \fill (B) circle (0.7pt);
    \draw (A)--(B);
\end{tikzpicture}}$, which is a hexagonal prism.
$\mathcal{Z}_\mathcal{G}$ has $60$ vertices, corresponding to the $60$ acyclic orientations of $\mathcal{G}$. $36$ of them are simple vertices, $20$ vertices lie at the intersection of $5$ facets, while $4$ vertices lie at $6$ different facets.

\paragraph{Causal representation}

Consider the following acyclic orientation of $\mathcal{G}$, whose corresponding vertex in $\mathcal{Z}_\mathcal{G}$ we call $v$.

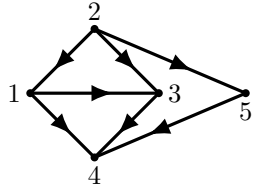
\begin{equation}
     \begin{tikzpicture}[scale=0.5]
    \coordinate (A) at (-1.7,0);
    \coordinate (B) at (0,1.7);
    \coordinate (C) at (1.7,0);
    \coordinate (D) at (0,-1.7);
    \coordinate (E) at (4,0);
    
    \fill (A) circle[radius=3pt];
    \fill (B) circle[radius=3pt];
    \fill (C) circle[radius=3pt];
    \fill (D) circle[radius=3pt];
    \fill (E) circle[radius=3pt];
    \node[anchor=east]  at (A) {1};
    \node[anchor=south] at (B) {2};
    \node[anchor=west]  at (C) {3};
    \node[anchor=north] at (D) {4};
    \node[anchor=north] at (E) {5};


    \draw[font=\large,->-,very thick] (B)--(C);
    \draw[font=\large,->-,very thick] (B)--(A);
    \draw[font=\large,->-,very thick] (A)--(D);
    \draw[font=\large,->-,very thick] (C)--(D);
    \draw[font=\large,->-,very thick] (A)--(C);
    \draw[font=\large,->-,very thick] (B)--(E);
    \draw[font=\large,->-,very thick] (E)--(D);
    \end{tikzpicture} \notag
\end{equation}
This orientation is compatible with causal thresholds $\lambda_{(2)},\lambda_{(1,2)},\lambda_{(1,2,3)},\lambda_{(2,5)},\lambda_{(1,2,5)}$ and $\lambda_{(1,2,3,5)}$. We consider them in this order. The circuits of hyperplanes (by abuse of language, we use $L_{(\dots)}$ to denote the corresponding hyperplane $\{\lambda_{(\dots)}=0\}$) are $\{L_{(2)}, L_{(1,2)}, L_{(2,5)}, L_{(1,2,5)}\}$, $\{L_{(1,2)}, L_{(1,2,3)}, L_{(1,2,5)}$, $ L_{(1,2,3,5)}\}$, and $\{L_{(2)}, L_{(1,2,3)}, L_{(2,5)}, L_{(1,2,3,5)}\}$, so the three broken circuits are obtained by removing the first element from each of these circuits.

The NBC bases are 
\begin{align}
B_1 &= \left\{L_{(2)},\,L_{(1,2)},\,L_{(1,2,3)},\,L_{(2,5)}\right\},\qquad B_2 = \left\{L_{(2)},\,L_{(1,2)},\,L_{(1,2,3)},\,L_{(1,2,5)}\right\},\notag\\
B_3 &= \left\{L_{(2)},\,L_{(1,2)},\,L_{(1,2,3)},\,L_{(1,2,3,5)}\right\}, \qquad B_4= \left\{L_{(2)},\,L_{(1,2)},\,L_{(2,5)},\,L_{(1,2,3,5)}\right\},\notag\\
B_5 &= \left\{L_{(2)},\,L_{(1,2)},\,L_{(1,2,5)},\,L_{(1,2,3,5)}\right\}, \qquad B_6= \left\{L_{(2)},\,L_{(1,2,3)},\,L_{(2,5)},\,L_{(1,2,5)}\right\},\notag \\
B_7 &= \left\{L_{(2)},\,L_{(2,5)},\,L_{(1,2,5)},\,L_{(1,2,3,5)}
\right\}.
\end{align}
Now we check whether the sequence of residues defined by these subsets of hyperplanes applied to $\Omega(\mathcal{Z}_\mathcal{G})$ vanishes or not. We look at the order of consecutive cuts, following rule 3 in Sec.~\ref{sec:RefinementFan}.
For example for $B_1$, $\mathcal{O}(\mathcal{C}(\lambda_{(2,5)})\cup\mathcal{C}(\lambda_{(1,2,3)}))=3$. Therefore
\begin{equation}                    \mathrm{Res}_{\lambda_{(2)}}\mathrm{Res}_{\lambda_{(1,2)}}\mathrm{Res}_{\lambda_{(1,2,3)}}\mathrm{Res}_{\lambda_{(2,5)}} \left( \tilde{\Omega}(\mathcal{Z}_\mathcal{G}) \right)=0.
\end{equation}
The only subsets of hyperplanes whose corresponding iterated residue does not vanish are $B_3$, $B_5$ and $B_7$. Finally, 
\begin{align}
    \tilde{\Omega}(\mathcal{S}_\mathcal{G})_v=\frac{1}{\prod_{e\in E(\mathcal{G})}  (2y_e)} \left( \frac{1}{\lambda_{(2)}\lambda_{(1,2)}\lambda_{(1,2,3)}\lambda_{(1,2,3,5)}} +\frac{1}{\lambda_{(2)}\lambda_{(1,2)}\lambda_{(1,2,5)}\lambda_{(1,2,3,5)}} \right. \nonumber \\ \left. +\frac{1}{\lambda_{(2)}\lambda_{(2,5)}\lambda_{(1,2,5)}\lambda_{(1,2,3,5)}}  \right),
    \label{eq:CausalExp3loop}.
\end{align}

From the point of view of subset compatibility studied in Sec.~\ref{sec:BondDecomposition}, we can see that, for all terms, the collection of causal thresholds in each of them correspond to nested subsets, and are thus compatible sets. Moreover, adding any other subset to any of these compatible sets would make them incompatible. For example, for the first term, adding the subset $\{2,5\}$ makes it incompatible with $\{1,2\}$, and adding $\{1,2,5\}$ makes it incompatible with $\{1,2,3\}$. These three compatible sets are maximal.

We can now get the unrefined form of Eq. (\ref{eq:CausalExp3loop}).
\begin{equation}
    \tilde{\Omega}(\mathcal{S}_\mathcal{G})_v=\frac{1}{\prod_{e\in E(\mathcal{G})}  (2y_e)}  \frac{\lambda_{(2,5)}\lambda_{(1,2,5)}+\lambda_{(2,5)}\lambda_{(1,2,3)}+\lambda_{(1,2)}\lambda_{(1,2,3)}}{\lambda_{(2)}\lambda_{(1,2)}\lambda_{(2,5)}\lambda_{(1,2,3)}\lambda_{(1,2,5)}\lambda_{(1,2,3,5)}} .
    \label{eq:Unrefined3loop}
\end{equation}

To illustrate the advantage of the unrefined expression, one can see that $\mathrm{Res}_{\lambda_{(2)}}\mathrm{Res}_{\lambda_{(1,2,3,5)}}$ applied individually on each term in \Eq{eq:CausalExp3loop} is non-vanishing, but, when applied to the whole sum in \Eq{eq:Unrefined3loop}, the residue vanishes because facets $L_{(2)}$ and $L_{(1,2,3,5)}$ meet at codimension $3$ in $\mathcal{S}_\mathcal{G}$. The unrefined form avoids the spurious higher-codimension singularities appearing in the sum over terms, and is thus better suited for numerical integration.

\paragraph{Spanning-tree representation}

We apply the linear algebra method introduced at the end of Sec.~(\ref{sec:JeffreyKirwan}) to illustrate the spanning-tree representation with the graph $\mathcal{G}$.
Consider the following edge ordering: $(12)$, $(23)$, $(14)$, $(34)$, $(13)$, $(25)$, $(45)$. Then, the $C$ matrix whose columns are the $\tilde{\mathcal{\beta}}_e^\kappa$ charges is:
\begin{equation}
C=
\left(
\begin{array}{ccccccc|ccccccc}
-1&-1& 0& 0&-1& 0& 0&
 1& 1& 0& 0& 1& 0& 0\\
 0& 0&-1&-1&-1& 0& 0&
 0& 0& 1& 1& 1& 0& 0\\
 0& 1& 0&-1& 0&-1&-1&
 0&-1& 0& 1& 0& 1& 1
\end{array}
\right).
\label{eq:Cmatrix3loop}
\end{equation}

We choose the generic vector $\eta=(-1,2,4)^T$. One could check that it is not on the boundary of any nonsingular basis cone.
Over all $14 \choose 3$ minors of $C$, there exist $24$ invertible minors $A$ such that $C_A^{-1} \eta$ has all its components strictly positive. Each of these minors corresponds to one of the $24$ spanning trees of $\mathcal{G}$. 
Take the minor
\begin{equation}
C_{(1,3,11)}=
\begin{pmatrix}
-1&0&0\\
0&-1&1\\
0&0&1
\end{pmatrix}.
\label{eq:CAmatrix3loop}
\end{equation}
where the subindex indicates the columns of $C$ appearing in the chosen minor. It corresponds to the charges $\{\tilde{\beta}_{(12)}^+,\tilde{\beta}_{(14)}^+,\tilde{\beta}_{(34)}^- \}$, so to the spanning tree $\mathcal{T}$ of $\mathcal{G}$ with edges $E(\mathcal{T})=\{(13),(23),(25),(45)\}$.

We have
\begin{equation}
C_{(1,3,11)}^{-1}\eta
=
\begin{pmatrix}
-1&0&0\\
0&-1&1\\
0&0&1
\end{pmatrix}^{-1}
\begin{pmatrix}
-1\\
2\\
4
\end{pmatrix}
=
\begin{pmatrix}
1\\
2\\
4
\end{pmatrix}
>0,
\label{eq:ConeTest3loop}
\end{equation}
so $\eta$ belongs to the cone generated by the charges $\{\tilde{\beta}_{(12)}^+,\tilde{\beta}_{(14)}^+,\tilde{\beta}_{(34)}^- \}$.
On the other hand,
\begin{align}
m^T=
\left(
y_{12},\,
y_{23}-x_2,\,
y_{14},\,
y_{34}-x_{45},\,
y_{13}+x_1,\,
y_{25},\,
y_{45}-x_5, \right.\\
\left. y_{12},\,
y_{23}+x_2,\,
y_{14},\,
y_{34}+x_{45},\,
y_{13}-x_1,\,
y_{25},\,
y_{45}+x_5
\right),
\label{eq:mvector3loop}
\end{align}
with $x_{45}=x_4+x_5$, and for the selected basis, 
\begin{equation}
m_{(1,3,11)}^T=
\left(
y_{12},\,
y_{14},\,
y_{34}+x_{45}
\right).
\label{eq:mA3loop}
\end{equation}

We can now solve for the value of the loop energies at the intersection of these three hyperplanes:
\begin{equation}
\ell_0^{(1,3,11)}=-\left(C_{(1,3,11)}^T\right)^{-1}m_{(1,3,11)} =-
\begin{pmatrix}
-1&0&0\\
0&-1&0\\
0&1&1
\end{pmatrix}^{-1}
\begin{pmatrix}
y_{12}\\
y_{14}\\
y_{34}+x_{45}
\end{pmatrix}
 =
\begin{pmatrix}
y_{12}\\
y_{14}\\
-y_{14}-y_{34}-x_{45}
\end{pmatrix}.
\label{eq:lA3loop}
\end{equation}
We can apply now \Eq{eq:JKRes2} to get the corresponding term in the spanning tree representation
\begin{align}
\frac{1}{(2y_{12})(2y_{14})(2y_{34})}&\frac{1}{y_{23}^2-\left(y_{12}+y_{14}+y_{34}+x_2+x_4+x_5\right)^2}\frac{1}{y_{13}^2-\left(y_{12}+y_{14}-x_1\right)^2}
\nonumber\\
\times&\frac{1}{y_{25}^2-\left(y_{14}+y_{34}+x_4+x_5\right)^2}\frac{1}{y_{45}^2-\left(
y_{14}+y_{34}+x_4\right)^2}.
\label{eq:JKterm3loop}
\end{align}
 
\end{appendices}

\bibliographystyle{JHEP}
\bibliography{references}

\end{document}